\documentclass[USenglish,onecolumn]{article}

\usepackage[utf8]{inputenc}
\usepackage[big]{dgruyter}
\usepackage{caption}
 \usepackage{amsmath}
\usepackage{amsfonts}
\usepackage{graphicx}
\usepackage{enumerate}
\usepackage{enumitem}
\usepackage{natbib}
\usepackage{tabularx}
\usepackage{url} 
\usepackage{multirow}
\usepackage{algorithm}
\usepackage{algpseudocode}
\usepackage{bibunits}
\usepackage[colorinlistoftodos,textwidth=2.2cm]{todonotes}

\usepackage{color, colortbl}
\definecolor{Gray}{gray}{0.85}
\usepackage{tikz}
\usetikzlibrary{shapes,decorations,arrows,calc,arrows.meta,fit,positioning}
\tikzset{
    -Latex,auto,node distance =1 cm and 2 cm,semithick,
    state/.style ={ellipse, draw, minimum width = 0.7 cm},
    point/.style = {circle, draw, inner sep=0.04cm,fill,node contents={}},
    bidirected/.style={Latex-Latex,dashed},
    el/.style = {inner sep=2pt, align=left, sloped}
}
\usepackage{float}
\usepackage{color,soul}

\newtheorem{theorem}{Theorem}

\newtheorem{lemma}[theorem]{Lemma}
\newtheorem{condition}{Condition}
\newtheorem{assumption}{Assumption}

\newcommand{\Real}{\mathbb{R}}

\newcommand{\Tra}{^{\sf T}} 
\newcommand{\Inv}{^{-1}}

\newcommand{\cs}{{\longrightarrow}} 
 
\newcommand{\cp}{\overset{p}{\longrightarrow}} 
\newcommand{\cd}{\overset{d}{\longrightarrow}}
\newcommand{\wc}{\mathrel{\leadsto}}

\newcommand{\op}{o_{{P}}}
\newcommand{\Op}{O_{{P}}}
\newcommand{\bbc}{\preceq}
\newcommand{\EP}{\mathbb{P}}
\newcommand{\EPn}{\mathbb{P}_{n}}
\newcommand{\EGn}{\mathbb{G}_{n}}

\newcommand{\ind}{\perp\!\!\!\!\perp}

\newcommand{\Olit}{\overline{L}_{i,t}}

\newcommand{\YpotDo}{Y_{i,t}(\overline{a}_{t-}^{0})}

\newcommand{\YkpotTo}{Y_{i,k,t}(\overline{a}_{t})}
\newcommand{\YkpotDo}{Y_{i,k,t}(\overline{a}_{t-}^{0})}

\newcommand{\Psiphatv}{\Psi(\psi_{0}, \hat{\vartheta})}

\begin{document}

  \articletype{Research Article{\hfill}Open Access}

  \author*[1]{Taekwon Hong}

\author[2]{Wenbin Lu}

\author[3]{Shu Yang}

\author[4]{Pulak Ghosh}

  \affil[1]{Department of Statistics, North Carolina State University,  Raleigh, NC 27695.; E-mail: taekwon$\_$hong@ncsu.edu}

  \affil[2]{Department of Statistics, North Carolina State University,  Raleigh, NC 27695.; E-mail: wlu4@ncsu.edu}

  \affil[3]{Department of Statistics, North Carolina State University,  Raleigh, NC 27695.; E-mail: syang24@ncsu.edu}
  
  \affil[4]{Decision Sciences \& Centre for Public Policy, Indian Institute of Management, Bangalore, India 560076.; E-mail: pulak.ghosh@iimb.ac.in}
  
  \title{\huge Multivariate Zero-Inflated Causal Model for Regional Mobility Restriction Effects on Consumer Spending}

  \runningtitle{}


  \begin{abstract}
{The COVID-19 pandemic presents challenges to both public health and the economy. Our objective is to examine how household expenditure, a significant component of private demand, reacts to changes in mobility. This investigation is crucial for developing policies that balance public health and the economic and social impacts. We utilize extensive scanner data from a major retail chain in India and Google mobility data to address this important question. However, there are a few challenges, including outcomes with excessive zeros and complicated correlations, time-varying confounding, and irregular observation times. We propose incorporating a multiplicative structural nested mean model with inverse intensity weighting techniques to tackle these challenges. Our framework allows semiparametric/nonparametric estimation for nuisance functions. The resulting rate doubly robust estimator enables the use of a conventional sandwich variance estimator without taking into account the variability introduced by these flexible estimation methods. We demonstrate the properties of our method theoretically and further validate it through simulation studies. Using the Indian consumer spending data and Google mobility data, our method reveals that the substantial reduction in mobility has a significant impact on consumers' fresh food expenditure.
}
\end{abstract}
  \keywords{Multiplicative structural nested mean model; Multivariate zero-inflated outcomes; Rate double robustness; Inverse intensity weighting}
   \classification[MSC 2020]{62D20}

  \journalname{Journal of Causal Inference}
\DOI{DOI}
  \startpage{1}
  \received{09-Apr-2024}
  \revised{21-Jan-2025, 02-Apr-2025, 16-Sep-2025}
  \accepted{01-Oct-2025}

  \journalyear{2025}
  \journalvolume{1}

\maketitle
\section{Introduction}
\label{sec:intro}

The COVID-19 pandemic has dual effects on public health and the economy \citep[]{Kraemer2020}, with research on measures like lockdowns yielding mixed findings. For example, while \cite{giffin2021} demonstrates the positive impact of reducing community mobility in curbing the spread of COVID-19, \cite{Anderson2020} emphasizes the direct and indirect economic consequences of such interventions. Most existing research focuses on overall changes in total spending \citep[e.g.,][]{Guerrieri2020, Eichenbaum2020}, but our interest lies in understanding how household spending across different categories has responded to lockdown measures, essential for devising policies that foster economic and social stability. Assessing the direct repercussions of pandemic interventions, however, is complex due to their evolving nature \citep[]{faulkner2021,shareef2021l} and varying degrees of stringency, as exemplified by India's four-phase nationwide lockdown in 2020 \citep{SAHA2021}. The diversity in regional characteristics and the severity of COVID-19 in different areas cause variations in the dynamics of specific restriction policies, e.g., state-level COVID-19 regulations before the nationwide lockdown in India \citep{state1, state2, state3}. This complexity, combined with regional variations in COVID-19 severity and response policies, suggests that a direct analysis based solely on nationwide interventions may be easily distorted.

We propose using observed mobility levels as the primary intervention variable, focusing on restrictions in mobility influenced by pandemic countermeasures, which better reflect the real-time situation in a specific location. This strategy aligns with the goal of such policies to limit people's mobility and reduce the potential for disease transmission \citep{Onyeaka2021}. Additionally, this framework is adaptable to various policies aimed at curtailing public mobility. Google mobility data \citep[]{googlemob} is a valuable source of data, covering various regions and five distinct categories of places. We specifically focus on the relative mobility change in retail, providing insights into changes in movement patterns relative to the pre-COVID baseline.

\begin{figure}\label{fig:mob} 
   \includegraphics[width=6in]{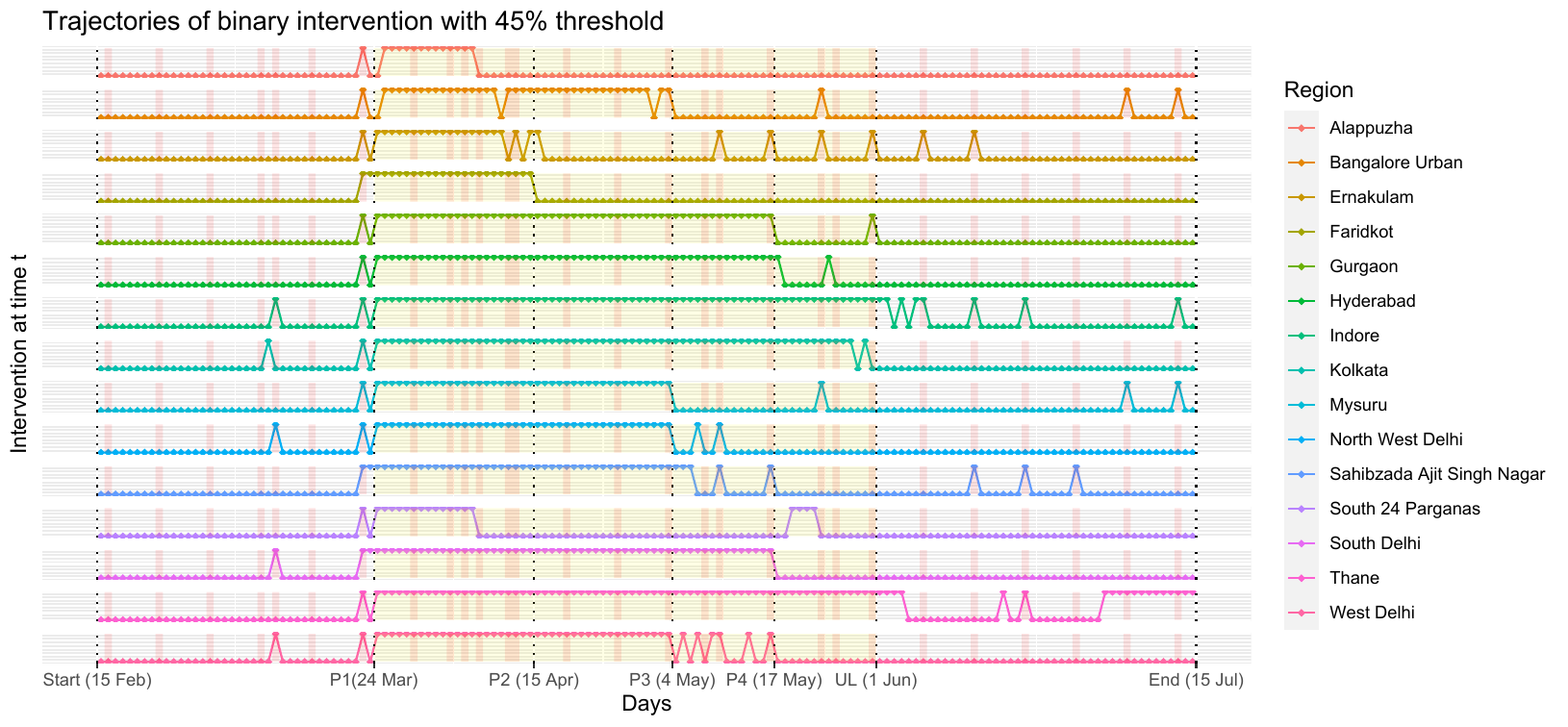}
\caption{\label{fig:mob} 
Binary-transformed intervention trajectories with $45\%$ threshold. The yellow-shaded region corresponds to the nationwide lockdown in India, and the red-shaded areas are gazetted holidays (10 Mar, Holi; 2 Apr, Ram Navami; 6 Apr, Mahavir Jayanti; 10 Apr, Good Friday; 7 May, Budha Purnima; 25 May, Id-ul-Fitr) and Sundays. Abbreviations: P1-P4, gradual phases in India's nationwide lockdown in 2020; UL, unlock phase.
}
\end{figure}

We introduce a binary intervention variable, denoted as $A_{t}\in\{0,1\}$, which signifies whether the reduction in relative percentage change exceeds $45\%$. This threshold was chosen based on specific aspects of the nationwide lockdown policy in India, such as the reopening of small shops with half staff \citep{aleem}. This policy motivated us to define a significant regional mobility restriction as falling within the range of $40\%$ to $50\%$ of mobility reduction, with $45\%$ serving as a representative value within this range. Refer to Figure \ref{fig:mob}; illustrating a dramatic decline in mobility at the beginning of the yellow-shaded interval, representing the nationwide lockdown, followed by a varied recovery across regions. The $45\%$ threshold captures significant mobility restrictions both before and after the nationwide lockdown. Using a higher threshold may overlook the prolonged effect during the unlock period post-June 2020. This variability confirms that the effect of social distancing policies is not constant across times and regions, emphasizing the importance of focusing on mobility restriction to measure the causal effect on consumer behavior. We also have conducted sensitivity analyses with multiple thresholds within $40\%$ to $50\%$ to ensure the robustness of our results.

This project has faced challenges due to the unique characteristics of the motivational data. Firstly, expenditures across different categories show a naturally occurring multivariate zero-inflated pattern, stemming from consumers' inconsistent purchases across categories during a single shopping trip. Distinguishing between the decision to purchase a specific item and the mechanism determining the spending amount for these items is complex. Various factors, such as category characteristics, external conditions (e.g., mobility restrictions), and individual consumer behavior patterns (e.g., preferences or timely necessities), influence these quantities, potentially having different structural processes. Additionally, the expenditure data is non-negative and right-skewed, with lower values being more common due to infrequent extreme spending.

Secondly, our study's observations are irregular and closely related to past variables including expenditure outcomes and the current intervention $A_{t}$. Unlike randomized clinical trials with fixed visitation schedules, consumers' shopping times are not predetermined. The timing of their next shopping trip is influenced by current intervention and various time-varying factors, such as past shopping behaviors, ongoing lockdown policies, and trends in area mobility levels. These factors can also affect expenditure outcomes, leading to distortion in simple regression methods \citep{robins2008,yang2022semiparametric, bae2024}. Correcting these associations is crucial for accurately measuring the causal effects of significant mobility restriction.

Previous research has focused on causal effects in the context of sequential interventions \citep[]{Robins1997}, but these models do not directly apply to our irregular and data-rich observation process closely tied to the current intervention value. Therefore, we adapted multiplicative structural nested mean models, estimating the causal effect of treatment sequence in the presence of time-varying confounders \cite{vansteelandt2014}, to accommodate multiple zero-inflated outcomes \cite{yu2022} and outcome-dependent observation times using the inverse intensity weighting technique \citep[e.g.,][]{Lin2004,Buz2007,Pull2013}. Our first contribution is a flexible framework for estimating the causal effect of current mobility restriction, considering time-varying confounding and outcome-dependent observation processes after treatment assignment, with clear assumptions outlined. Second, we introduce a rate doubly robust estimation estimator that allows for a semiparametric proportional model for the observation process and nonparametric estimation for other nuisance functions. A rate doubly robust estimator is beneficial in the sense that it asymptotically follows a normal distribution even with flexible nonparametric nuisance function estimations and the conventional sandwich variance estimator. \cite{Farrell2022} The proposed estimator has been shown to be a doubly robust rate estimator through theoretical derivation and simulation studies.

The rest of the article is organized as follows. Section \ref{sec:meth} introduces the basic setup, including notation, assumptions, and models, followed by the proposed estimators and their asymptotic distribution. Simulation studies demonstrating the empirical performance of our proposed estimators are provided in Section \ref{sec:sim}. An application to 2020 market transactions in India is given in Section \ref{sec:app} to study the causal effect of the significant mobility restrictions on consumer behavior. Concluding remarks and potentially promising future studies are discussed in Section \ref{sec:disc}.

\section{Proposed Method}
\label{sec:meth}
\subsection{Notation}

\begin{figure}
\begin{center}
\begin{tikzpicture}
    \draw [dashed] (2.9,-7) -- (2.9,0) -- (2.9,-7.1);
    \draw [dashed] (5.55,-7) -- (5.55,0) -- (5.55,-7.1);
    \draw [dashed] (8.2,-7) -- (8.2,0) -- (8.2,-7.1)  ;
    \draw [dashed] (11,-7) -- (11,0) -- (11,-7.1) ;
    \filldraw 
    (1,-7) circle (0.1pt) node[align=left,   below] {Time (t)} --
    (2.9,-7) circle (2pt) node[align=center,   below] {t=1} --
    (5.55,-7) circle (2pt) node[align=center, below] {t=2}     -- 
    (8.2,-7) circle (2pt) node[align=center,  below] {t=3} -- 
    (11,-7) circle (2pt) node[align=center,  below] {t=4} --
    (12.5,-7) circle (0.1pt) node[align=center,   below] {};
    \node (g0)[xshift=1cm, yshift=-0.5cm] {$\overline{G}_{0}$};
    \node (g1) [right =of g0] {$\overline{G}_{1}$};
    \node (g2) [right =of g1] {$\overline{G}_{2}$};
    \node (g3) [right =of g2] {$\overline{G}_{3}$};
    \node (g4) [right =of g3] {$\overline{G}_{4}$};
    \node (a0) [below right =of g0, xshift=-2.3cm] {$A_{0}$};
    \node (a1) [right =of a0] {$A_{1}$};
    \node (a2) [right =of a1] {$A_{2}$};
    \node (a3) [right =of a2] {$A_{3}$};
    \node (a4) [right =of a3] {};
    \node[state,rectangle] (reg) [above left of = a0, xshift=-1cm] {Regional};
    \node (l0) [below right =of a0, xshift=-3cm] {$\overline{I}_{i,0}$};
    \node (l1) [below right =of a1, xshift=-3cm] {$\overline{I}_{i,1}$};
    \node (l2) [below right =of a2, xshift=-3cm] {$\overline{I}_{i,2}$};
    \node (l3) [below right =of a3, xshift=-3cm] {$\overline{I}_{i,3}$};
    \node (l4) [right =of l3] {$\overline{I}_{i,4}$};
    \node (dn1) [below right =of l1, xshift=-2cm] {$\shortstack{$\Delta N_{i,1}$ \\ $Y_{i,1}$}$};
    \node (dn2) [below right =of l2, xshift=-2cm] {$\shortstack{$\Delta N_{i,2}$ \\ $Y_{i,2}$}$};
    \node (dn3) [below right =of l3, xshift=-2cm] {$\shortstack{$\Delta N_{i,3}$ \\ $Y_{i,3}$}$};
    \node (dn4) [below right =of l4, xshift=-2.5cm] {};
    \node[state,rectangle] (ind) [above left of = dn1, xshift=-4.5cm] {Individual};
    \path (g0) edge (g1);
    \path (g1) edge (g2);
    \path (g2) edge (g3);
    \path (g3) edge (g4);
    \path (g0) edge (a0);
    \path (g1) edge (a1);
    \path (g2) edge (a2);
    \path (g3) edge (a3);
    \path (g1) edge[bend right=35] (dn1);
    \path (g2) edge[bend right=35] (dn2);
    \path (g3) edge[bend right=35] (dn3);
    \path (a0) edge (g1);
    \path (a1) edge (g2);
    \path (a2) edge (g3);
    \path (a3) edge (g4);
    \path (a1) edge[bend left=35] (dn1);
    \path (a2) edge[bend left=35] (dn2);
    \path (a3) edge[bend left=35] (dn3);
    \path (l1) edge[bend left=25] (dn1);
    \path (l2) edge[bend left=25] (dn2);
    \path (l3) edge[bend left=25] (dn3);
    \path (l0) edge (l1);
    \path (l1) edge (l2);
    \path (l2) edge (l3);
    \path (l3) edge (l4);
    \path (dn1) edge (l2);
    \path (dn2) edge (l3);
    \path (dn3) edge (l4);
\end{tikzpicture}
\end{center}
\caption{ \label{causdiag} Diagrams within the envisioned framework to illustrate the order of the observed variables. In the initial stage, cumulative regional factors denoted as $\overline{G}_{t}$ exert an influence on the regional intervention $A_{t}$. The set encompassing all prognostic factors up to time $t$ preceding $A_{t}$, comprising cumulative information about individual consumers $\overline{I}_{i,t}$ and regions $\overline{G}_{t}$, is represented as $\Olit=\{\overline{G}_{t},\overline{I}_{i,t}\}$. The resulting intervention $A_{t}$, in conjunction with the cumulative information $\Olit$, generates the current visiting moment $\Delta N_{i,t}$ and the corresponding expenditure $Y_{i,t}$. Note that all the variables generated serve as past trajectories for the subsequent time point.
}
\end{figure}
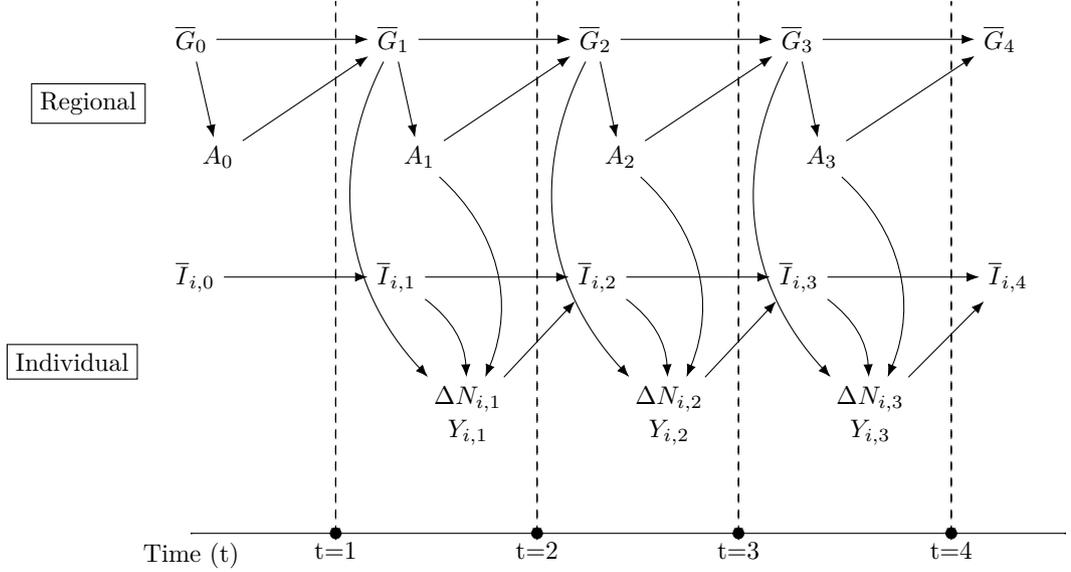

Consider $K$ categories of expenditures over a research period from $t_{0}=0$ to $T$ with $T>0$. We are interested in discrete time points, such as daily customer visits to a shop. For each subject $i\in \{1,\dots, n\}$, we define the counting process for visits $\{{N}_{i,t}\}_{t\geq 0}$, where $N_{i,t}$ represents the total number of visits up to time $t>0$, with $N_{i,0}=0$. We use an overline to indicate the history of an observed variable up to time $t$, e.g., $\overline{A}_{t}=\{A_{j}:j\leq t\}$, and $t-$ to indicate the last time point before $t$, e.g., $\overline{A}_{t-}=\{A_{j}:j<t\}$. An indicator for a visit occurring at time $t$ is denoted by $\Delta N_{i,t}=N_{i,t}-N_{i,t-}\in \{0,1\}$, where $\Delta N_{i,t}=1$ if the subject $i$ had opportunity to visit at time $t$, and $0$ otherwise. When a consumer visits, we observe zero-inflated expenditures across $K$ categories, denoted as $\{Y_{i,t}=(Y_{i,1,t},\dots,Y_{i,K,t})\Tra\in\Real_{0+}^{K}\}$ for $t$, where $\Real_{0+}$ indicates positive numbers including $0$. We further posit the presence of an underlying non-negative expenditure on the entire granularity of our interest; however, we only have access to some of these expenses when $\Delta N_{i,t}=1$.

Consumers in the region share the same intervention sequences, suggesting a lack of individualization. We represent the cumulative regional-level covariates before the intervention $A_{t}$ as $\overline{G}_{t}$. The set $\overline{G}_{t}$ is allowed to encompass region-specific time-varying variables (e.g., holiday indicators) and prior intervention sequences preceding $A_{t}$, denoted as $\overline{A}_{t-}$. It is important to note that $\overline{G}_{t}$ is available prior to the assignment of the intervention $A_{t}$ at time $t$, i.e., $A_{t}\notin \overline{G}_{t}$. Additionally, we define baseline covariate $X_{i}$ for each subject $i$ and further denote the individualized information up to time $t$ as $\overline{I}_{i,t} = \{X_{i}, \overline{Y}_{i,t-}, \overline{N}_{i,t-}\}$. Combining these factors as $\Olit = \{\overline{G}_{t}, \overline{I}_{i,t}\}$ affects $\Delta N_{i,t}$ and $Y_{i,t}$. We assume that the consumer's visit timing or expenditure scale does not influence the intervention assignment, i.e., $A_{t}$ depends solely on the regional factors $\overline{G}_{t}$. Assuming that subject $i$ visits the shop for $t_{1}<\dots< t_{m}\leq T$, the observation order is: $\{\overline{I}_{i,t_{1}},\overline{G}_{t_{1}}\}$, ${A}_{t_{1}}$, $\{\Delta N_{i,t_{1}}, {Y}_{i,t_{1}}\}$, $\dots$, $\overline{G}_{t_{m}}$, ${A}_{t_{m}}$, $\{\Delta N_{i,t_{m}},{Y}_{i,t_{m}}\}$, $\overline{G}_{T}$. See Figure \ref{causdiag} to overview the mechanism. We assume no informative censoring in the visiting process is suitable for analyzing repeated transactions. As $\overline{N}_{i,t}$ represents all visits up to time $t$ for customer $i$, the history of observed data for the individual through time $t$ is denoted as $\overline{O}_{i,t}=\{X_{i},\overline{G}_{t},\overline{N}_{i,t},\overline{A}_{t}, \overline{Y}_{i,t}\}$, assumed to be identically distributed across subjects. The main notations presented and those to be introduced in the following sections of this manuscript are detailed in Table 1.

\begin{table}
    \centering \footnotesize
    \begin{tabularx}{\textwidth}{c|X}
        \hline
        \multicolumn{2}{c}{Regional-level variables} \\ \hline
        $A_{t}$ & Binary intervention at time $t$ \\
        $G_{t}$ & Regional-level prognostic factors of the intervention\\
        \hline
        \multicolumn{2}{c}{Individual-level variables for subject $i$} \\ \hline
        $X_{i}$ & Individual baseline covariate (e.g., demographic information) \\
        $N_{i,t}$ & Total number of visits up to time $t$ \\
        $\Delta N_{i,t}$ & Indicator for a visit occurring at time $t$ \\
        $Y_{i,t,k}$ & Zero-inflated outcome at category $k$ and time $t$\\
        $I_{i,t}$ & Aggregation of all individualized information at time $t$\\
        \hline
        \multicolumn{2}{c}{Combined variables for subject $i$} \\ \hline
        $L_{i,t}$ & Combined information including regional $G_{t}$ and individual $I_{i,t}$ before assigning $A_{t}$\\
        $O_{i,t}$ & Aggregation of all the observed data at time $t$\\
        \hline
        \multicolumn{2}{c}{General notations} \\ \hline
        $n$ & Number of subjects in the study\\
        $T$ & Length of follow-up time \\
        $\psi$ & Target causal parameter \\
        $(\gamma,\lambda)$ & Visiting intensity parameters\\
        $\pi$ & Propensity score\\
        $h$ & Conditional mean outcome model $E\{H(\overline{O}_{i,t},\psi)\mid\overline{L}_{i,t}\}$\\
        $m_{0}^{p}(\overline{O}_{i,t};\beta^{p})$ & Unknown baseline function of $\overline{O}_{i,t}$ with parameter $\beta^{p}$ related to zero-inflation probability  \\
        $m_{0}^{y}(\overline{O}_{i,t};\beta^{y})$ & Unknown baseline function of $\overline{O}_{i,t}$ with parameter $\beta^{y}$ related to non-zero outcome \\
        $m_{a}(\overline{O}_{i,t})$ & Known function of $\overline{O}_{i,t}$ in outcome model associated with $A_{t}$ and $\psi$\\
        \hline
    \end{tabularx}
    \caption{List of notations. }
    \label{tab:not}
\end{table}

\subsection{Assumptions and models}

Under potential outcome framework \citep{rubin1974}, let $Y_{it}^{(\overline{a}_{t})}$ denote the potential outcome across the $K$ categories that would be seen at time $t$ had the subject $i$ received the intervention $\overline{a}_{t}=\overline{a}_{t}$. Our interest lies in comparing the target outcome under $\overline{a}_{t}$ versus $\overline{a}_{t-}^{0}=\{\overline{a}_{t-},0\}$, where $\overline{a}_{t-}^{0}$ denotes the scenario where subject $i$ received $\overline{a}_{t-}$ before time $t$ and $a_{t}=0$ at time $t$, with $a_{0}=0$ for clarity. Therefore, we compare $Y_{it}^{(\overline{a}_{t})}$ with $Y_{it}^{(\overline{a}_{t-},0)}$, where $Y_{it}^{(\overline{a}_{t-},0)}$ is indicating the potential outcome that would be seen at time $t$ had the subject $i$ received the sequence of treatment $\overline{a}_{t-}$ and $a_{t}=0$ at time $t$, where ${a}_{i0}$ is given as an initial value. We presume that future information of each variable process cannot affect the potential outcomes, i.e., $Y_{it}^{(\overline{a}_{T})}=Y_{it}^{(\overline{a}_{t})}$ for any $t$. 

Recall that our main interest is estimating population-level causal effects on each category $k$ as mean ratios; the visiting process is not our main interest. In further derivations, we rely on widely accepted assumptions for potential outcomes in related literature.
\begin{assumption}[Consistency]\label{ass:cons}
For any individual $i$ and time $t$, the observed outcome is equal to the potential outcome under the observed sequence of interventions $\overline{A}_{t}=\overline{a}_{t}$, i.e., $Y_{i,t}=Y_{i,t}(\overline{a}_{t})$.
\end{assumption} 
\begin{assumption}[Positivity]\label{ass:pos}
For any time $t$, there exists non-zero probability of receiving any level of intervention $A_{t}$ given any regional information $\overline{G}_{t}$, i.e., $0<P(A_{t}=1|\overline{G}_{t})<1$.
\end{assumption} 
\begin{assumption}[Sequential ignorability]\label{ass:trt}
For any individual $i$, intervention at time $t$ is conditionally independent of the potential outcome had received the sequence of intervention $\overline{a}_{t-}$ and $a_{t}=0$ at time $t$ given the prior regional information, i.e., $A_{t}\ind \YpotDo|\overline{G}_{t}$ for all $\overline{a}_{t-}^{0}$.
\end{assumption} 
Assumption \ref{ass:cons} states that the observed outcome aligns with the target potential expenditure under the given intervention sequences. Assumption \ref{ass:pos} ensures that, conditional on the available regional information, every level of intervention has a nonzero probability of occurring at any given time. Assumption \ref{ass:trt} suggests that the assignment of the intervention policy $A_{t}$ is independent of the default potential outcome $\YpotDo$ given the available regional information $\overline{G}_{t}$ before time $t$, eliminating any remaining hidden confounders between the intervention and the outcome at time $t$ from $\overline{G}_{t}$.

Considering the irregular visiting patterns of customers, where the time gaps between purchases vary and can influence current purchases, we adopt the outcome-dependent process model with the visiting at random (VAR) assumption \citep[e.g.,][]{yang2019bmk,yang2022semiparametric,sang2022functional}. We assume that the visiting process at time $t$ follows a proportional hazard model:
\begin{assumption}[History-dependent irregular visits]\label{ass:vis}
For each time $t$, visiting time is independent of the potential outcome given the past available information and the intervention, i.e., $Y_{i,t}(\overline{a}_{t}) \ind \Delta N_{i,t}|\Olit,A_{t}$. Furthermore, the visiting intensity $\lambda(t|\Olit,A_{t})$ follows a proportional hazard models as follow: 
 \begin{align*} 
    \lambda(t|\Olit,A_{t})=\exp\big\{ m_{v}(\Olit,A_{t})\Tra\gamma_{0} \big\}\lambda_{0}(t),
\end{align*}
where $\lambda_{0}(t)$ is a baseline intensity function, and $m_{v}(\cdot)$ is a function of $\{\Olit,A_{t}\}$ with true intensity parameter $\gamma_{0}$. 
\end{assumption}
The visiting intensity model for Assumption \ref{ass:vis} is flexible due to unspecified baseline intensity $\lambda_{0}(t)$ while the curse of dimensionality is alleviated by using proportional regression parameters $\gamma_{0}$. Assumption \ref{ass:vis} can be seen as an additional sequential ignorability assumption further clarifying that there is no more unobserved confounder connecting the current visit and the corresponding expenditure out of $\{\Olit, A_{t}\}$. 

Recall that we are only interested in the observed expenditure. We can formulate the causal effect by the following multiplicative structural nested model \citep{Robins1994} 
\begin{align}\label{snmm}
    \frac{\mathbb{E}\big\{\YkpotTo|\Olit,{A}_{t}={a}_{t}\big\}}{\mathbb{E}\big\{\YkpotDo|\Olit,{A}_{t}={a}_{t}\big\}}=\exp \big\{m_{a}(\Olit;\psi_{k0})a_{t} \big\},
\end{align}
where $m_{a}(\Olit;{\psi_{k0}})\in\Real^{d}$ is a known function of $\{\Olit\}$ with $p$-dimensional vector type of parameters $\psi_{k0}$ for each $k$th category, satisfying $m_{a}(\;\cdot\;;\;0)=0$. The multiplicative model (\ref{snmm}) is a natural choice for the transaction data, considering that the expenditure for a given category is naturally a zero-inflated outcome with unknown correlations. The parameter $\psi_{k0}$ effectively captures the conditional causal effect of the intervention on the mean shift of potential outcomes in the $k$th category, as described in equation (\ref{snmm}). This causal estimand quantifies the direct effect of the intervention sequence modification from the default $\overline{a}_{t-}^{0}=\{\overline{a}_{t-},0\}$ to $\overline{a}_{t}=\{\overline{a}_{t-},a_{t}\}$, conditioning on available individual and regional information $\Olit$. 

Let us illustrate how the multiplicative model (\ref{snmm}) works by providing an example. Assume there exist arbitrary baseline random effects for subject $i$ $\nu_{i}=\big(\nu^{p}_{i},\nu^{y}_{i})$ where $\nu^{p}_{i}$ and $\nu^{y}_{i}$ are assigned for zero-inflated probability and the mean of log-normal distribution, respectively. More specifically, for given category $k\in\{1,\dots,K\}$, time $t$, and $\Delta N_{i,t}=1$, we can imagine $\YkpotTo|\Olit,A_{t}=a_{t},\nu^{y}_{i}$ has a probability $1-p_{k}(\psi^{p}_{k0}|\Olit,A_{t}=a_{t},\nu^{p}_{i})$ to be zero and follows log-normal distribution $log N\big\{\mu_{k}(\psi^{y}_{k0}|\Olit,A_{t}=a_{t},\nu^{y}_{i}),\sigma^{2}\big\}$ otherwise;
\begin{align*}
    & p_{k}(\beta^{p}_{k0},\psi^{p}_{k0}|\Olit, A_{t}=a_{t},\nu^{p}_{i})=\exp \big\{ \nu^{p}_{i}+m_{0}^{p}(\Olit;\beta^{p}_{k0})+m_{a}(\Olit)\Tra \psi^{p}_{k0}a_{t} \big\}, \\
    & \mu_{k}(\beta^{y}_{k0},\psi^{y}_{k0}|\Olit, A_{t}=a_{t},\nu^{y}_{i})= \nu^{y}_{i}+ m_{0}^{y}(\Olit;\beta^{y}_{k0})+ m_{a}(\Olit)\Tra \psi^{y}_{k0}a_{t},
\end{align*}
where $m_{0}(\Olit)$ and $m_{a}(\Olit)$ are baseline functions of prior information $\Olit$ regardless of $A_{t}$. Then, it can be shown that
\begin{align*}
    \mathbb{E}\big\{ \YkpotTo|\Olit,A_{t}=a_{t}\big\}& = 
    \int \big[p_{k}(\beta^{p}_{k0},\psi^{p}_{k0}|\Olit,A_{t}=a_{t},\nu^{p}_{i})\\
    & \qquad \times \exp\big\{ \mu_{k}(\beta^{y}_{k0},\psi^{y}_{k0}|\Olit,A_{t}=a_{t},\nu^{y}_{i})+\frac{1}{2}\sigma^2\big\}\big]d{P}(\nu_{i}),\\
    & = \exp\big[\{ m_{a}(\Olit)\Tra \psi^{p}_{k0}+m_{a}(\Olit)\Tra \psi^{y}_{k0}\}a_{t}\big]\times b(\Olit),
\end{align*}
where $b(\Olit)$ is the remaining function of $\Olit$ with complicated form. Therefore, model (\ref{snmm}) is satisfied as
\begin{align}\label{examp}
    \frac{\mathbb{E}\big\{\YkpotTo|\Olit,A_{t}=a_{t}\big\}}{\mathbb{E}\big\{\YkpotDo|\Olit,A_{t}=a_{t}
    \big\}}&=\exp\big[\{ m_{a}(\Olit)\Tra (\psi^{p}_{k0}+\psi^{y}_{k0})\}a_{t}\big],
\end{align}
which implies that $\psi_{k0}=(\psi^{p}_{k0}+\psi^{y}_{k0})$ is a target causal parameter capturing the average controlled direct effect while fixing $\overline{L}_{i,t}$. Note that $m_{a}(\cdot)$ function should be specified based on the expertise in the field, though $m^{p}_{0}(\cdot)$ or $m^{y}_{0}(\cdot)$ can be arbitrary functions of $\Olit$.

\subsection{Identification and estimation}\label{subsec:id}

To estimate the parameters of the structural nested mean model under the causal inference framework, we establish the G-estimator \cite{vansteelandt2014} of the parameter $\psi_{0}=(\psi_{10},\dots,\psi_{K0})\Tra\in\Real^{p\times K}$ for model (\ref{snmm}):
\begin{align*}
   H_{k}(\overline{O}_{i,t},\psi_{0})\equiv Y_{i,k,t} \exp \big\{-m_{a}(\Olit;{\psi_{k0}})A_{t} \big\}\in \Real_{0+},
\end{align*}
mimicking the quantity $\YkpotDo$ given $\{\Olit,A_{t}\}$. Let us denote a combined vector of $H_{k}(\overline{O}_{i,t},\psi_{0})$ across all $K$ categories as $H(\overline{O}_{i,t},\psi_{0})=\{H_{1}(\overline{O}_{i,t},\psi_{0}),\dots,H_{K}(\overline{O}_{i,t},\psi_{0})\}\Tra\in\Real_{0+}^{K}$. Previous assumptions can derive the following useful lemma.
\begin{lemma} \label{lem:H}
Under model (\ref{snmm}), provided Assumptions \ref{ass:cons} - \ref{ass:trt} {are satisfied}, the following equality holds for all $k\in\{1,\dots,K\}$ and $t$,
    \begin{align*}
        \mathbb{E}\big\{H_{k}(\overline{O}_{i,t},\psi_{0})|\Olit,{A}_{t}={a}_{t} \big\} 
        & =  \mathbb{E}\{H_{k}(\overline{O}_{i,t},\psi_{0}) |\Olit\},
    \end{align*}
\end{lemma}
which implies that $\mathbb{E}\{H_{k}(\overline{O}_{i,t},\psi_{0}) |\Olit,A_{t}\}=\mathbb{E}\{H_{k}(\overline{O}_{i,t},\psi_{0}) |\Olit\}\overset{let}{=}h_{k0}(\Olit)$.  
We denote the combination of true $h_{k0}(\Olit)$ as $h_{0}(\Olit)=\{h_{10}(\Olit),\dots, h_{K0}(\Olit)\}\Tra$. Let $\pi_{0}(\overline{G}_{t})={P}(A_{t}=1|\overline{G}_{t})$ be true propensity score, based on Assumption \ref{ass:trt}. 

Toward the consistent estimation of the causal parameter $\psi_{0}$, we construct the following doubly-robust estimating function:
\begin{align} \label{estf_iiw}
    \sum_{t=1}^{T}\Psi_{i,t}(\psi_{0},\pi,h,\gamma) \Delta N_{i,t}\equiv\sum_{t=1}^{T} c(\Olit)\big\{H(\overline{O}_{i,t},\psi_{0})-h(\Olit)\big\}\frac{\{{A_{t}}-\pi(\overline{G}_{t})\}\Delta N_{i,t} }{\exp\big\{ m_{v}(\Olit,A_{t})\Tra \gamma\big\}},
\end{align}
where its mean can be derived as
\begin{align*} 
    & \mathbb{E}\bigg[ \sum_{t=1}^{T} c(\Olit)\big\{H(\overline{O}_{i,t},\psi_{0})-h(\Olit)\big\}\frac{\{{A_{t}}-\pi(\overline{G}_{t})\}\Delta N_{i,t} }{\exp\big\{ m_{v}(\Olit,A_{t})\Tra \gamma\big\}}\bigg]\nonumber \\
    & = \mathbb{E}\bigg[\sum_{t=1}^{T}c(\Olit)\big[\mathbb{E}\{H(\overline{O}_{i,t},\psi_{0})|\Olit,A_{t}\}-h(\Olit)\big]  \frac{\{{A_{t}}-\pi(\overline{G}_{t})\}\mathbb{E}(\Delta N_{i,t}|\Olit,A_{t})}{\exp\big\{ m_{v}(\Olit,A_{t})\Tra \gamma\big\}}\bigg] \nonumber\\
    & = \mathbb{E}\bigg[ \sum_{t=1}^{T}c(\Olit)\big\{h_{0}(\Olit)-h(\Olit)\big\} {\{\pi_{0}(\overline{G}_{t})-\pi(\overline{G}_{t})\}\lambda_{0}(t)}\bigg],
\end{align*}
where the last equality holds for $\gamma=\gamma_{0}$. Consider that the choice of $c(\Olit)$ impacts the estimator's efficiency but not its consistency. An efficient selection for $c(\Olit)$ requires estimating the outcomes' covariance structure \citep{Robins1994, goetghebeur1997}. However, the optimal choice may not significantly enhance efficiency over a simpler alternative; see \citep{yu2022}. Therefore, in practice, we set $c(\Olit) = \partial m_{a}(\Olit;{\psi})/\partial \psi$; which can be naturally obtained while deriving partial derivative of the estimating equation \ref{estf_iiw} respect to $\psi$. The mean of estimating equation (\ref{estf_iiw}) equals 0 for $\psi = \psi_{0}$ if the conditional outcome mean function is correct ($h = h_{0}$), or if both the intensity parameter ($\gamma = \gamma_{0}$) and propensity score ($\pi = \pi_{0}$) are correct. Estimation of these nuisance functions should precede solving estimating function (\ref{estf_iiw}). For the visiting intensity parameter $\gamma_{0}$, we can use the traditional Cox proportional model following established conventions \citep{Andersen1982,Anderson1993} to obtain consistent estimates $\hat{\gamma}$. For the propensity model $\pi_{0}(\overline{G}_{t})$, both parametric methods (e.g., logistic regression) and nonparametric methods (e.g., generalized additive models) can be employed. However, the parametric approach for the outcome model $h_{0}(\Olit) = \mathbb{E}\{H(\overline{O}_{i,t},\psi_{0})|\Olit\}$ is challenging due to its complex form resulting from the intricate correlation structure of longitudinal multivariate outcomes and the irregular visiting process. It also depends on unknown functions $m^{p}_{0}(\cdot)$ or $m^{y}_{0}(\cdot)$. Decomposing the function $h_{0}(\Olit)$  with respect to $A_{t}$ at time $t$ for its estimation, suggested by \cite{yu2022}, has a flaw in the sense that it requires consistent estimates $\hat{\pi}(\Olit)$. To detour this risk, if we have enough number of subjects and time $t$ satisfying $A_{t}=0$, we can use the equality in the previous Lemma \ref{lem:H}:
\begin{align*}
     h_{k0}(\Olit) & = \mathbb{E}\big[\exp\big\{ -m_{a}(\Olit;{\psi_{k0}})a_{t}\big\} Y_{i,k,t}^{(\overline{a}_{t})} |\Olit,{A}_{t}={a}_{t}\big]\\
        & = \mathbb{E}\{\YkpotDo |\Olit,{A}_{t}={a}_{t}\}\\
        & = \mathbb{E}( Y_{i,k,t} |\Olit,{A}_{t}=0),
\end{align*}
without ${\pi}(\overline{G}_{t})$ or ${\mu}_{0}(\Olit)$. In practice, the parametric correctness of any of these nuisance functions is very difficult to achieve; therefore, researchers may favor more flexible models using nonparametric methods. When we are applying nonparametric or semiparametric estimation methods for these nuisance functions $\pi$, $\gamma$, and $h$, our proposed estimator still has root-T consistency for $\psi_{0}$ under a certain rate assumption for their estimation. A practical, step-by-step estimation guideline is included in the Appendix for ease of implementation.

\subsection{Asymptotic properties}

We first introduce the notation $\wc$ to represent weak convergence and define $A\bbc B$ to indicate that $A$ is bounded by a constant time $B$. Denote
\begin{align*}
    \EP g\{\Psi(\psi,\vartheta)\} &= \lim_{T\rightarrow \infty}\mathbb{E} \bigg[\frac{1}{T} \sum_{t=1}^{T}g\{\Psi_{i,t}(\psi, \vartheta) \} \Delta N_{i,t} \bigg]\\
    \EPn g\{\Psi(\psi,\vartheta)\} &= \frac{1}{n} \sum_{i=1}^{n}  \frac{1}{T}\sum_{t=1}^{T}g\{\Psi_{i,t}(\psi, \vartheta) \}  \Delta N_{i,t},
\end{align*}
for any function $g$ of $\Psi_{i,t}(\psi,\vartheta)$. We denote Euclidean norm $\|v\|_{2}=(v\Tra v)^{1/2}$ for some vector $v$ and $L_{2}(P)$ norm $\|\vartheta\|^{2}_{2,P}=\lim_{T\rightarrow \infty}T^{-1}\sum_{t=1}^{T}\int \|\vartheta(\Olit,A_{t})\|_{2}^{2}dP(\Olit,A_{t})$. We further denote the set of nuisance functions as $\vartheta = (\pi,\gamma,h)$ and assume the target causal parameter $\psi\in\Real^{p}$. Let $\Theta$ be the regression parameter space, which is a compact set in the Euclidean space. Denote a set of nuisance parameters near true $\vartheta$ as $\Xi_{\vartheta}=\{\vartheta:\|\vartheta-\vartheta_{0}\|_{2,P}<\delta\}$ for some $\delta>0$ and let $l^{\infty}(\Xi_{\vartheta_{0}})$ as the collection of all bounded functions $f:\Xi_{\theta_{0}}\rightarrow \Real^{p}$. Define a Cartesian product space of our interest $\mathcal{U}=\Theta\times \Xi_{\vartheta_{0}}$. Let $\dot{\Psi}_{i,t,\psi}(\psi,\vartheta)=\partial \Psi_{i,t}(\psi,\vartheta)/ \partial \psi$ and $\dot{\Psi}_{i,t,\vartheta}(\psi,\vartheta)=\partial \Psi_{i,t}(\psi,\vartheta)/ \partial \vartheta$ be the derivatives with respect to each parameter. Next we provide the regularity conditions required to establish the asymptotic distribution of the proposed estimator.

\begin{condition}
    \label{ass:rc1} There exists unique $\psi_{0}$ satisfying $\EP\Psi(\psi_{0},\vartheta_{0})=0$. If $\|\EP\Psi(\psi_{n},\vartheta_{0})\|_{2}\cs 0$, then $\|\psi_{n}-\psi_{0}\|_{2}\cs 0$ for any sequence of $\{\psi_{n}\}\in \Theta$.
\end{condition}
\begin{condition}
    \label{ass:rc2} For all $i$ and on the prespecified granularity $t\leq T$,
    \begin{itemize}
        \item[(i)] There exists a finite $\epsilon-$net $\mathcal{U}_{\epsilon}$ of $\mathcal{U}=\Theta\times \Xi_{\vartheta_{0}}$ for any $\epsilon>0$.
        \item[(ii)] $\Xi_{\vartheta_{0}}$ has uniformly integrable entropy.
        \item[(iii)] $\|c(\Olit)\|_{2}$, $\|Y_{i,t}\|_{2}$, $\|m_{a}(\Olit)\|_{2}$, and $\|m_{v}(\Olit)\|_{2}$ are bounded almost surely.
        \item[(iv)] $\|\vartheta(\Olit)\|_{2}$ is bounded almost surely for all $\vartheta\in\Xi_{\vartheta_{0}}$. 
        \item[(v)] $\|\dot{\Psi}_{i,t,\vartheta}(\psi,\vartheta)\|_{2}$ is bounded almost surely for all $\psi\in\Theta$ and $\vartheta\in\Xi_{\vartheta_{0}}$.
        \item[(vi)] $\Sigma_1 \equiv  \lim_{T\rightarrow \infty}  \frac{1}{T}\sum_{t=1}^{T} \mathbb{E} \Bigl\{\dot{\Psi}_{i,t,\psi}({\psi_{0}},\vartheta_{0})\Delta N_{i,t}\Bigr\}$ is invertible.
    \end{itemize}
\end{condition}
\begin{condition}
    \label{ass:rc3} Assume $\lim_{T\rightarrow \infty}n/T=c$, where $c$ is either a positive constant or infinity. In addition, the estimators of the nuisance functions satisfy:
    \begin{align*}
        &  \|\hat{h}-h_{0}\|_{2,P}\times \big( \|\hat{\gamma}-{\gamma}_{0}\|_{2}+\|\hat{\pi}-\pi_{0}\|_{2,P}\big)=\op(T^{-1/2}).
    \end{align*}

\end{condition}

\begin{theorem}[Rate-double robustness] 
Under the assumed regularity conditions, we have, as $T$ goes to infinity, $\hat{\psi}_{}$ is a consistent estimator of $\psi_{0}$, and  
    \begin{align*}
    T^{1/2}(\hat{\psi}_{}-\psi_{0})\cd N(0, \Sigma_1\Inv \Sigma_2\Sigma_1\Inv),
    \end{align*}
    where 
    \begin{align*}
    \Sigma_2 = \lim_{n,T\rightarrow \infty}  \frac{1}{T}\sum_{t=1}^{T}\mathbb{E}\bigg[\bigg\{\frac{1}{n}\sum_{i=1}^{n}\Psi_{i,t}(\psi_{0},\vartheta_{0})\Delta N_{i,t}\bigg\}^{\otimes 2} 
    \bigg].
    \end{align*}
    Note that this asymptotic covariance matrix $\Sigma^{*}=\Sigma_1\Inv \Sigma_2\Sigma_1\Inv$ can be consistently estimated by $\hat{\Sigma}^{*} =  \hat{\Sigma}_1\Inv \hat{\Sigma}_2\hat{\Sigma}_1\Inv$, where $\hat{\Sigma}_1 = (nT)^{-1}\sum_{i=1}^{n}\sum_{t=1}^{T}\dot{\Psi}_{i,t,\psi}(\hat{\psi},\hat{\vartheta})\Delta N_{i,t}$ and
    \begin{align*}
    \hat{\Sigma}_2 = (nT)^{-1}\sum_{t=1}^{T}\bigg\{\sum_{i=1}^{n}\Psi_{i,t}(\hat{\psi},\hat{\vartheta})\Delta N_{i,t}\bigg\}^{\otimes 2}.
    \end{align*}
\end{theorem}

We make a few remarks about the assumed conditions. First, Condition \ref{ass:rc1} ensures that the true regression parameter $\psi_{0}$ can be identified locally. Condition \ref{ass:rc2} generally imposes constraints on the complexity of nuisance functions to establish the necessary Lemmas for uniform and weak convergence. Specifically, Conditions \ref{ass:rc2} (i) and \ref{ass:rc2} (ii) are prerequisites for weak convergence, as demonstrated in \cite{bae2010}. For further elaboration on such classes of functions, refer to Section 2.6 of \cite{Vaart1996}. Conditions \ref{ass:rc2} (iii)-(v) are necessary for establishing convergence, typically holding when variables and parameters are bounded. Similar conditions to Condition \ref{ass:rc2} are commonly encountered in Z-estimation and empirical processes literature \citep[e.g.,][]{Vaart1996, vaart1998}. For instance, $\Psi(\psi,\vartheta)$ and $\dot{\Psi}_{\psi}(\psi,\vartheta)$ belong to weak Glivenko-Cantelli classes and Donsker classes \citep[e.g.,][]{yang2022semiparametric}. A more comprehensive exploration of Donsker classes can be found in \cite{Kennedy2016}. In \cite{Cher2018}, it has been shown that Condition \ref{ass:rc2} can be relaxed by employing data-splitting techniques. The invertibility of $\EP\dot{\Psi}({\psi_{0}},\vartheta_{0})$ in Condition \ref{ass:rc2} (vi) is standard in literature and required to derive asymptotic properties of the estimator.

Last, Condition \ref{ass:rc3} requires a proper rate for estimating the nuisance parameters of to ensure the rate-double robustness of the proposed estimator of the causal parameters of interest. In this work, we utilize nonparametric/semiparametric estimators for the nuisance parameters $(\pi,h,\gamma)$, such as generalized additive models. In general,  for such flexible estimation methods, we can achieve $\|\hat{h}-h_{0}\|_{2,P}=\op(n^{-1/4})$, $\|\hat{\gamma}-{\gamma}_{0}\|_{2}=\Op(n^{-1/2})$, and $\|\hat{\pi}-\pi_{0}\|_{2,P}=\op(T^{-1/4})$. Note that the follow-up time $T$ is assumed to go to infinity to ensure the consistency and proper rate for estimating the propensity score model $\pi_0$, which only depends on regional-level time-related factors but not on individual-related factors. 
In addition, Condition \ref{ass:rc3} assumes that the sample size $n \rightarrow \infty$ at the same rate as or faster than $T \rightarrow \infty$. This is usually true in many applications, e.g., in our motivational data, $n=22666$ and $T=150$. 

\section{Simulation}\label{sec:sim}

We further conducted simulations to examine and compare the finite sample performance of the following estimating equations:
\begin{align*}
    \Psi_{\pi,\gamma}&\equiv\frac{1}{n}\sum_{i=1}^{n}\frac{1}{T}\sum_{t=1}^{T} c(\Olit)\frac{H(\overline{O}_{i,t},\psi_{0}) \{{A_{t}}-\pi(\overline{G}_{t})\} \Delta N_{i,t}}{\exp\big\{ m_{v}(\Olit,A_{t})\Tra \gamma\big\}},\\
    \Psi_{h}&\equiv\frac{1}{n}\sum_{i=1}^{n}\frac{1}{T}\sum_{t=1}^{T} c(\Olit) \frac{\big\{H(\overline{O}_{i,t},\psi_{0})-h(\Olit)\big\}A_{t}\Delta N_{i,t}}{\exp\big\{ m_{v}(\Olit,A_{t})\Tra \gamma\big\}},
\end{align*}
and
\begin{align*}
    \Psi_{\pi,\gamma,h}&\equiv\frac{1}{n}\sum_{i=1}^{n}\frac{1}{T}\sum_{t=1}^{T} c(\Olit) \frac{\big\{H(\overline{O}_{i,t},\psi_{0})-h(\Olit)\big\}\{{A_{t}}-\pi(\overline{G}_{t})\}\Delta N_{i,t}}{\exp\big\{ m_{v}(\Olit,A_{t})\Tra \gamma\big\}}.
\end{align*}
It should be noted that the estimator presented in the final estimating equation is our proposed estimator. The first two estimators, on the other hand, are devised by deliberately excluding either the conditional mean outcome $h$ or the propensity $\pi$. Consequently, the consistency of these estimators hinges significantly on the accurate specification of the remaining nuisance functions.

We conduct $1000$ Monte Carlo iterations for all simulations. The estimating equations can be solved using gradient-based root-finding algorithms, such as the \texttt{multiroot} function in the R package \texttt{rootSolve}. We assess the bias and sample variance of the estimates from the estimators, while the derivation of theoretical standard errors and coverage is not feasible for the estimators solving $\Psi_{\pi,\gamma}$ and $\Psi_{h}$ due to nonparametric methods. However, we provide the sandwich variance estimator and its coverage for the rate doubly robust estimator based on $\Psi_{\pi,\gamma,h}$, leveraging the rate-double robustness.

\subsection{Data generating procedures and estimations}

Consider $n=\{200,600\}$ number of individuals where each is contained in one of the $10$ regions. Assume there is a time-varying covariate $I_{\text{LkDn},t}$ (e.g., nationwide lockdown indicator) taking $1$ only if $t\in(T/2,T]$ where $T\in\{100,200\}$. We further define ${D}_{\text{LkDn},t}$, the cumulative duration of the nationwide lockdown, to capture the linear-in-time effect originating from the initiation of the lockdown. Denote $\overline{G}_{t}=\{\overline{I}_{\text{LkDn},t},\overline{D}_{\text{LkDn},t},\overline{A}_{t-}\}$ be the combined regional covariates affecting the intervention assignment. Target time granularity is the natural number to focus on day-by-day shopping behavior. Regional-level intervention trajectory can be sequentially generated using a simple logistic model for $t\in\{1,\dots,T\}$:
\begin{align*}
    \pi(A_{t}|\overline{G}_{t})& =\frac{\exp(\theta_{1}+\theta_{2}I_{\text{LkDn},t}+\theta_{3}{D}_{\text{LkDn},t} )}{1+\exp(\theta_{1}+\theta_{2}I_{\text{LkDn},t}+\theta_{3}{D}_{\text{LkDn},t})},
\end{align*}
where $\theta=(-0.5,0.25,1)\Tra$ and $A_{-1}=A_{0}=R_{0}=0$. We will use a flexible gam model \cite{hastie2017} for propensity estimation, assuming rate-double robustness. We consider two propensity models: (i) a correctly specified $\pi_{C}(\cdot)$ including all factors $\{I_{\text{LkDn},t},D_{\text{LkDn},t}\}$, and (ii) a wrongly specified $\pi_{W}(\cdot)$ without any specified factors except for an intercept.

After generating the intervention sequences in the region, we alternately generate visits and corresponding expenditures. We assume each consumer's last visit before the research period is $t=0$, with an initial non-zero expenditure $Y^{+}_{0}$. The initial visiting time point is generated based on $\overline{L}_{i,0}=\{Y^{+}_{0},\overline{G}_{0}\}$, and subsequent visits are generated based on observed trajectories of the variables $\Olit$. For irregular visits generation, we use the method of \cite{Daley2002} in the spirit of \cite{Lin2004}. We assume proportional hazards for time $t$ on the prespecified granularity: 
\begin{align*}
        \lambda(t|\Olit,A_{t} )& = 0.075\exp( 0.125 I_{\text{LkDn},t}) \exp\big\{0.5E_{\text{Short},i,t-}-0.2A_{t}(1+D_{\text{LkDn},t} )\big\}  \label{Sim_I} , 
\end{align*}
where $E_{\text{Short},i,t-}$ is the ratio of zero expenditures across categories in the previous shopping, reflecting short-term consumer behavior. The current intervention $A_{t}$ and its modifier based on $D_{\text{LkDn},t}$ also exist in the intensity model. Obtaining a consistent visiting intensity parameter $\hat{\gamma}$ for $\gamma_{0}$ can be easily implemented by \texttt{coxph} in R. We obtain the intensity parameter $\gamma$ from two models: (i) $\gamma_{C}$ from a correctly specified intensity model including all factors $\{E_{\text{Short},i,t-},A_{t},A_{t}\times D_{\text{LkDn},t}\}$, and (ii) $\gamma_{W}$ from a partial intensity model with $\{E_{\text{Short},i,t-},A_{t}\}$, ignoring the interaction term $A_{t}\times D_{\text{LkDn},t}$.

To generate expenditures corresponding to the first visit, we assume there are 4 different categories ($K=4$), making the outcome expenditure of an individual $i$ at time $t$ a length-4 vector $Y_{i,t}=(Y_{i,t1},\dots, Y_{i,t4})\Tra\in\Real^{4}$. The target model (\ref{snmm}) of interest is specified by
\begin{align*}
    \exp \{m_{a}(\Olit;\psi_{k})a_{t}\}=&\exp\big[\{\underbrace{I_{\text{$k$th Category}}(\psi_{k0}+ \psi_{k1} D_{\text{LkDn},t})}_{\text{Category-specific Intervention Effects}} +\underbrace{\psi_{(1)}I_{\text{LkDn},t}+\psi_{(2)}E_{\text{Short},i,t-}}_{\text{Modifiers Shared by All Categories}}\}a_{t}\big].
\end{align*}
$\psi_{0}=(\psi_{10},\psi_{11},\psi_{20},\psi_{21},\psi_{30},\psi_{31},\psi_{40},\psi_{41},\psi_{(1)},\psi_{(2)})\Tra$, capturing the default intervention effect per category and its modifications over time after the lockdown initiation. We also consider the ongoing nationwide lockdown and previous shopping patterns of each consumer as common factors shared by all categories. Each element of $\psi_{0}$ is a summation of the parameters from the probability of observing positive outcomes $\psi^{p}_{0}=(\psi^{p}_{10},\dots,\psi^{p}_{(2)})\Tra$ and the mean outcome $\psi^{y}_{0}=(\psi^{y}_{10},\dots,\psi^{y}_{(2)})\Tra$, i.e., $\psi_{0}=\psi^{p}_{0}+\psi^{y}_{0}$. In this simulation, we set the following parameters:
\begin{equation*}\label{P1}
    \tag{P1}\left\{\begin{split}
    \psi^{p}_{0}&=(0.5,0.5,-0.5,-0.5,0.15,0.15,-0.15,-0.15,0,0)\Tra\\
    \psi^{y}_{0}&=(-0.15,-0.15,0.15,0.15,-0.5,-0.5,0.5,0.5,0.35,0.35)\Tra \\
    \psi_{0}&= \psi^{p}_{0}+ \psi^{y}_{0}=(0.35,0.35,-0.35,-0.35,-0.35,-0.35,0.35,0.35,0.35,0.35)\Tra
    \end{split}\right\}
\end{equation*}
\begin{equation*}\label{P2}
    \tag{P2}\left\{\begin{split}
    \psi^{p}_{0}&=(0.15,0.15,-0.1,-0.1,0.05,0.05,-0.2,-0.2,0,0)\Tra\\
    \psi^{y}_{0}&=(0.2,0.2,-0.25,-0.25,0.3,0.3,-0.15,-0.15,0.35,0.35)\Tra \\
    \psi_{0}&= \psi^{p}_{0}+ \psi^{y}_{0}=(0.35,0.35,-0.35,-0.35,0.35,0.35,-0.35,-0.35,0.35,0.35)\Tra
    \end{split}\right\}
\end{equation*}

\begin{figure}[H]
\begin{center}
   \includegraphics[width=5in]{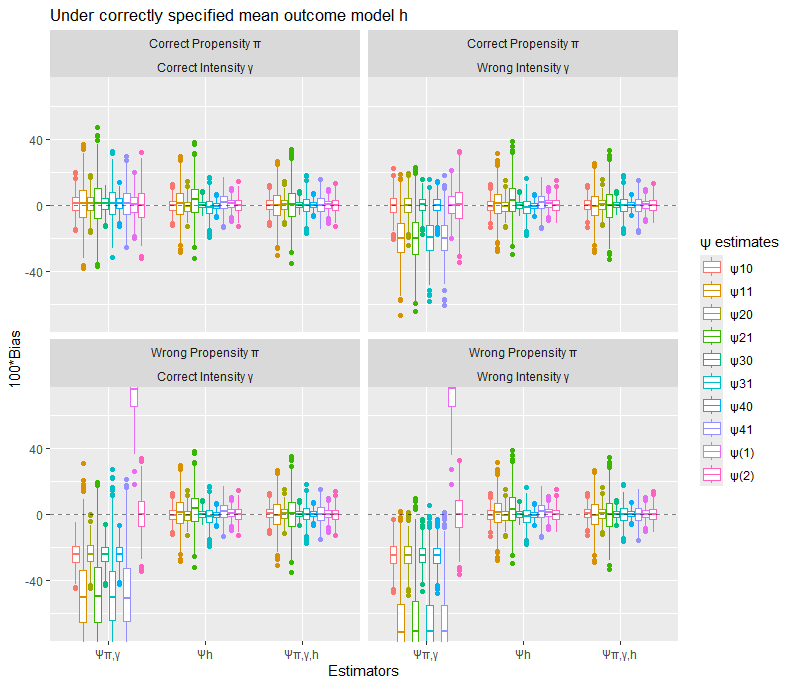} 
   \includegraphics[width=5in]{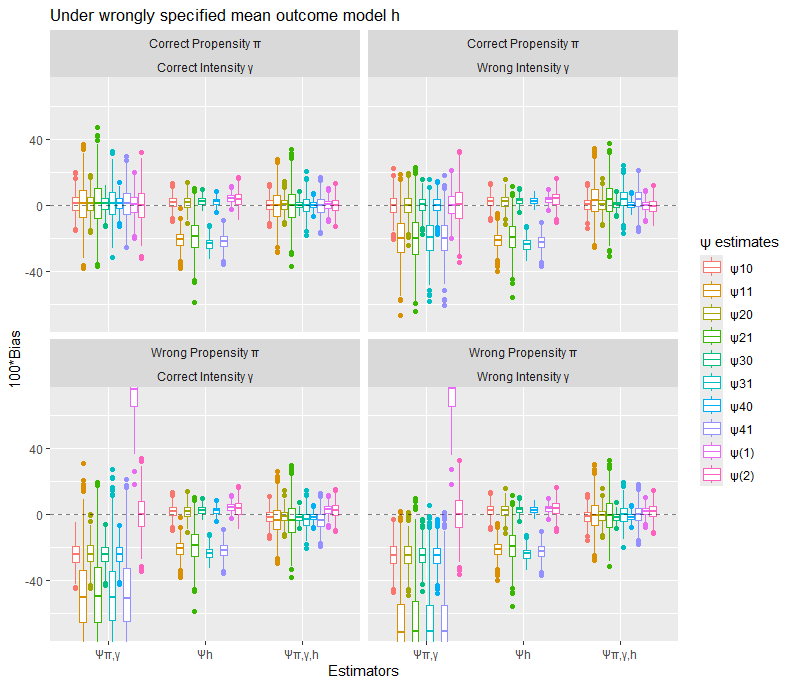}
\end{center}
\caption{\label{simfig:revm1n600tau200} Boxplots of the estimators under (\ref{P1}), $n=600$ and $T=200$. 
Our proposed estimator is $\Psi_{\pi,\gamma,h}$. If both propensity model $\pi$ and intensity parameter $\gamma$ are correctly specified, $\Psi_{\pi,\gamma}$ and $\Psi_{\pi,\gamma,h}$ are consistent. If $h$ is correctly specified, $\Psi_{h}$ and $\Psi_{\pi,\gamma,h}$ are consistent. 
}
\end{figure}
\subsection{Result}

We have assigned coefficients with the same magnitude to $\psi_{0}$ for simplicity and fair comparison across settings (\ref{P1}) and (\ref{P2}). In (\ref{P1}), where the signs of $\psi^{p}_{0}$ and $\psi^{y}_{0}$ are different, or equivalently, the intervention affects the zero-inflation probability and the conditional mean outcome in the same direction. Under (\ref{P2}), the sign of each element in $\psi^{p}_{0}$ and $\psi^{y}_{0}$ is consistent, implying that the intervention influences the zero-inflation probability and the mean outcomes in opposite directions. Other settings are feasible if they do not compromise computational stability.

We use the conditional normal distribution to generate expenditures that satisfy not only the mean model but also both within-time and between-time correlations for the positive values. Due to the complexity of the data generation and unknown $m_{0}^{y}(\Olit)$ and $m_{0}^{p}(\Olit)$, parametric models cannot be used to estimate $h(\Olit)$. We employ the gam method again to estimate $h(\Olit)$ nonparametrically with the following models: (i) a correctly specified $h_{C}(\Olit)\sim 1+Y^{+}_{t-}+s(I_{\text{LkDn},t},D_{\text{LkDn},t}, E_{\text{Short},i,t-})$ including all factors, where $Y^{+}_{t-}\in\Real^{4}$ is the previously observed non-zero expenditure, and (ii) a wrongly specified propensity $h_{W}(\Olit)\sim 1+s(E_{\text{Short},i,t-},k=5)$. For some categories with non-extreme zero-inflation probability, we further decompose $h(\Olit)$ into $\mathbb{E}( Y_{i,k,t}>0 |\Olit,{A}_{t}=0)\mathbb{E}( Y_{i,k,t} |\Olit,{A}_{t}=0, Y_{i,k,t}>0)$ for stability.

\begin{table}[H]
\caption{Causal parameter $\psi$ estimation results for parameter settings (\ref{P1}) and (\ref{P2}) under the correctly specified nuisance functions $(\pi,\gamma,h)$. Note: Bias (mean of the estimates minus true value), SSD (sample standard deviation of the estimates), ESE (mean of estimated standard error), and CP (empirical 95\% coverage probability). Note that all quantities are multiplied by $10^2$.  \label{simtab}}
\begin{center}
\scriptsize
\begin{tabular}{c|cc|cccc|ccccc}
  
  \multicolumn{3}{c|}{\multirow{2}{*}{$\Psi_{\pi,\gamma,h}$}}  & \multicolumn{4}{c|}{{n=200 and T=100}} & \multicolumn{4}{c}{{n=600 and T=200}} \\ 
  \multicolumn{3}{c|}{} & \shortstack{\textcolor{black}{BIAS}\\ $\times 10^2$} & \shortstack{\textcolor{black}{SSD}\\ $\times 10^2$} & \shortstack{\textcolor{black}{ESE}\\ $\times 10^2$} & \shortstack{CP\\ $\times 10^2$} & \shortstack{\textcolor{black}{BIAS}\\ $\times 10^2$} & \shortstack{\textcolor{black}{SSD}\\ $\times 10^2$} & \shortstack{\textcolor{black}{ESE}\\ $\times 10^2$} & \shortstack{CP\\ $\times 10^2$} \\ 
  \hline
  \multirow{10}{*}{(\ref{P1})} & \multirow{8}{*}{\shortstack{Category \\ Specific} } &$\hat{\psi}_{10}$ & 0 & 9.1 & 9.2 & 94.4 & 0.1 & 3.8 & 3.8 & 94.1 \\ 
  && $\hat{\psi}_{11}$ & -0.3 & 21.5 & 20.8 & 94 & -0.3 & 8.8 & 8.4 & 94.5 \\ 
  && $\hat{\psi}_{20}$ & 0.5 & 9.5 & 9.8 & 95.7 & 0 & 3.8 & 4 & 96.8 \\ 
  && $\hat{\psi}_{21}$ & -1.5 & 28.6 & 27.3 & 93.9 & 0.1 & 10.7 & 11.1 & 95.5 \\ 
  && $\hat{\psi}_{30}$ & 0 & 5.4 & 5.3 & 94 & 0 & 2.3 & 2.3 & 94.6 \\ 
  && $\hat{\psi}_{31}$ & 0.1 & 13 & 12.9 & 94.5 & -0.2 & 5.3 & 5 & 92.7 \\ 
  && $\hat{\psi}_{40}$ & 0 & 5.1 & 5 & 93.7 & 0 & 2.2 & 2.2 & 95.1 \\ 
  && $\hat{\psi}_{41}$ & 0.8 & 13 & 12.9 & 94.5 & 0.1 & 5.4 & 5.3 & 93.5 \\ \cline{2-11}
  &\multirow{2}{*}{\shortstack{Shared \\ Modifiers} }& $\hat{\psi}_{(1)}$ & -0.3 & 7.4 & 7.1 & 93.5 & 0.2 & 3.1 & 2.9 & 93.3 \\ 
  && $\hat{\psi}_{(2)}$ & -0.3 & 10.1 & 9.7 & 94.3 & -0.3 & 4.1 & 4.1 & 94.7 \\ 
  \hline
  
  \multirow{10}{*}{(\ref{P2}) } & \multirow{8}{*}{\shortstack{Category \\ Specific} } &$\hat{\psi}_{10}$ & 0.1 & 106 & 10.3 & 94.4 & 0.1 & 4.2 & 4.2 & 94.6 \\ 
  && $\hat{\psi}_{11}$ & -2.3 & 25.4 & 24.5 & 92.8 & -0.1 & 9.9 & 9.8 & 95 \\ 
  && $\hat{\psi}_{20}$ & -0.3 & 8.1 & 7.9 & 93 & 0 & 3.4 & 3.3 & 93.1 \\ 
  && $\hat{\psi}_{21}$ & -0.6 & 20.8 & 19.6 & 93.5 & 0.1 & 8.2 & 7.9 & 93.3 \\ 
  && $\hat{\psi}_{30}$ & 0 & 5.9 & 5.6 & 93.3 & 0.1 & 2.3 & 2.4 & 95.8 \\ 
  && $\hat{\psi}_{31}$ & -0.4 & 14.8 & 13.5 & 92.5 & 0 & 5.4 & 5.5 & 95 \\ 
  && $\hat{\psi}_{40}$ & 0 & 4.9 & 4.9 & 94.3 & -0.1 & 2.1 & 2.1 & 95.9 \\ 
  && $\hat{\psi}_{41}$ & -0.8 & 13.7 & 13.1 & 93.3 & 0.1 & 5.3 & 5.3 & 93.7 \\ \cline{2-11}
  &\multirow{2}{*}{\shortstack{Shared \\ Modifiers} }& $\hat{\psi}_{(1)}$ & 0.5 & 7.2 & 6.8 & 92.6 & 0 & 2.7 & 2.8 & 95.3 \\ 
  && $\hat{\psi}_{(2)}$ & 0 & 9.3 & 9 & 94.7 & 0.2 & 3.8 & 3.8 & 94.5 \\ 
   
\end{tabular}
\end{center}
\end{table}
We first establish that our proposed estimator offers additional safeguards for consistent estimates, as previously asserted; refer to Figure \ref{simfig:revm1n600tau200}. Due to space constraints, we only present results from setting (\ref{P1}) with $n=600$ and $T=200$ in this subsection, while additional results are available in the Appendix. When $h(\Olit)$ is correctly specified using the gam model with all prognostic factors included, both estimators using $\Psi_{h}$ and $\Psi_{\pi,\gamma,h}$ produce unbiased estimates for the true causal parameter $\psi_{0}$, irrespective of whether the propensity $\pi(\overline{G}_{t})$ or intensity parameter $\gamma$ is correctly specified. However, the estimator using $\Psi_{\pi,\gamma}$ necessitates the correct specification of both $\pi(\overline{G}_{t})$ and $\gamma$ since it excludes $h(\Olit)$. Conversely, when $h(\Olit)$ is incorrectly specified, estimates obtained by solving $\Psi_{h}$ consistently exhibit bias, whereas the doubly robust estimator based on $\Psi_{\pi,\gamma,h}$ continues to provide consistent estimates for $\psi_{0}$ if both $\pi(\overline{G}_{t})$ and $\gamma$ are correctly specified. Additionally, our proposed estimator $\Psi_{\pi,\gamma,h}$ demonstrates lower bias levels under incorrect $h(\Olit)$ specifications, even if either $\pi(\overline{G}_{t})$ or $\gamma$ is incorrectly specified.

We then verify the rate-double robustness of the proposed estimator, assuming the correct specification of all nuisance functions $(\pi, \gamma, h)$ through the use of flexible nonparametric/semiparametric methods for their estimation, as shown in Table \ref{simtab}. It is important to note that the convergence of the proposed estimator depends on both the sample size $n$ and the time period $T$. Even with smaller values such as $n=200$ and $T=100$, the bias level is not substantial and the coverage probability remains around $95\%$, albeit with some instability. However, as both $n$ and $T$ increase to $n=600$ and $T=200$, there is a significant improvement in both bias and coverage probability. This indicates that the proposed estimator allows for the use of a conventional sandwich variance estimator without accounting for the variability introduced by the flexible estimation methods for the nuisance functions. Additional simulation results across varied settings and corresponding R code are available in the Appendix, along with a GitHub repository.

In summary, the proposed estimator demonstrates consistency if either $h(\Olit)$ or $\{\pi(\overline{G}_{t}),\gamma\}$ is correctly specified, making it the preferred choice across all simulation scenarios. Furthermore, it is a rate-double robust estimator, enabling the use of a straightforward sandwich-type variance estimator for its asymptotic variance. The simulation results support the use of our proposed estimator to investigate the impact of reduced mobility during the pandemic in the subsequent section.

\section{Application to the Indian Consumer Spending Data}\label{sec:app}

\subsection{Data and variables}
We access 22,666 eligible consumers between February 15 and July 14, 2020, within a single retail chain market in India. Our analysis is focused on in-store transactions particularly, spanning 15 different branches located in Alappuzha, Bangalore Urban (BGL Urban), Ernakulam, Faridkot, Gurgaon, Hyderabad, Indore, Kolkata, Mysuru, North West Delhi (NW Delhi), Sahibzada Ajit Singh Nagar (SAS Nagar), South 24 Parganas (PGS 24 (S)), South Delhi (S Delhi), Thane, and West Delhi (W Delhi). We denote the set of all regions, including Alappuzha as the default region, as $\mathcal{R}$ and the set without Alappuzha as $\mathcal{R}_{-1}$. The study incorporates temporal variables such as the nationwide lockdown in India from March 24th to May 31st \citep{Jeffrey2020}, as well as gazetted holidays (10 Mar, Holi; 2 Apr, Ram Navami; 6 Apr, Mahavir Jayanti; 10 Apr, Good Friday; 7 May, Budha Purnima; 25 May, Id-ul-Fitr) with Sundays during the research period. Additional details and exploratory data analysis results can be found in the Appendix.

While our framework theoretically accommodates the incorporation of all past information as time progresses, it is impractical to include all such information during model fitting due to the increasing number of arguments. Researchers can strategically work within finite-dimensional factors by employing strategies such as utilizing summary measures or imposing restrictions on the inclusion of lagged variables in practice. In this analysis, the set of covariates utilized for modeling comprises individualized covariates, such as $I_{\text{Male},i}$ (an indicator taking the value 1 for males), $E_{\text{Long},i,t-}$ (the logarithm of the average observed expenditures up to time $t-$, reflecting long-term consumer behavior), and $E_{\text{Short},i,t-}$ (the ratio of zero expenditures across categories in the previous shopping, reflecting short-term consumer behavior). Additionally, regional covariates include $I_{\text{Hm}}$ (an indicator for hypermarkets), $R(\overline{A}_{t-})$ (the ratio of significant mobility restriction in the past 7 days), $I_{\text{Hday}, t}$ (an indicator for gazetted holidays and Sundays), $I_{\text{LkDn}, t}$ (an indicator of nationwide lockdown), and $D_{\text{LkDn}, t}$ (proportion of the cumulative days after the initiation of the nationwide lockdown), with indicators denoted as $\{I_{r}:r\in\mathcal{R}\}$ for the 15 regions. Most factors have scales ranging from $0$ to $1$ for further computational stability.

\subsection{Specification of models}
Let us briefly introduce models for the nuisance parameters. The propensity $\pi(\overline{G}_{t})$ estimation is implemented based on the following gam model below:
\begin{align*}
    A_{t}|\overline{G}_{t} \sim  \big(\sum_{r\in\mathcal{R}}{I_{r}}\big) +  I_{\text{Hday},t} + I_{\text{LkDn},t}+ s(D_{\text{LkDn},t})+ s(R(\overline{A}_{t-})) . 
\end{align*}
We have not taken into account the interaction between the regional indicator and other variables due to its complexity. The goodness of fit appears to be satisfactory, as evidenced by the adjusted R-squares around (0.862) and the proportion of deviance explained around (81.3$\%$). It's worth noting that while the presence of smoothing terms makes the interpretation of the model less straightforward, this is acceptable as our primary focus is on ensuring the correct specification of the model rather than its interpretability.

Visiting intensity parameters are estimated using the \texttt{coxph} in R. All the available information $\Olit$, current $A_{t}$, and the intervention modifiers are plugged into the visiting intensity fitting. More specifically, the intensity model in Assumption \ref{ass:vis} is specified as 

\begin{align*}
     m_{v}(\Olit,A_{t})\Tra \gamma_{0} = \big\{ &\underbrace{\{I_{r}:r\in\mathcal{R}_{-1}\},  R(\overline{A}_{t-}), I_{\text{LkDn},t}}_{\text{Regional covariates}}, \underbrace{I_{\text{Male},i},  E_{\text{Long},i,t-}, E_{\text{Short},i,t-}}_{\text{Individual covariates}},\\
     & \underbrace{A_{t}\times I_{\text{Hm}},A_{t}\times R(\overline{A}_{t-}), A_{t}\times I_{\text{LkDn},t}, A_{t}\times I_{\text{Male},i}, A_{t}\times E_{\text{Long},i,t-}, A_{t}\times E_{\text{Short},i,t-}}_{\text{Intervention modifiers}},  \\
     & \underbrace{A_{t}\times 1, A_{t}\times I_{\text{Hday},t}, A_{t}\times D_{\text{LkDn},t}}_{\text{Main intervention effects}}\} \gamma_{0}
\end{align*}
For conditional mean function $h$, fully multidimensional interactions in gam model are not feasible due to the higher number of covariates, i.e., the sheer volume of potential combinations among the variables becomes impractical. We have established the specific gam model used to fit $h(\Olit)$ for category $k$ within the set ${1,\dots, 7}$ as follows:
\begin{align*}
    Y_{i,k,t}|\Olit,A_{t}=0 \sim\;& \big(\sum_{r\in\mathcal{R}}{I_{r}}\big)+s\big\{R(\overline{A}_{t-})\big\}+I_{\text{LkDn},t}+I_{\text{Hday},t}+s(D_{\text{LkDn},t}) \\
    & +I_{\text{Male},i}+s(E_{\text{Long},i,t-})+ s(E_{\text{Short},i,t-}). 
\end{align*}
where $A_{t}=0$. Note that we have decomposed $h(\Olit)$ into two parts, zero-inflation probability and positive expenditure, based on a large number of consumers. We impose a linear model for $m_{a}(\Olit;\psi_{k})$ in (\ref{snmm}) with a conformable causal parameter vector $\psi_{k}$:
\begin{align*}
    m_{a}(\Olit;\psi_{k})=&\underbrace{I_{\text{$k$th Category}}\times (\psi_{k0}+ \psi_{k1}I_{\text{Hday},t}+ \psi_{k2} D_{\text{LkDn},t})}_{\text{Category-specific intervention effects}}\\
     & +\underbrace{\psi_{(1)}I_{\text{Hm}}+\psi_{(2)} R(\overline{A}_{t-})+\psi_{(3)}I_{\text{LkDn},t}+\psi_{(4)}I_{\text{Male},i}+\psi_{(5)}E_{\text{Long},i,t-}+\psi_{(6)}E_{\text{Short},i,t-}}_{\text{Intervention modifiers shared by all categories}},
\end{align*}
for $k\in\{1,\dots, 7\}$. This approach enables the segmentation of the model into components specific to each category and components shared across categories. Specifically, the primary intervention effects resulting from mobility restriction for the $k$-th category are disaggregated into three subcomponents: (i) the overall causal effect $\psi_{k0}$, (ii) an additional effect during holidays $\psi_{k1}$, and (iii) an extra linear-in-time effect originating from the initiation of lockdown $\psi_{k2}$. This specification allows for a nuanced understanding of the multifaceted causal impact of significant mobility restriction on spending within each categorical domain. Furthermore, $\{I_{\text{Hm}}, R(\overline{A}_{t-}),I_{\text{LkDn},t},I_{\text{Male},i},E_{\text{Long},i,t-},E_{\text{Short},i,t-}\}$, the factors shared across categories, are incorporated to refine the causal models for enhanced precision.

\subsection{Results}

\begin{figure}
\begin{center}
\includegraphics[width=6in]{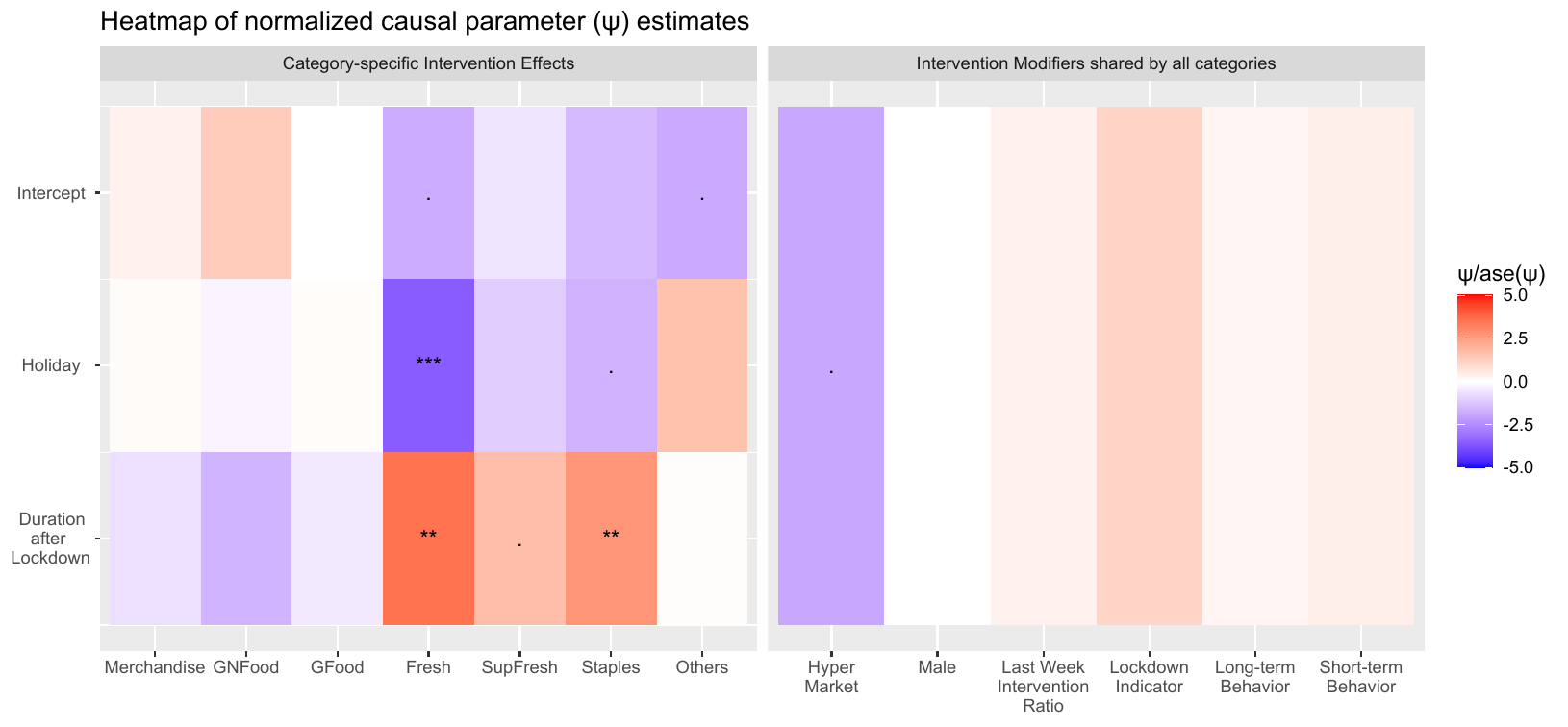}
\end{center}
\caption{\label{fig:heatmap} Heatmap of causal parameter $\psi$ estimates normalized by their asymptotic standard error estimates. The left-hand-sided plot is for category-specific factors, and the right-hand-sided plot is for common factors shared by all categories. 
}
\end{figure}

Substantial decreases in mobility have a profound impact on how consumers allocate their fresh food spending, as illustrated in Table \ref{tab:Psi}. To maintain a familywise error rate below $5\%$ across 27 multiple tests for causal parameter estimates, we have restricted our analysis to factors with p-values of $0.0018$ or less, using the Bonferroni Correction; these factors are shaded in gray in Table \ref{tab:Psi}. Notably, a significant decrease in mobility by $45\%$ results in a decrease in average spending on fresh food during holiday periods. However, after the nationwide lockdown is implemented, spending gradually increases, indicating the possibility of inflation beyond the baseline over time. This trend underscores an increased demand for fresh food items when mobility is significantly reduced, likely due to people spending more time at home during interventions. Figure \ref{fig:heatmap} offers a clear visual representation of the estimates detailed in Table \ref{tab:Psi}. While the desired level of significance has not been attained due to the multiplicity correction, except for the fresh food category; however, analogous trends to those observed in fresh food expenditure are evident in the super fresh food and staples categories. Specifically, regional mobility restriction causally induces a slight reduction in holiday spending but an increase in expenditure as time progresses after the initiation of the nationwide lockdown. This lends further credence to the notion that the new normal has altered people's consumption patterns in the food-related category. 

\begin{table}[H]
\caption{Causal parameter $\psi$ estimates, their standard errors, and the corresponding p-value with stars. Note: *** if the p-value is less than 0.001, ** if it is less than 0.01, * if it is less than 0.05, and . if it is less than 0.1. In order to maintain a family error rate below $5\%$ while conducting 27 multiple tests for causal parameter estimates, we exclusively consider factors with p-values less than $0.0018$ in our analysis. This criterion is satisfied by gray-shaded factors. Abbreviations: LkDn, Lockdown; Hm, Hypermarket; Hday, Holiday; RC, Relative change in Google mobility in the area from baseline. \label{tab:Psi}}
\begin{center}
\scriptsize
 \begin{tabular}{ccccccc}
  
  \multicolumn{7}{c}{Causal Parameter $\psi$ Estimation}\\
  \hline
  \multicolumn{7}{c}{Category-specific Intervention Effects} \\
  Category & \multicolumn{2}{c}{$I_{\text{$k$th Category}}$} & \multicolumn{2}{c}{$I_{\text{$k$th Category}}\times I_{\text{Hday},t}$ }  & \multicolumn{2}{c}{$I_{\text{$k$th Category}}\times D_{\text{LkDn},t}$ } \\
    & $\hat{\psi}$ (SE) & p-value & $\hat{\psi}$ (SE) & p-value  & $\hat{\psi}$ (SE) & p-value \\
  \hline
  Merchandise &  0.13 (0.39) & 0.735()  &  0.04 (0.34) & 0.903()  &  -0.41 (0.61) & 0.5() \\ 
  GNFood & 0.28 (0.21) & 0.194() & -0.04 (0.17) & 0.805() & -0.42 (0.26) & 0.103()  \\ 
  GFood & 0 (0.17) & 0.996() & 0.01 (0.11) & 0.918() & -0.07 (0.15) & 0.639() \\ 
  Fresh & -0.43 (0.24) & 0.077(.) & \cellcolor{Gray}-0.88 (0.25) & \cellcolor{Gray}$<$0.001(***) & \cellcolor{Gray}1.44 (0.41) & \cellcolor{Gray} 0.001(**) \\ 
  SupFresh & -0.11 (0.2) & 0.571() & -0.17 (0.16) & 0.295() & 0.44 (0.26) & 0.084(.) \\ 
  Staples & -0.28 (0.19) & 0.141() & -0.25 (0.15) & 0.097(.) & 0.65 (0.24) & 0.007(**) \\ 
  Others & -0.49 (0.27) & 0.066(.) & 1.25 (0.78) & 0.11() & 0.04 (0.66) & 0.953() \\ 
  \hline
  \multicolumn{7}{c}{Remaining Intervention Modifiers} \\  
  & & $\hat{\psi}$ (SE) & p-value & & $\hat{\psi}$ (SE) & p-value  \\
  \hline
  \multirow{3}{*}{\shortstack{Shared by\\ All Categories}}& $I_{\text{Hm}}$ & -0.15 (0.08) & 0.058(.) & 
  $R(\overline{A}_{t-})$  & 0.03 (0.09) & 0.748() \\ & $I_{\text{LkDn},t}$  & 0.12 (0.11) & 0.242() & 
  $I_{\text{Male},i}$ & 0 (0.08) & 0.953() \\
  & $E_{\text{Long},i,t-}$ & 0.01 (0.04) & 0.709() & 
  $E_{\text{Short},i,t-}$ & 0.07 (0.17) & 0.674()\\
  
\end{tabular}
\end{center}
\end{table}

Recall that we define $A_{t}=1$ when the absolute relative percentage change exceeds $45\%$ and $0$ in the main manuscript. We present results for adjacent thresholds around $45\%$, i.e., $\{40\%,42.5\%,47.5\%,50\%\}$, as part of a sensitivity analysis to ensure the robustness and consistency of our findings. Table \ref{tab:senpsi} shows the corresponding estimation results for the target causal parameter $\psi$. The enduringly significant causal parameters in fresh food expenditures, notably the holiday indicator and cumulative time following the nationwide lockdown, are readily apparent across various intervention thresholds. These findings corroborate the main analysis's conclusions based on the $45\%$ threshold in Table \ref{tab:Psi}, demonstrating their general validity across adjacent thresholds and indicating the analysis's robustness.
\begin{table}[H]
\caption{Causal parameter $\psi$ estimates with $A_{t}\in\{I_{RC<-40\%},I_{RC<-42.5\%},I_{RC<-47.5\%},I_{RC<-50\%}\}$, their standard errors, and the p-values. Note: *** if the p-value is less than 0.001, ** if it is less than 0.01, * if it is less than 0.05, and . if it is less than 0.1. The gray-shaded row corresponds to the identified factors in the main analysis using $A_{t}=I_{RC<-45\%}$. Abbreviations: RC, Relative change in Google mobility in the area from baseline. \label{tab:senpsi}}
\renewcommand{\arraystretch}{0.73}
\begin{center}
\scriptsize
\begin{tabular}{ccccccccc}
  
  \multicolumn{9}{c}{Causal Parameter $\psi$ Estimation}\\
   & \multicolumn{2}{c}{$A_{t}=I_{RC<-40\%}$}  & \multicolumn{2}{c}{$A_{t}=I_{RC<-42.5\%}$}  & \multicolumn{2}{c}{$A_{t}=I_{RC<-47.5\%}$} & \multicolumn{2}{c}{$A_{t}=I_{RC<-50\%}$}   \\
  & $\hat{\psi}$ (SE) & p-value & $\hat{\psi}$ (SE) & p-value  & $\hat{\psi}$ (SE) & p-value & $\hat{\psi}$ (SE) & p-value \\
  \hline
  \multicolumn{9}{c}{Category-specific Intervention Effects: Merchandise (Merchandise)} \\
  \hline  
  Intercept & 0.29 (0.38) & 0.444() & -0.11 (0.3) & 0.726() & 0.21 (0.34) & 0.537() & 0.7 (0.81) & 0.39() \\ 
  $I_{\text{Hday},t}$ & 0.1 (0.32) & 0.761() & 0.16 (0.26) & 0.536() & 0.14 (0.31) & 0.66() & 0.15 (0.48) & 0.75() \\ 
  $D_{\text{LkDn},t}$  & -0.7 (0.6) & 0.245() & -0.35 (0.44) & 0.421() & -1.09 (0.55) & 0.049(*) & -1.6 (1.33) & 0.23() \\ 
  \hline
  \multicolumn{9}{c}{Category-specific Intervention Effects: Non-food Grocery (GNFood)} \\
  \hline
  Intercept & 0.3 (0.24) & 0.209() & 0.23 (0.27) & 0.384() & 0.22 (0.21) & 0.297() & 0.24 (0.23) & 0.304() \\ 
  $I_{\text{Hday},t}$ & -0.2 (0.19) & 0.3() & -0.13 (0.22) & 0.545() & 0.14 (0.19) & 0.465() & 0.17 (0.25) & 0.496() \\ 
  $D_{\text{LkDn},t}$ & -0.4 (0.31) & 0.194() & -0.49 (0.35) & 0.166() & -0.59 (0.24) & 0.013(*) & -0.41 (0.3) & 0.165() \\  
  \hline
  \multicolumn{9}{c}{Category-specific Intervention Effects: Food Grocery (GFood)} \\
  \hline 
  Intercept & 0.19 (0.19) & 0.315() & 0.17 (0.21) & 0.411() & 0.01 (0.19) & 0.972() & 0.03 (0.2) & 0.879() \\ 
  $I_{\text{Hday},t}$ & -0.25 (0.13) & 0.057(.) & -0.13 (0.12) & 0.29() & 0.12 (0.14) & 0.385() & 0.04 (0.14) & 0.786() \\ 
  $D_{\text{LkDn},t}$ & -0.08 (0.16) & 0.641() & -0.28 (0.18) & 0.117() & -0.09 (0.17) & 0.593() & 0.2 (0.22) & 0.367() \\ 
    \hline
  \multicolumn{9}{c}{Category-specific Intervention Effects: Fresh Food (Fresh)} \\
  \hline
  Intercept &  -0.34 (0.27) & 0.213() & -0.53 (0.24) & 0.029(*) & -0.34 (0.22) & 0.128() & -0.06 (0.21) & 0.772() \\     \rowcolor{Gray}
  $I_{\text{Hday},t}$ & -0.73 (0.34) & 0.03(*) & -0.67 (0.29) & 0.022(*) & -0.91 (0.29) & 0.002(**) & -0.72 (0.37) & 0.05(.) \\     \rowcolor{Gray}
  $D_{\text{LkDn},t}$ & 1.29 (0.48) & 0.008(**) & 1.48 (0.47) & 0.002(**) & 1.23 (0.45) & 0.006(**) & 1.06 (0.51) & 0.04(*) \\ 
    \hline
  \multicolumn{9}{c}{Category-specific Intervention Effects: Super Fresh Food (SupFresh)} \\
  \hline
  Intercept & 0.19 (0.26) & 0.483() & 0.02 (0.28) & 0.953() & -0.06 (0.2) & 0.754() & -0.01 (0.18) & 0.956() \\ 
  $I_{\text{Hday},t}$ & -0.1 (0.19) & 0.603() & -0.06 (0.2) & 0.768() & -0.11 (0.2) & 0.57() & -0.42 (0.16) & 0.006(**) \\ 
  $D_{\text{LkDn},t}$ & -0.02 (0.36) & 0.966() & 0.11 (0.38) & 0.768() & 0.33 (0.32) & 0.295() & 0.82 (0.3) & 0.006(**) \\  
    \hline
  \multicolumn{9}{c}{Category-specific Intervention Effects: Staples (Staples)} \\
  \hline
  Intercept & -0.09 (0.2) & 0.661() & -0.16 (0.22) & 0.469() & -0.26 (0.19) & 0.167() & -0.13 (0.19) & 0.476() \\ 
  $I_{\text{Hday},t}$ & -0.43 (0.16) & 0.008(**) & -0.37 (0.15) & 0.018(*) & -0.27 (0.17) & 0.122() & -0.28 (0.22) & 0.203() \\ 
  $D_{\text{LkDn},t}$ & 0.53 (0.23) & 0.018(*) & 0.46 (0.24) & 0.052(.) & 0.63 (0.26) & 0.013(*) & 0.72 (0.31) & 0.019(*) \\ 
    \hline
  \multicolumn{9}{c}{Category-specific Intervention Effects: Others (Others)} \\
  \hline
  Intercept & -0.06 (0.25) & 0.803() & -0.2 (0.3) & 0.511() & -0.13 (0.29) & 0.649() & 0.22 (0.47) & 0.64() \\ 
  $I_{\text{Hday},t}$ & 0.76 (0.46) & 0.099(.) & 0.69 (0.51) & 0.174() & 0.13 (0.47) & 0.778() & 0.04 (0.56) & 0.942() \\ 
  $D_{\text{LkDn},t}$ & -0.34 (0.47) & 0.477() & -0.21 (0.58) & 0.717() & 0.52 (0.65) & 0.423() & 0.27 (0.84) & 0.748() \\  
  \hline
  \multicolumn{9}{c}{Intervention Modifiers shared by all categories} \\
  \hline
  $I_{\text{Hm}}$ & 0.01 (0.03) & 0.649() & -0.01 (0.03) & 0.834() & 0.07 (0.04) & 0.049(*) & 0.03 (0.04) & 0.465() \\ 
  $R(\overline{A}_{t-})$ & -0.09 (0.08) & 0.234() & 0.03 (0.08) & 0.722() & -0.11 (0.08) & 0.157() & -0.06 (0.07) & 0.375() \\ 
  $I_{\text{LkDn},t}$ & -0.05 (0.06) & 0.454() & -0.17 (0.06) & 0.004(**) & -0.08 (0.07) & 0.242() & -0.09 (0.06) & 0.124() \\ 
  $I_{\text{Male},i}$ & 0.03 (0.04) & 0.441() & 0.01 (0.04) & 0.864() & -0.04 (0.04) & 0.387() & 0.01 (0.04) & 0.794() \\ 
  $E_{\text{Long},i,t-}$ & -0.01 (0.02) & 0.738() & -0.02 (0.02) & 0.343() & -0.06 (0.03) & 0.04(*) & 0.01 (0.03) & 0.765() \\ 
  $E_{\text{Short},i,t-}$ & -0.05 (0.09) & 0.569() & -0.11 (0.09) & 0.228() & -0.15 (0.09) & 0.121() & -0.03 (0.09) & 0.708() \\ 
\end{tabular}
\end{center}
\end{table}

\section{Discussion}\label{sec:disc}

In this study, we propose a multiplicative structural nested mean model for causal modeling of zero-inflated outcomes, enabling the analysis of multivariate outcomes with irregularly spaced observation times. Our approach accommodates time-varying confounders and informative observations. We demonstrate the performance of our proposed estimator through simulations incorporating flexible models for propensity, observation intensity, and conditional mean outcome models, leveraging rate-double robustness. 

Applying our methodology to mobility data provides a deeper understanding of the causal relationships between mobility restrictions and consumer expenditures across various categories. This analysis highlights significant variations in the effects based on category characteristics, holidays, and the time elapsed since lockdowns. Specifically, our findings reveal that fresh food expenditures during holidays were initially negatively impacted by stringent mobility restrictions. This result, which reported short-term reductions in fresh food consumption, aligns with findings from existing studies in various countries (e.g., Italy \cite{bracale2020}, France \cite{deschasaux2021}, Denmark, Germany, and Slovenia \cite{janssen2021}). However, our approach extends this understanding by demonstrating that the observed reductions in fresh food consumption predominantly occurred during holidays, with expenditures rebounding as time progressed.

The versatility of our proposed framework extends to various fields of application. First, it can naturally be extended to commercial analyses measuring the impact of restricting the floating population's magnitude. For instance, it can be applied to investigate the optimal consumer density in a service setting, considering the number of present consumers to influence customers' emotions and behavioral responses in a shop \citep{Michael1991, Rompay2008}. Second, our framework is applicable to any field assessing causal effects on longitudinal outcomes with excessive zeros, irregular, and informative observations. An example is the analysis of longitudinal semicontinuous outcomes, such as examining irregularly but rarely observed insurance claims under a discount strategy.

Despite its merits, our approach also has some limitations. Methodologically, our framework focuses on treatment effects on the mean shift of potential outcomes and does not separately quantify treatment effects on zero-inflation probability and mean positive outcomes. Imposing an additional assumption of total separability of prognostic factors for these quantities can be a solution but may not be practical in real analysis and heavily depends on researchers' expertise. Furthermore, in our current model, we only consider binary treatment, which does not cover multiple treatments or continuous treatments in the model. For example, in the considered data application, we can use the relative mobility reduction in a given area as the continuous treatment variable rather than a dichotomized intervention. However, this may present challenges in specifying the reference group and applying semiparametric or nonparametric methods for estimating nuisance functions due to the limited data.

Although we have conducted extensive simulations to examine various model misspecification scenarios and performed sensitivity analyses concerning different intervention threshold sets in application study, the assessment of fundamental modeling assumptions, such as unmeasured confounding, remains limited within our current framework. In a single-time point setting with an outcome of interest at the end of the study, the existing literature offers several approaches \citep{bross1966, schlesselman1978, rosenbaum1983, yanagawa1984, lin1998, imbens2003, ichino2008}; however, these methods often depend on untestable assumptions or are confined to narrowly defined contexts. Ding and VanderWeele \citep{ding2016} further developed the method by introducing an assumption-free sensitivity analysis using only two sensitivity parameters with straightforward interpretability; however, still it pertains to single-time point analyses. In the longitudinal setting, the available sensitivity analysis literature is quite limited (e.g., \citep{ brumback2004, yang2018}), and none of these approaches is directly applicable to our current framework, which accommodates structural nested models for repeated and irregularly observed outcomes with non-monotone intervention sequences. Further research on sensitivity analysis tailored to structural nested mean models is warranted.

From an application standpoint, another limitation of the study lies in the lack of analysis of demographic subgroups due to the unavailability of private consumer information, such as gender or socioeconomic status. While we observed a general decline in fresh food consumption, there exists contrary evidence suggesting increased consumption of fresh food among females \cite{janssen2021}. Moreover, there is a study reporting the reductions in the demand for fresh products were driven by nationwide income declines induced by COVID-19 interventions \cite{hasan2020}. Future studies exploring the behavior of potential demographic subgroups, assuming improved data accessibility, would allow for more precise and detailed conclusions.

\begin{description}[style=unboxed,leftmargin=0cm,itemsep=1\baselineskip]
  \item[Funding information] S.Y. is partially supported by the National Science Foundation grant SES 2242776.
  \item[Author contributions] All authors have accepted responsibility for the entire content of this manuscript and approved its submission. All authors reviewed the results and approved the final version. All contributed to the conceptualization and methodology. T.H. developed the code, and performed the simulations and data analysis. T.H. drafted the manuscript with input from all coauthors.
  \item[Conflict of interest] Authors state no conflict of interest.
  \item[Data availability statement] The data that support the findings of this study are available from P.G., but restrictions apply to the availability of these data due to consumer privacy, and so are not publicly available. Data are however available from the authors upon reasonable request and with permission of P.G.
\end{description}

\bibliographystyle{vancouver}
\bibliography{bib}

\appendix
\section*{Appendix}
This Appendix comprises three sections. Section \ref{secA} presents the proofs related to Theorem 2. Section \ref{SecB} offers insights into conditional mean outcome generation and additional simulation results accompanied by Figures and Tables. Lastly, Section \ref{SecC} delves into further details concerning Google mobility data and motivational data. It also includes the results of intensity parameter estimation.

\section{Proofs} \label{secA}
\subsection{Proof of Lemma 1}
Under model (\ref{snmm}), provided that Assumptions \ref{ass:cons} - \ref{ass:trt} are satisfied, the following equality holds for all $k\in\{1,\dots,K\}$ and $t$, 
    \begin{align*}
        \mathbb{E}\big\{H_{k}(\overline{O}_{i,t},\psi_{0})|\Olit,{A}_{t}={a}_{t} \big\} & = \mathbb{E}\big[ Y_{i,k,t} \exp\big\{ -m_{a}(\Olit;{\psi_{k0}})A_{t}\big\}|\Olit,{A}_{t}={a}_{t} \big] \\
        & = \exp\big\{ -m_{a}(\Olit;{\psi_{k0}})a_{t}\big\} \mathbb{E}( Y_{i,k,t} |\Olit,{A}_{t}={a}_{t})\\
        & = \exp\big\{ -m_{a}(\Olit;{\psi_{k0}})a_{t}\big\} \mathbb{E}\{ Y_{i,k,t}(\overline{a}_{t}) |\Olit,{A}_{t}={a}_{t}\}\\
        & = \mathbb{E}\{ \YkpotDo |\Olit,{A}_{t}={a}_{t}\}\quad \\
        & = \mathbb{E}\{  \YkpotDo |\Olit\}\\
        & = \mathbb{E}\big[ \mathbb{E}\{ \YkpotDo |\Olit,A_{t})\}|\Olit\big] \\
        & = \mathbb{E}\big [ \mathbb{E}\{H_{k}(\overline{O}_{i,t},\psi_{0}) |\Olit,A_{t}\}|\Olit \big] \quad \\ 
        & =  \mathbb{E}\{H_{k}(\overline{O}_{i,t},\psi_{0}) |\Olit\}.
    \end{align*}

\subsection{Proof of Theorem 2}

As the number of regions is fixed, we assume that a single region exists and contains all subjects for simplicity. Recall that  we can achieve $\|\hat{h}-h_{0}\|_{2,P}=\op(n^{-1/4})$, $\|\hat{\gamma}-{\gamma}_{0}\|_{2}=\Op(n^{-1/2})$, and $\|\hat{\pi}-\pi_{0}\|_{2,P}=\op(T^{-1/4})$ utilizing nonparametric/semiparametric estimators for the nuisance functions $(\pi,h,\gamma)$, such as generalized additive models. Under the Condition \ref{ass:rc3}, we can establish that $\|\hat{h}-h_{0}\|_{2,P}=\op(T^{-1/4})$ and $\|\hat{\gamma}-{\gamma}_{0}\|_{2}=\Op(T^{-1/2})$ as $n$ diverges to infinity at a rate comparable to that of $T$ or much larger than $T$. Let $\delta>0$ be an arbitrary positive value. Since $\|\hat{h}-h_{0}\|_{2,P}=\op(n^{-1/4})$, for any $\epsilon>0$, there exists a large $M>0$ such that
\begin{align*}
    P(\|\hat{h}-h_{0}\|_{2,P}>\delta n^{-1/4})<\epsilon,
\end{align*}
for any $n>M$. Using $\lim_{T\rightarrow \infty}n/T=c>0$, we find $N_{1}\in \mathbb{N}$ such that
\begin{align*}
    \delta T^{-1/4}>\delta \bigg( \frac{2n}{c}\bigg)^{-1/4},
\end{align*}
for all $T>N_{1}$ where $c$ is replaced as an arbitrary large number if $\lim_{n\rightarrow \infty}n/T=\infty$. In addition, we find $N_{2}\in\mathbb{N}$ satisfying
\begin{align*}
    \frac{2n}{c}>M,
\end{align*}
for all $T>N_{2}$ as $n\nearrow \infty$. Then, it completes the proof by showing
\begin{align*}
       P(\|\hat{h}-h_{0}\|_{2,P}>\delta T^{-1/4})<P(\|\hat{h}-h_{0}\|_{2,P}>\delta n^{-1/4})<\epsilon ,
\end{align*}
for any $T>\max(N_{1},N_{2})$, i.e., $\|\hat{h}-h_{0}\|_{2,P}=\op(T^{-1/4})$. It can be shown that $\|\hat{\gamma}-\gamma_{0}\|_{2,P}=\op(T^{-1/4})$ under $\lim_{T\rightarrow \infty}n/T=c>0$ in the same way. We can demonstrate that $\|\hat{\pi}-\pi_{0}\|_{2,P}=\op(n^{-1/4})$ in a similar manner. Therefore, we now have $\|\hat{h}-h_{0}\|_{2,P}=\op(T^{-1/4})$ and $\|\hat{\gamma}-\gamma_{0}\|_{2,P}=\op(T^{-1/4})$. 

Assuming Condition \ref{ass:rc2} and \ref{ass:rc3} hold, we have the following Lemmas. The proof of these Lemmas is provided in the subsequent subsections.
\begin{lemma}\label{lem1}
    Assuming the Condition \ref{ass:rc2} and \ref{ass:rc3} hold, we have
    \begin{align*}
        & \sup_{\psi\in\Theta, \vartheta\in\Xi_{\vartheta_{0}}}\|\EPn\Psi(\psi,\vartheta)-\EP\Psi(\psi,\vartheta)\|_{2}= \op(1), \text{ and}\\
        & \sup_{\psi\in\Theta, \vartheta\in\Xi_{\vartheta_{0}}}\|\EPn \dot{\Psi}_{\psi}(\psi,\vartheta)-\EP\dot{\Psi}_{\psi}(\psi,\vartheta)\|_{2}= \op(1),
    \end{align*}
    as $T \rightarrow \infty$. 
\end{lemma}
\begin{lemma}\label{lem2}
    Assuming the Condition \ref{ass:rc2} and \ref{ass:rc3} hold, we have
    \begin{align*}
        & \EGn\Psi(\psi_{0},\vartheta)\wc Z\in l^{\infty}(\Xi_{\vartheta_{0}}),
    \end{align*}
\end{lemma}
as $T\rightarrow \infty$ where the limiting process $Z=\{Z(\vartheta):\vartheta\in \Xi_{\vartheta_{0}}\}$ is mean-zero multivariate Gaussian process, and the sample paths of $Z$ belong to a set $S\subset l^{\infty}(\Xi_{\vartheta_{0}})$ of uniformly continuous functions with respect to $\|\cdot \|_{2,P}$. 

We denote $\hat{\psi}_{}$ as the estimates of the proposed rate-double robust estimator. Observe that
\begin{align*}
    \|\EP\Psi({\hat{\psi}},{\vartheta}_{0})\|_{2}& \leq \|\EP\Psi({\hat{\psi}},{\vartheta}_{0})-\EP\Psi({\hat{\psi}},\hat{\vartheta})\|_{2}+\|\EP\Psi({\hat{\psi}},\hat{\vartheta})\|_{2}\\
    & =\|\EP\Psi({\hat{\psi}},{\vartheta}_{0})-\EP\Psi({\hat{\psi}},\hat{\vartheta})\|_{2}+\|\EP\Psi(\hat{\psi},\hat{\vartheta})-\EPn\Psi({\hat{\psi}},\hat{\vartheta})\|_{2}\\
    & \leq \|\EP\Psi({\hat{\psi}},{\vartheta}_{0})-\EP\Psi({\hat{\psi}},\hat{\vartheta})\|_{2}\\
    & \quad + \sup_{\psi\in\Theta, \vartheta\in\Xi_{\vartheta_{0}}}\|\EP\Psi({\psi},{\vartheta})-\EPn\Psi({{\psi}},{\vartheta})\|_{2}\\
    & = \|\EP\Psi({\hat{\psi}},{\vartheta}_{0})-\EP\Psi({\hat{\psi}},\hat{\vartheta})\|_{2}+\op(1),
\end{align*}
by Lemma \ref{lem1}. Using Taylor expansion for the first term, we have
\begin{align*}
   \|\EP\Psi({\hat{\psi}},{\vartheta}_{0})-\EP\Psi({\hat{\psi}},\hat{\vartheta})\|_{2}& = \|\EP\{\Psi({\hat{\psi}},{\vartheta}_{0})-\Psi({\hat{\psi}},\hat{\vartheta})\}\|_{2}\\
    & = \| \EP\{ \dot{\Psi}_{\vartheta}(\hat{\psi},\widetilde{\vartheta})(\hat{\vartheta}-\vartheta_{0})   \}\|_{2}\\
    & \leq \EP \|\dot{\Psi}_{\vartheta}(\hat{\psi},\widetilde{\vartheta})(\hat{\vartheta}-\vartheta_{0})  \|_{2}\\
    & \leq \bigg\{\EP \|\dot{\Psi}_{\vartheta}(\hat{\psi},\widetilde{\vartheta})\|_{2}^{2}\times\EP\|(\hat{\vartheta}-\vartheta_{0})\|_{2}^{2}\bigg\}^{1/2}\\
    & \bbc \|\hat{\vartheta}-\vartheta_{0}\|_{2,P}\\
    & =\op(1),
\end{align*}
by Cauchy-Schwarz inequality and Condition \ref{ass:rc2} (v) where $\widetilde{\vartheta}$ lies between the true $\vartheta_{0}$ and the estimate $\hat{\vartheta}$. Combining it with  Condition \ref{ass:rc1}, we can derive the consistency of $\hat{\psi}$ to $\psi_{0}$. Now, using Taylor expansion for the estimating equation $\EPn\{\Psi(\hat{\psi},\hat{\vartheta})\}$ around $\psi_{0}$:
\begin{align*}\label{eq:S2}
    0 & = \EPn\{\Psi(\hat{\psi},\hat{\vartheta})\}\\
    & = \EPn\{\Psi(\psi_{0},\hat{\vartheta})\}+\EPn\{\dot{\Psi}_{{\psi}}(\widetilde{\psi},\hat{\vartheta})\}(\hat{\psi}-\psi_{0}), \tag{S2}
\end{align*}
where $\widetilde{\psi}$ is located between the true $\psi_{0}$ and the estimate $\hat{\psi}$. Due to Lemma \ref{lem1}
\begin{align*}
    \sup_{\psi\in\Theta, \vartheta\in\Xi_{\vartheta_{0}}}\|\EPn \dot{\Psi}_{\psi}(\psi,\vartheta)-\EP\dot{\Psi}_{\psi}(\psi,\vartheta)\|_{2}\cp 0 \text{ as $T$ goes infinity},
\end{align*}
Combined with $\widetilde{\psi}\cp \psi_{0}$ and the consistency of $\hat{\vartheta}$ to $\vartheta_{0}$ in Condition \ref{ass:rc3}, we have
\begin{align*}
    \EPn\{\dot{\Psi}_{{\psi}}(\widetilde{\psi},\hat{\vartheta})\}\cp \EP\{\dot{\Psi}_{{\psi}}({\psi}_{0},{\vartheta}_{0})\},
\end{align*}
as $T$ goes infinity. Therefore,
\begin{align*}
    T^{1/2}(\hat{\psi}-\psi_{0})=-\EP\{\dot{\Psi}_{{\psi}}({\psi}_{0},{\vartheta}_{0})\}\Inv T^{1/2} \EPn\{\Psi(\psi_{0},\hat{\vartheta})\}+\op(1),
\end{align*}
as invertibility specified in Condition \ref{ass:rc2} (vi). Thus, it suffices to analyze $\EPn\{\Psi(\psi_{0},\hat{\vartheta})\}$ to derive asymptotic distribution of $\hat{\psi}$. Observe that
\begin{align}\label{eq:S3}
    T^{1/2}\EPn\{\Psi(\psi_{0},\hat{\vartheta})\}=\EGn\{\Psi(\psi_{0},\hat{\vartheta})\}+T^{1/2}\EP\{\Psi(\psi_{0},\hat{\vartheta})\}. \tag{S3}
\end{align}
Start with the first term in (\ref{eq:S3}). Recall Lemma \ref{lem2} and $\hat{\vartheta}\cp \vartheta_{0}$. Then define a continuous mapping $S:l^{\infty}(\Xi_{{\theta}_{0}})\times \Xi_{{\theta}_{0}} \rightarrow \Real^{p}$ as $S(f,\vartheta)=f(\vartheta)-f(\vartheta_{0})$. As all sample paths of $Z$ are continuous on $\Xi_{{\theta}_{0}}$, we can use the Continuous-Mapping Theorem as
\begin{align*}
    S(Z,\hat{\vartheta})=\EGn\Psi(\psi_{0},\hat{\vartheta})-\EGn\Psi(\psi_{0},\vartheta_{0})\wc  S(Z,{\vartheta}_{0})=0, 
\end{align*}
or equivalently, 
\begin{align*}
    \EGn\{\Psi(\psi_{0},\hat{\vartheta})\}=\EGn\{\Psi(\psi_{0},{\vartheta}_{0})\}+\op(1).
\end{align*}
Denote $T^{-1}\sum_{t=1}^{T}=\lim_{T\rightarrow \infty}T^{-1}\sum_{t=1}^{T}$ for $T=\infty$. We can show that the second term in (\ref{eq:S3}) as follows:
\begin{align*} 
     \EP \Psiphatv & =\mathbb{E} \bigg[ \frac{1}{T}\sum_{t=1}^{T} c(\Olit)\big\{H(\overline{O}_{it},\psi_{0})-\hat{h}(\Olit)\big\}\frac{\{{A_{t}}-\hat{\pi}(\overline{G}_{t})\}\Delta N_{it} }{\exp\big\{ m_{v}(\Olit,A_{t})\Tra \hat{\gamma}\big\}}  \bigg]\\
    & = \mathbb{E}\bigg[ \frac{1}{T}\sum_{t=1}^{T}c(\Olit)\big[\mathbb{E}\{H(\overline{O}_{it},\psi_{0})|\Olit,A_{t}\}-\hat{h}(\Olit)\big]  \frac{\{{A_{t}}-\hat{\pi}(\overline{G}_{t})\}\mathbb{E}(\Delta N_{it}|\Olit,A_{t})}{\exp\big\{ m_{v}(\Olit,A_{t})\Tra \hat{\gamma}\big\}}\bigg] \nonumber\\
    & = \mathbb{E}\bigg[\frac{1}{T} \sum_{t=1}^{T}c(\Olit)\big[\mathbb{E}\{H(\overline{O}_{it},\psi_{0})|\Olit\}-\hat{h}(\Olit)\big] \frac{\{{A_{t}}-\hat{\pi}(\overline{G}_{t})\}\lambda_{0}(t)}{ \underbrace{\exp\big\{ m_{v}(\Olit,A_{t})\Tra (\hat{\gamma}-\gamma_{0})\big\}}_{\overset{let}{=} \nu_{\hat{\gamma}}(\Olit,A_{t}) }}\bigg]\\
    & = \mathbb{E}\bigg( \frac{1}{T}\sum_{t=1}^{T}c(\Olit)\big\{h_{0}(\Olit)-\hat{h}(\Olit)\big\}\\
    & \qquad \qquad \qquad \times { \bigg[ \frac{\{{A_{t}}-\hat{\pi}(\overline{G}_{t})\}}{ \nu_{\hat{\gamma}}(\Olit,A_{t})  }-\{A_{t}-\hat{\pi}({\overline{G}_{t}})\}+A_{t}-\hat{\pi}({\overline{G}_{t}}) \bigg] \lambda_{0}(t)}\bigg)\\
    & = \mathbb{E}\bigg[ \frac{1}{T}\sum_{t=1}^{T}c(\Olit)\big\{h_{0}(\Olit)-\hat{h}(\Olit)\big\}\{A_{t}-\hat{\pi}({\overline{G}_{t}})\} { \bigg\{ \frac{1- \nu_{\hat{\gamma}}(\Olit,A_{t})}{ \nu_{\hat{\gamma}}(\Olit,A_{t})}\bigg\} \lambda_{0}(t)}\bigg]\\
    & \quad +\mathbb{E}\bigg[\frac{1}{T} \sum_{t=1}^{T}c(\Olit)\big\{h_{0}(\Olit)-\hat{h}(\Olit)\big\}\{\pi_{0}(\overline{G}_{t})-\hat{\pi}({\overline{G}_{t}})\} \lambda_{0}(t)\bigg]\\
    & = \mathbb{E}\bigg( \frac{1}{T}\sum_{t=1}^{T}c(\Olit)\big\{h_{0}(\Olit)-\hat{h}(\Olit)\big\}\\
    & \qquad \qquad \qquad \times { \bigg[  \frac{\{A_{t}-\hat{\pi}({\overline{G}_{t}})\}\{1- \nu_{\hat{\gamma}}(\Olit,A_{t})\}}{ \nu_{\hat{\gamma}}(\Olit,A_{t})}}+\pi_{0}(\overline{G}_{t})-\hat{\pi}({\overline{G}_{t}})\bigg]\lambda_{0}(t)\bigg).
\end{align*}
Combining Cauchy-Schwarz inequality and uniform boundedness of functions from Condition \ref{ass:rc2} (iii), the quantity can be bounded as
\begin{align*}
    \| \EP \Psiphatv  \|_{2}&\leq  \frac{1}{T}\sum_{t=1}^{T}\mathbb{E}\bigg(\bigg\| c(\Olit)\big\{h_{0}(\Olit)-\hat{h}(\Olit)\big\} \\
    & \qquad \qquad \qquad \times { \bigg[  \frac{\{A_{t}-\hat{\pi}({\overline{G}_{t}})\}\{1- \nu_{\hat{\gamma}}(\Olit,A_{t})\}}{ \nu_{\hat{\gamma}}(\Olit,A_{t})}}+\pi_{0}(\overline{G}_{t})-\hat{\pi}({\overline{G}_{t}})\bigg]\bigg\|_{2}\bigg)\lambda_{0}(t)\\
    &\leq  \frac{1}{T}\sum_{t=1}^{T}\mathbb{E}\bigg(\bigg\| c(\Olit)\big\{h_{0}(\Olit)-\hat{h}(\Olit)\big\} {  \frac{\{A_{t}-\hat{\pi}({\overline{G}_{t}})\}\{1- \nu_{\hat{\gamma}}(\Olit,A_{t})\}}{ \nu_{\hat{\gamma}}(\Olit,A_{t})}}\bigg\|_{2}\\
    & \qquad \qquad \qquad + \| c(\Olit)\big\{h_{0}(\Olit)-\hat{h}(\Olit)\big\} 
    \{\pi_{0}(\overline{G}_{t})-\hat{\pi}({\overline{G}_{t}})\}\|_{2}\bigg)\lambda_{0}(t)\\
    &\bbc  \frac{1}{T}\sum_{t=1}^{T}\mathbb{E}\bigg(\| c(\Olit)\big\{h_{0}(\Olit)-\hat{h}(\Olit)\big\} {  \{1- \nu_{\hat{\gamma}}(\Olit,A_{t})\}}\|_{2}\\
    & \qquad \qquad \qquad + \| c(\Olit)\big\{h_{0}(\Olit)-\hat{h}(\Olit)\big\} 
    \{\pi_{0}(\overline{G}_{t})-\hat{\pi}({\overline{G}_{t}})\}\|_{2}\bigg)\lambda_{0}(t)\\
    & \bbc \frac{1}{T}\sum_{t=1}^{T}\mathbb{E}\big[\|\big\{h_{0}(\Olit)-\hat{h}(\Olit)\big\}\|_{2} \times \{\|\gamma_{0}-\hat{\gamma}\|_{2}+|\pi_{0}(\overline{G}_{t})-\hat{\pi}(\overline{G}_{t})|\}\big]\lambda_{0}(t)\\
\end{align*}
\begin{align*}
    & \bbc \frac{1}{T}\sum_{t=1}^{T}\mathbb{E}\big[\|\big\{h_{0}(\Olit)-\hat{h}(\Olit)\big\}\|_{2}^{2}\big]^{1/2}\times \big[ \|\gamma_{0}-\hat{\gamma}\|_{2}+\mathbb{E}\big\{|\pi_{0}(\overline{G}_{t})-\hat{\pi}(\overline{G}_{t})|^{2}\big\}^{1/2}\big]\lambda_{0}(t) \\
    & \bbc \bigg(\frac{1}{T}\sum_{t=1}^{T}\mathbb{E}\big[\|\big\{h_{0}(\Olit)-\hat{h}(\Olit)\big\}\|_{2}^{2}\big]^{1/2} \lambda_{0}(t)\bigg)\\
    & \qquad \times\bigg[\|\gamma_{0}-\hat{\gamma}\|_{2}+ \frac{1}{T}\sum_{t=1}^{T}\mathbb{E}\big\{|\pi_{0}(\overline{G}_{t})-\hat{\pi}(\overline{G}_{t})|^{2}\big\}^{1/2}\big]\lambda_{0}(t)\bigg] \\
    & = \|\hat{h}-h_{0}\|_{2,P}\times \big( \|\hat{\gamma}-{\gamma}_{0}\|_{2}+\|\hat{\pi}-\pi_{0}\|_{2,P}\big)=\op(T^{-1/2})
\end{align*}
following the Condition \ref{ass:rc3} for $T\rightarrow \infty$. It implies that equation (\ref{eq:S3}) is equivalent to 
\begin{align*}
    T^{1/2}\EPn\{\Psi(\psi_{0},\hat{\vartheta})\}& =\EGn\{\Psi(\psi_{0},\hat{\vartheta})\}+T^{1/2}\EP\{\Psi(\psi_{0},\hat{\vartheta})\} \\
    & = \EGn\{\Psi(\psi_{0},{\vartheta}_{0})\}+\op(1).
\end{align*}
Now if suffices to show that $\EGn\{\Psi(\psi_{0},{\vartheta}_{0})\}$ follows asymptotic normal distribution:
\begin{align*}\label{eq:S4}
   \EGn\{\Psi(\psi_{0},{\vartheta}_{0})\}\cd N(0,\Sigma)\tag{S4},
\end{align*}
as $T$ goes infinity. For notational convenience, denote that
\begin{align*}
    \Psi_{n(\bullet),t}(\psi,\vartheta)\equiv\frac{1}{n}\sum_{i=1}^{n}\Psi_{i,t}(\psi,\vartheta)\Delta N_{i,t}.
\end{align*}
(\ref{eq:S4}) can be proved by showing that
\begin{align*}\label{eq:S5}
    T^{-1/2}\sum_{t=1}^{T} \Sigma^{-1/2}\Psi_{n(\bullet),t}(\psi_{0},\vartheta_{0})\cd N(0,I),\tag{S5}
\end{align*}
as $T\rightarrow \infty$, due to the assumption that we are working on the prespecified granularities $\{1,\dots,T\}$ rather than continuous time grids. Let $\mathcal{F}_{t}=\{\cup_{i=1}^{n}\overline{L}_{i,t}\}$, then we have $\mathbb{E}\{\Psi_{n(\bullet),t}(\psi_{0},\vartheta_{0})|\mathcal{F}_{t-1} \}=0$, i.e., (\ref{eq:S5}) forms martingale difference sequence with respect to the filtration $\{\sigma(\mathcal{F}_{t})\}_{t\geq 1}$ where $\sigma(\mathcal{F}_{t})$ is the $\sigma$-algebra generated by $\mathcal{F}_{t}$. For asymptotic normality, we utilize the martingale central limit theorem for triangular arrays \citep{Mcleish1974}; therefore, it suffices to confirm the following Conditions \ref{cons1} and \ref{cons2}.
\begin{condition}
    \label{cons1} $\max_{1\leq t \leq T} \| T^{-1/2}\Sigma^{-1/2}\Psi_{n(\bullet),t}(\psi_{0},\vartheta_{0}) \|_{2}\cp 0$ as $T\rightarrow \infty$.
\end{condition}
Condition \ref{cons1} holds as
\begin{align*}
     &\max_{1\leq t \leq T} \|(T\Sigma)^{-1/2}\Psi_{n(\bullet),t}(\psi_{0},\vartheta_{0}) \|_{2}\\
     &\qquad  \leq \|(T\Sigma)^{-1/2}\|_{2} \max_{1\leq t \leq T}\|\Psi_{n(\bullet),t}(\psi_{0},\vartheta_{0}) \|_{2}\\
     &\qquad \leq \|(T\Sigma)^{-1/2}\|_{2} \max_{1\leq t \leq T} \frac{1}{n}\sum_{i=1}^{n}\frac{\|c(\Olit)\|_{2}\|H(\overline{O}_{it},\psi_{0})-{h}_{0}(\Olit)\|_{2}\|A_{t}-\pi_{0}(\overline{G}_{t})\|_{2}}{\|\exp\big\{ m_{v}(\Olit,A_{t})\Tra {\gamma}_{0}\big\}\|_{2}}\\
     &\qquad  =\Op (T^{-1/2}),
\end{align*}
by Condition \ref{ass:rc2} (iii). 
\begin{condition}
    \label{cons2} $T^{-1} \sum_{t=1}^{T}\{\Sigma^{-1/2}\Psi_{n(\bullet),t}(\psi_{0},\vartheta_{0})\}^{\otimes 2}\cp I$ as $T\rightarrow \infty$.
\end{condition}
Let us define $M_{t}= \Psi_{n(\bullet),t}(\psi_{0},\vartheta_{0})^{\otimes 2}-\mathbb{E}[ \Psi_{n(\bullet),t}(\psi_{0},\vartheta_{0})^{\otimes 2}|\mathcal{F}_{t-1}]$. Then $\{M_{t}\}_{i\geq 1}$ constructs a martingale difference sequence with respect to the filtration $\{\sigma(\mathcal{F}_{t})\}_{t\geq 1}$. Because $\mathbb{E}(M_{l}M_{k}\Tra)=0$ for $l\neq k$ and $\mathbb{E}(M_{l}M_{l}\Tra)$ is bounded for all $l$, we have
\begin{align*}\label{eq:S6}
    \bigg\| T^{-1}\sum_{t=1}^{T}\big[\Psi_{n(\bullet),t}(\psi_{0},\vartheta_{0})^{\otimes 2}-\mathbb{E}\{ \Psi_{n(\bullet),t}(\psi_{0},\vartheta_{0})^{\otimes 2}|\mathcal{F}_{t-1}\}\big]  \bigg\|=\op(1) \tag{S7}
\end{align*}
Using (\ref{eq:S6}), we can show that
\begin{align*}
     \bigg\|T^{-1} \sum_{t=1}^{T}\{\Sigma^{-1/2}\Psi_{n(\bullet),t}(\psi_{0},\vartheta_{0})\}^{\otimes 2}-I \bigg\|_{2} & =\bigg\|\Sigma^{-1/2} \bigg\{T^{-1}\sum_{t=1}^{T}\Psi_{n(\bullet),t}(\psi_{0},\vartheta_{0})^{\otimes 2}-\Sigma\bigg\}(\Sigma^{-1/2})\Tra \bigg\|_{2}\\
     &\leq  \|\Sigma^{-1/2}\|_{2}^{2}\bigg\| \bigg\{T^{-1}\sum_{t=1}^{T}\Psi_{n(\bullet),t}(\psi_{0},\vartheta_{0})^{\otimes 2}-\Sigma\bigg\}\Tra \bigg\|_{2}\\
     & =\op(1),
\end{align*}
which proves Condition \ref{cons2}. Therefore, we have shown that 
\begin{align*}
    T^{1/2}(\hat{\psi}-\psi_{0})& =-\EP\{\dot{\Psi}_{{\psi}}({\psi}_{0},{\vartheta}_{0})\}\Inv   \EGn\{\Psi(\psi_{0},{\vartheta}_{0})\}\\
    & \cd N[0, \EP\{\dot{\Psi}_{{\psi}}({\psi}_{0},{\vartheta}_{0})\}\Inv \Sigma\EP\{\dot{\Psi}_{{\psi}}({\psi}_{0},{\vartheta}_{0})\}\Inv],
\end{align*}
as $T$ goes infinity.

\subsection{Proof of Lemma \ref{lem1}}
We set $\mathcal{F}_{t}=\{\cup_{i=1}^{n}\overline{L}_{i,t}\}$. We further define a martingale difference sequence adapted to $\{\sigma(\mathcal{F}_{t})\}_{t\geq 1}$ as follow: for any $(\psi, \vartheta)\in\mathcal{U}$ and $1\leq t \leq T$,
\begin{align*} 
    M_{t}(\theta,\vartheta)= \Psi_{n(\bullet),t}(\psi,\vartheta)-\mathbb{E}[ \Psi_{n(\bullet),t}(\psi,\vartheta)|\mathcal{F}_{t-1}],
\end{align*}
where $\|M_{t}(\theta,\vartheta)\|\leq R$ for some constant $R$ almost surely due to Condition \ref{ass:rc2} (iii). Additionally define two predictable quadratic variation processes for the martingale $M_{t}(\theta,\vartheta)$: for each $1\leq t' \leq T$,
\begin{align*}
    W_{col,t'}=\sum_{t=1}^{t'}\mathbb{E}\{M_{t}(\theta,\vartheta)^{\otimes 2}|\mathcal{F}_{t-1}\}\quad \text{and} \quad W_{row,t'}=\sum_{t=1}^{t'}\mathbb{E}[\{M_{t}(\theta,\vartheta)\Tra\}^{\otimes 2}|\mathcal{F}_{t-1}].
\end{align*}
Again, by Condition \ref{ass:rc2}, 
\begin{align*}
    \max_{1\leq t\leq T} \{ \|W_{col,t}\|_{2},\|W_{row,t}\|_{2}\}\leq TM \quad a.s.,
\end{align*}
for some constant $M$. Then, by the Freedman's inequality for rectangular matrix based on Corollary 1.3 of \cite{tropp2011}, for any $\tau\geq 0$ and $TM>0$,
\begin{align*}
    \EP\bigg\{ \bigg\|\sum_{t=1}^{T}M_{t}(\psi,\vartheta)\bigg\|_{2}\geq \tau \bigg\}\leq (1+p)\exp\bigg\{\frac{\tau^{2}/2}{TM+R\tau/3} \bigg\},
\end{align*}
where $p$ is the length of the target parameter vector $\psi$. This inequality implies that
\begin{align*}\label{eq:S8}
    \bigg\|\sum_{t=1}^{T}M_{t}(\psi,\vartheta)\bigg\|_{2}=\Op(T), \tag{S8}
\end{align*}
when we set $\tau=\sqrt{T\log T}$ for any $(\psi,\vartheta)\in\mathcal{U}$. Due to the continuity of $M_{t}(\psi,\vartheta)$ and Condition \ref{ass:rc2} (i), for any $\lambda>0$, there exists a corresponding finite $\epsilon-$net $\mathcal{U}_{\epsilon}$ satisfying
\begin{align*} \label{eq:S9}
    \sup_{(\psi,\vartheta)\in\mathcal{U}}\frac{1}{T}\bigg\|\sum_{t=1}^{T}M_{t}(\psi,\vartheta)\bigg\|_{2}\leq  \sup_{(\psi,\vartheta)\in\mathcal{U}_{\epsilon}}\frac{1}{T}\bigg\|\sum_{i=1}^{T}M_{t}(\psi,\vartheta)\bigg\|_{2}+\delta. \tag{S9}
\end{align*}
Combining (\ref{eq:S8}) with finite $\mathcal{U}_{\epsilon}$, we derive
\begin{align*}
    \sup_{(\psi,\vartheta)\in\mathcal{U}_{\epsilon}}\frac{1}{T}\bigg\|\sum_{i=1}^{T}M_{t}(\psi,\vartheta)\bigg\|_{2}=\op(1),
\end{align*}
and therefore, as $T\rightarrow \infty$,
\begin{align*}
     \sup_{\psi\in\Theta, \vartheta\in\Xi_{\vartheta_{0}}}\|\EPn\Psi(\psi,\vartheta)-\EP\Psi(\psi,\vartheta)\|_{2}= \op(1),    
\end{align*}
by (\ref{eq:S9}). Similarly, we can obtain
\begin{align*}
    \sup_{\psi\in\Theta, \vartheta\in\Xi_{\vartheta_{0}}}\|\EPn \dot{\Psi}_{\psi}(\psi,\vartheta)-\EP\dot{\Psi}_{\psi}(\psi,\vartheta)\|_{2}= \op(1),
\end{align*}
as $T\rightarrow \infty$. 

\subsection{Proof of Lemma \ref{lem2}}

It is equivalent to show that 
\begin{align*}
    \frac{1}{T}\sum_{t=1}^{T}M_{t}(\psi_{0},\vartheta)\wc Z\in l^{\infty}(\Xi_{\vartheta_{0}}) \text{ as } T\rightarrow \infty,
\end{align*}
where $M_{t}(\psi_{0},\vartheta)=\Psi_{n(\bullet),t}(\psi_{0},\vartheta)-\mathbb{E}[ \Psi_{n(\bullet),t}(\psi_{0},\vartheta)|\mathcal{F}_{t-1}]\in\Real^{p}$ is a martingale difference sequence adapted to $\{\sigma(\mathcal{F}_{t})\}_{t\geq 1}$. Denote $j$th element for $M_{t}(\psi_{0},\vartheta)$ as $M_{t}^{(j)}(\psi_{0},\vartheta)$, i.e., $M_{t}(\psi_{0},\vartheta)=\{M_{t}^{(1)}(\psi_{0},\vartheta),\dots,M_{p}^{(j)}(\psi_{0},\vartheta) \}\Tra$ for $\psi_{0}\in\Real^{p}$. We utilize the Corollary 1 of \cite{bae2010}, the uniform central limited theorem for a martingale-difference sequence. It suffices to show that the following assumptions are satisfied:
\begin{assumption} \label{lemass1}
    $\Xi_{\vartheta_{0}}$ has uniformly integrable entropy.
\end{assumption}
\begin{assumption} \label{lemass2}
    For $j=1,\dots,p$, there exists a constant $L$ such that
    \begin{align*}
        P\bigg( \sup_{\vartheta,\vartheta'\in\Xi_{\vartheta_{0}}}\frac{\sum_{t=1}^{T}\mathbb{E}\big[ \{M_{t}^{(j)}(\psi_{0},\vartheta)-M_{t}^{(j)}(\psi_{0},\vartheta')\}^{2}|\mathcal{F}_{t-1} \big] }{T\|\vartheta-\vartheta'\|_{2,P}^{2}}\geq L \bigg)\rightarrow 0 \text{ as } T\rightarrow \infty.
    \end{align*}
\end{assumption}
\begin{assumption} \label{lemass3}
    As $T\rightarrow \infty$,
    \begin{align*}
        \frac{1}{T}\sum_{t=1}^{T}\mathbb{E}\{ M_{t}(\psi_{0},\vartheta)^{\otimes 2}|\mathcal{F}_{t-1}\}\cp \Sigma_{\vartheta} \text{ for each }\vartheta \in \Xi_{\vartheta_{0}},
    \end{align*}
    where $\Sigma_{\vartheta}$ is a positive definite matrix.
\end{assumption}
\begin{assumption} \label{lemass4}
    For every $\epsilon>0$ and $j=1,\dots,p$,
    \begin{align*}
        \frac{1}{T}\sum_{t=1}^{T}\mathbb{E}\big[\{M_{t}^{(j)}(\psi_{0},\boldsymbol{\vartheta})\}^{2}\times I_{M_{t}^{(j)}(\psi_{0},\boldsymbol{\vartheta})>\epsilon\sqrt{T}}  \big]\cp 0 \text{ as } T\rightarrow \infty,
    \end{align*}
    where $\boldsymbol{\vartheta}$ is an envelope function for $\vartheta \in \Xi_{\vartheta_{0}}$. 
\end{assumption}
Assumption \ref{lemass1} is satisfied by Condition \ref{ass:rc2} (ii), and Assumption \ref{lemass3} can be proved by the law of the large number of dependent non-identical samples given by Theorem 1 of \citep{andrew}, $T$ goes infinity. We provide additional proof for Assumption \ref{lemass2} (sufficient condition to the Lipschitz condition) and \ref{lemass4} (Lindeberg condition) based on the prespecified regularity conditions \ref{ass:rc1} - \ref{ass:rc3}.

Here is the proof for Assumption \ref{lemass2} first. Using Taylor expansion, we obtain
\begin{align*}
    \mathbb{E}\big[ \{M_{t}^{(j)}(\psi_{0},\vartheta)-M_{t}^{(j)}(\psi_{0},\vartheta')\}^{2}|\mathcal{F}_{t-1} \big]&=\mathbb{E}\bigg(\bigg[ \bigg\{ \frac{\partial M_{t}^{(j)}(\psi_{0},\vartheta)}{\partial \vartheta}\bigg|_{\vartheta=\tilde{\vartheta}}\bigg\}\Tra(\vartheta-\vartheta') \bigg]^{2}\bigg|\mathcal{F}_{t-1}\bigg)\\
    &=(\vartheta_{t}-\vartheta'_{t})\Tra \mathbb{E}\bigg[ \bigg\{ \frac{\partial M_{t}^{(j)}(\psi_{0},\vartheta)}{\partial \vartheta}\bigg|_{\vartheta=\tilde{\vartheta}}\bigg\}^{\otimes 2} \bigg|\mathcal{F}_{t-1}\bigg] (\vartheta_{t}-\vartheta'_{t})\\
    & \leq L \|\vartheta_{t}-\vartheta'_{t}\|_{2,P}^{2},
\end{align*}
where $\vartheta_{t}=\vartheta(\mathcal{F}_{t-1})$, $\vartheta'_{t}=\vartheta'(\mathcal{F}_{t-1})$, and $\tilde{\vartheta}$ is located between the $\vartheta$ and $\vartheta'$. The last inequality holds due to Condition \ref{ass:rc2} (v), i.e., uniform boundedness across $t$ and $j$ for $\vartheta\in\Xi_{\vartheta_{0}}$. Therefore,  
\begin{align*}
       \sup_{\vartheta,\vartheta'\in\Xi_{\vartheta_{0}}}\frac{\sum_{t=1}^{T}\mathbb{E}\big[ \{M_{t}^{(j)}(\psi_{0},\vartheta)-M_{t}^{(j)}(\psi_{0},\vartheta')\}^{2}|\mathcal{F}_{t-1} \big] }{T\|\vartheta-\vartheta'\|_{2,P}^{2}}&\leq L\sup_{\vartheta,\vartheta'\in\Xi_{\vartheta_{0}}}\frac{T^{-1}\sum_{t=1}^{T} \|\vartheta_{t}-\vartheta'_{t}\|_{2,P}^{2} }{\|\vartheta-\vartheta'\|_{2,P}^{2}},
\end{align*}
where $T^{-1}\sum_{t=1}^{T} \|\vartheta_{t}-\vartheta'_{t}\|_{2,P}^{2}\cp \|\vartheta-\vartheta'\|_{2,P}^{2}$ by the law of the large number of dependent non-identical samples given by Theorem 1 of \citep{andrew}, which completes the proof.

Lastly, we prove the Assumption \ref{lemass4}. By uniform boundedness for $\|\vartheta\|_{2}$ almost surely for all $\vartheta\in \Xi_{\vartheta_{0}}$ in Condition \ref{ass:rc2} (iv), we can choose a constant envelope function $\boldsymbol{\vartheta}$. Furthermore, $\|M_{t}^{(j)}(\psi_{0},\boldsymbol{\vartheta})\|_{2}$ is also bounded almost surely due to Condition \ref{ass:rc2}. Therefore,
\begin{align*}
    \frac{1}{T}\sum_{t=1}^{T}\mathbb{E}\big[\{M_{t}^{(j)}(\psi_{0},\boldsymbol{\vartheta})\}^{2}\times I_{M_{t}^{(j)}(\psi_{0},\boldsymbol{\vartheta})>\epsilon\sqrt{T}}  \big] \bbc P\big\{ T^{-1/2}M_{t}^{(j)}(\psi_{0},\boldsymbol{\vartheta})>\epsilon \big\}    \cp 0 \text{ as } T\rightarrow \infty,
\end{align*}
which completes the proof.

\section{Simulation}\label{SecB}

The simulation code for data generation, estimation, and the production of Tables and Figures is available at \url{https://github.com/tkh5956/JCI_2025_Hong}.

\subsection{Estimation guideline}

The estimation guideline described in Section \ref{subsec:id} is summarized below.

\begin{enumerate}
    \item \textbf{Estimate the Nuisance Functions} (i.e., intensity parameter $\gamma$, propensity $\pi$, and mean outcome function $h$): 
    \begin{enumerate}
        \item \textbf{Intensity}: Utilize the Cox proportional hazard model \citep{lin1989, therneau2015} to estimate the intensity parameter $\gamma$, based on Assumption \ref{ass:vis}.
        \item \textbf{Propensity}: Fit the model for $\pi$ using either parametric methods (e.g., logistic regression) or nonparametric methods (e.g., GAM \citep{hastie2017, wood2017}). Nonparametric methods are generally preferred for their flexibility.
        \item \textbf{Mean Outcome}: Employ nonparametric methods (e.g., GAM) to estimate the mean outcome function $h$, given its potentially complex and unknown structure. If there are sufficient samples with $A_{t}=0$ and a dense timeline, directly estimating $h_{k0}(\overline{L}_{i,t}) = \mathbb{E}(Y_{i,k,t}|\Olit, A_{t}=0)$ is recommended for better accuracy.
    \end{enumerate}
    \item \textbf{Predict Nuisance Functions}: Use the fitted models to predict the values of $\hat{\gamma}$, $\hat{\pi}$, and $\hat{h}$.
    \item \textbf{Plug-In Step}: Substitute the predicted values into 
    \[
    c(\Olit) = \frac{\partial m_{a}(\Olit;\psi)}{\partial \psi}
    \]
    and into the estimating equation (Equation \ref{estf_iiw}).
    \item \textbf{Solve the Estimating Equations}: Solve the equations derived in the previous step using gradient-based root-finding algorithms. For instance, the \texttt{multiroot} function from the R package \texttt{rootSolve} \citep{soetaert2014} can be used to achieve this.
\end{enumerate}

\subsection{Details on the generating expenditures}

We employ the conditional normal distribution to generate expenditures that satisfy not only the mean model but also both within-time and between-time correlations. For convenience, we skip the individual index $i$. Suppose we are interested in the within-time correlation and between-time correlations among the outcomes $Y_{t}=(Y_{1,t},\dots, Y_{4,t})\in\Real^{4}$, given by
\begin{align*}
    {\Sigma}_{w}=\left(\begin{array}{cccc}
        1 & 0.05 & -0.05 & 0.1 \\
        0.05 & 1 & 0.3 & 0.45\\
        -0.05 & 0.3 & 1 & 0.50\\
        0.1 & 0.45 & 0.50 & 1
    \end{array}\right),
\end{align*}
with a temporal correlation parameter ${\rho}=0.9$ for non-zero adjacent expenditure. To create non-zero $Y_{t+l}$ given non-zero $Y_{t}$, we utilize the following multivariate normal variables:
\begin{align*}
    \left(\begin{array}{c}
        Z_{t}^{y} \\
        Z_{t+l}^{y}
    \end{array}\right) \sim N\left\{\left(\begin{array}{c}
        \mu_{t}^{y} \\
        \mu_{t+l}^{y}
    \end{array}\right),\left(\begin{array}{cc}
        \Sigma_{w}^{y} & \rho^{y} \Sigma_{w}^{y} \\
        \rho^{y} \Sigma_{w}^{y} & \Sigma_{w}^{y}
    \end{array}\right)\right\},
\end{align*}
where $\mu_{t}^{y}$ and $\Sigma_{w}^{y}$ are in $\Real^{4}$ and $\Real^{4\times 4}$, respectively. We exponentiate $Z_{t}$ to obtain $Y_{t}=\exp(Z_{t}^{y})$. Given $Z_{t}^{y}$ and the corresponding $Y_{t}$, we can successively generate $Z_{t+1}^{y}$ and the corresponding $Y_{t+1}$ using the conditional multivariate normal distribution above. For category $k\in\{1,\dots,4\}$, we assume
\begin{align*}
  \mu_{kt}^{y}& = \mu^{y}_{k0}+ m_{0}^{y}(\Olit) \beta^{y}_{k0}+m_{a}(\Olit)\Tra \psi^{y}_{k0} a_{t},
\end{align*}
where the baseline mean outcome for $Z_{t}$ is $\mu^{y}_{0}=(1,1,1,1)\Tra$, $m_{0}^{y}(\Olit)$ is an arbitrary function not depending on the current intervention $A_{t}$, and $\beta_{0}^{y}=(-0.25,-0.25, -0.25,-0.25)\Tra$ is the corresponding parameter. We have set $m_{0}^{y}(\Olit)$ to be a scalar for simplicity, but it can be a vector like $m_{a}(\Olit)$ without affecting the conclusion of this simulation, unless it causes computational instability. The following useful facts apply:
\begin{align*}
    \mathbb{E}(Y_{t})&=\exp\bigg\{\mu_{t}^{y}+\frac{1}{2}D(\Sigma_{w}^{y})1_{4}\bigg\},\\
    Cov(Y_{t})&= \{\mathbb{E}(Y_{t})\mathbb{E}(Y_{t})\Tra\}\circ \{\exp(\Sigma_{w}^{y})-1_{4}1_{4}\Tra\},
\end{align*}
where $D(\cdot):\Real^{4\times 4}\rightarrow\Real^{4\times 4}$ keeps diagonal elements only, $\circ$ denotes the element-wise product (Hadamard product) in matrix algebra. We can derive the following equalities: 
\begin{align*}
    (\Sigma_{w}^{y})_{k,l}&=\log \bigg\{ ({\Sigma}_{w})_{k,l}\sqrt{\exp(\Sigma_{w}^{y})_{k,k}-1}\sqrt{\exp(\Sigma_{w}^{y})_{l,l}-1}+1 \bigg\}, \tag{C1} \label{Sim_C1} \\
    (\rho^{y} \Sigma_{w}^{y})_{k,l}&=\log \bigg\{ {\rho}({\Sigma}_{w})_{k,l}\sqrt{\exp(\Sigma_{w}^{y})_{k,k}-1}\sqrt{\exp(\Sigma_{w}^{y})_{l,l}-1}+1 \bigg\}, \tag{C2} \label{Sim_C2}
\end{align*}
which require initial values for the diagonal elements of $\Sigma_{w}^{y}$ to set the remaining correlations. In our simulation, $(\Sigma_{w}^{y})_{k,k}=|\mu^{y}_{k0}|/4=0.25$.

To introduce `zero-inflation' into the expenditures, we have created zero-inflated probabilities $1_{4}-p_{t}\in\Real^{4}$ as follows:
\begin{align*}
   \left(\begin{array}{c}
        Z_{t}^{p} \\
        Z_{t+l}^{p}
    \end{array}\right) \sim N\left\{\left(\begin{array}{c}
        \mu_{t}^{p} \\
        \mu_{t+l}^{p}
    \end{array}\right),\left(\begin{array}{cc}
        \Sigma_{w}^{p} & \rho^{p} \Sigma_{w}^{p} \\
        \rho^{p} \Sigma_{w}^{p} & \Sigma_{w}^{p}
    \end{array}\right)\right\},
\end{align*}
using the following marginal mean
\begin{align*}
    \mu_{kt}^{p}& = \mu^{p}_{k0}+ m^{p}_{0}(\Olit) \beta^{p}_{k0}+m_{a}(\Olit)\Tra \psi^{p}_{k0} a_{t},
\end{align*}
where the baseline mean outcome for $Z_{t}$ is $\mu^{p}_{0}=\{\log(0.3),\log(0.5),\log(0.7),\log(0.9)\}\Tra$, $m_{0}^{p}(\Olit)=m_{0}^{y}(\Olit)$ is a function not depending on the current intervention $A_{t}$, and its corresponding parameter is $\beta_{0}^{p}=(-0.05,0.05, -0.05,0.02)\Tra$. We assume $(\Sigma_{w}^{p})_{k,k}=|\mu_{k0}^{p}|/8$ and set the remaining correlations can be specified using (\ref{Sim_C1}) and (\ref{Sim_C2}). These settings are cautiously chosen to satisfy $p_{t}=\exp(Z_{t}^{p})\in (0,1)$ as we obtain $p_{t}=\exp(Z_{t}^{p})$ so that $1-p_{kt}$ is zero-inflation probability for category $k$. Recall that there is no restriction on the form of $m_{0}^{p}(\Olit)$ unless it destroys computational instability, i.e., it does not need to be equivalent to $m_{0}^{y}(\Olit)\in \Real$. We have set $m_{0}^{p}(\Olit)=(I_{\text{LkDn},t} E_{\text{Short},i,t-})/(1+ D_{\text{LkDn},t})$ so that it has a different form from $m_{0}^{y}(\Olit)=(1+ D_{\text{LkDn},t})/(I_{\text{LkDn},t} E_{\text{Short},i,t-}+1)$. 

\subsection{Additional simulation results}
Recall that we have set the following parameters:
\begin{equation*}\label{P1}
    \tag{P1}\left\{\begin{split}
    \psi^{p}_{0}&=(0.5,0.5,-0.5,-0.5,0.15,0.15,-0.15,-0.15,0,0)\Tra,\\
    \psi^{y}_{0}&=(-0.15,-0.15,0.15,0.15,-0.5,-0.5,0.5,0.5,0.35,0.35)\Tra,\\
    \psi_{0}&= \psi^{p}_{0}+ \psi^{y}_{0}=(0.35,0.35,-0.35,-0.35,-0.35,-0.35,0.35,0.35,0.35,0.35)\Tra.
    \end{split}\right\},
\end{equation*}
and
\begin{equation*}\label{P2}
    \tag{P2}\left\{\begin{split}
    \psi^{p}_{0}&=(0.15,0.15,-0.1,-0.1,0.05,0.05,-0.2,-0.2,0,0)\Tra,\\
    \psi^{y}_{0}&=(0.2,0.2,-0.25,-0.25,0.30,0.30,-0.15,-0.15,0.35,0.35)\Tra,\\
    \psi_{0}&= \psi^{p}_{0}+ \psi^{y}_{0}=(0.35,0.35,-0.35,-0.35,0.35,0.35,-0.35,-0.35,0.35,0.35)\Tra.
    \end{split}\right\}.
\end{equation*}
We have assigned coefficients with the same magnitude to $\psi_{0}$ for simplicity and fair comparison across the factors.

\begin{figure}[H]
   \includegraphics[width=5in]{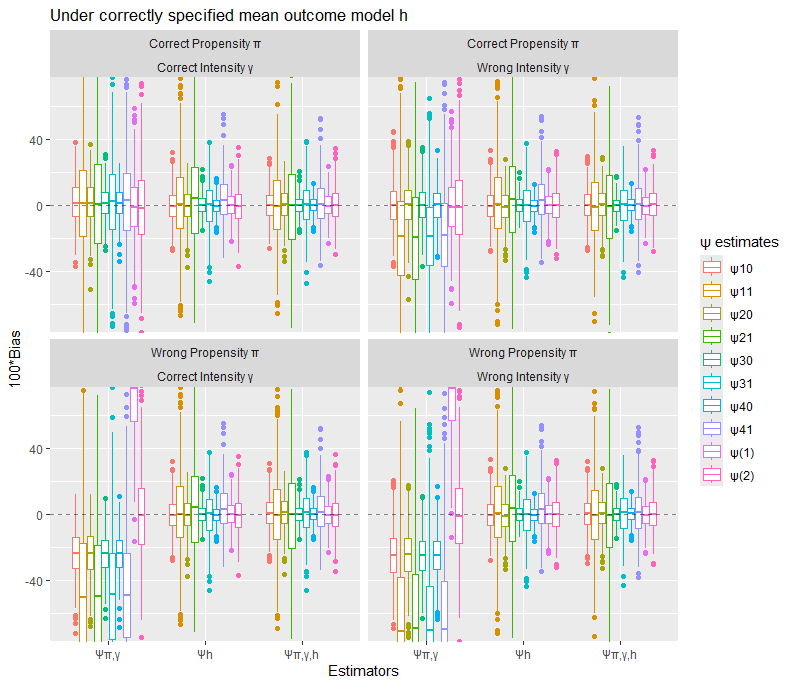} 
   \includegraphics[width=5in]{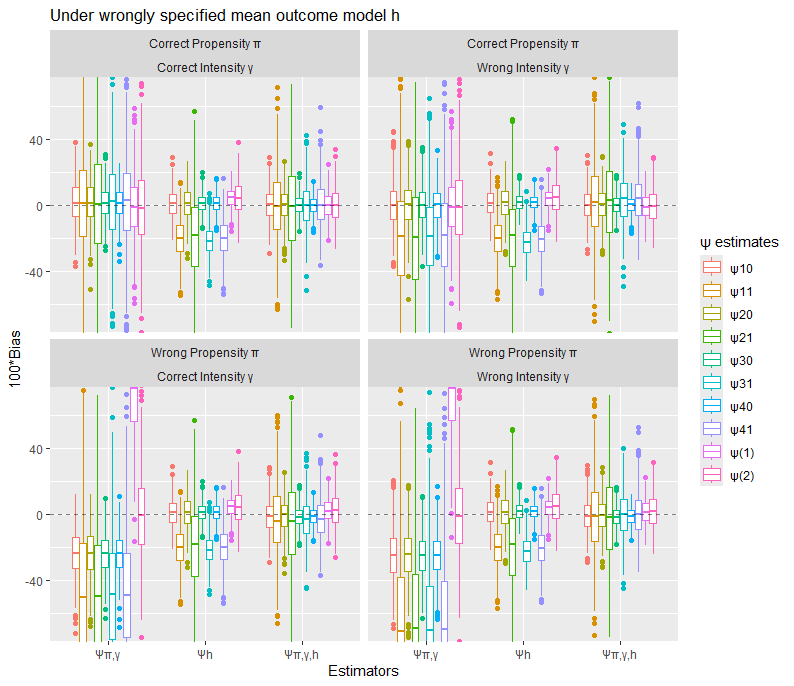} 
\caption{\label{simfig:revm1n200tau100} Boxplots of the estimators for (\ref{P1}), $n=200$ and $T=100$. 
Our proposed estimator is $\Psi_{\pi,\gamma,h}$. If both propensity model $\pi$ and intensity parameter $\gamma$ are correctly specified, $\Psi_{\pi,\gamma}$ and $\Psi_{\pi,\gamma,h}$ are consistent. If $h$ is correctly specified, $\Psi_{h}$ and $\Psi_{\pi,\gamma,h}$ are consistent.
}
\end{figure}

\begin{figure}[H]
   \includegraphics[width=5in]{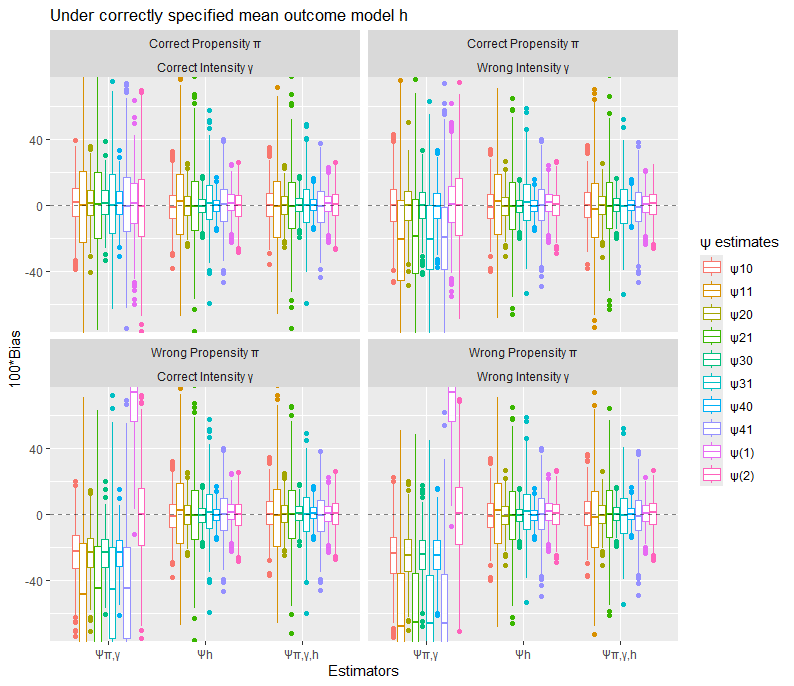} 
   \includegraphics[width=5in]{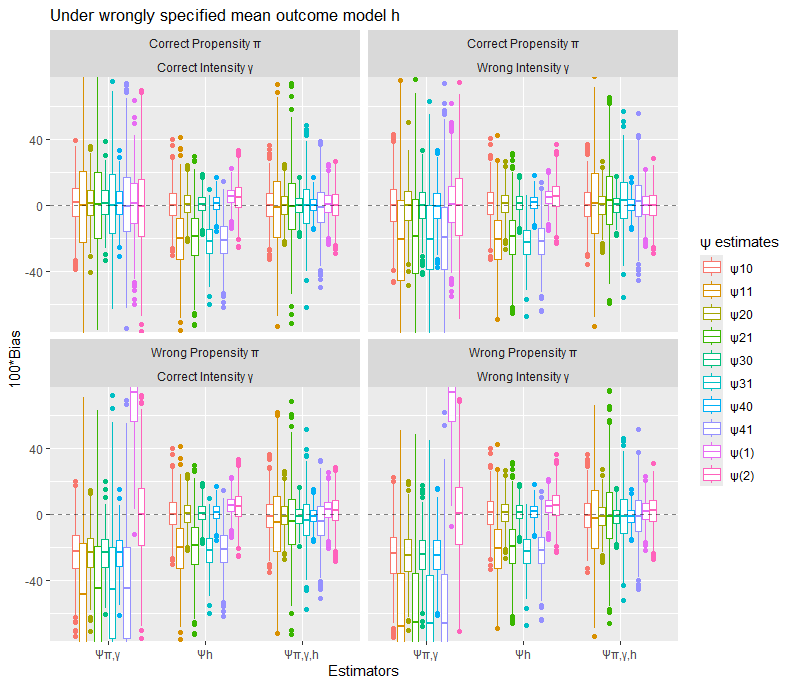} 
\caption{\label{simfig:revm2n200tau100} Boxplots of the estimators for (\ref{P2}), $n=200$ and $T=100$. 
Our proposed estimator is $\Psi_{\pi,\gamma,h}$. If both propensity model $\pi$ and intensity parameter $\gamma$ are correctly specified, $\Psi_{\pi,\gamma}$ and $\Psi_{\pi,\gamma,h}$ are consistent. If $h$ is correctly specified, $\Psi_{h}$ and $\Psi_{\pi,\gamma,h}$ are consistent.
}
\end{figure}

\begin{figure}[H]
   \includegraphics[width=5in]{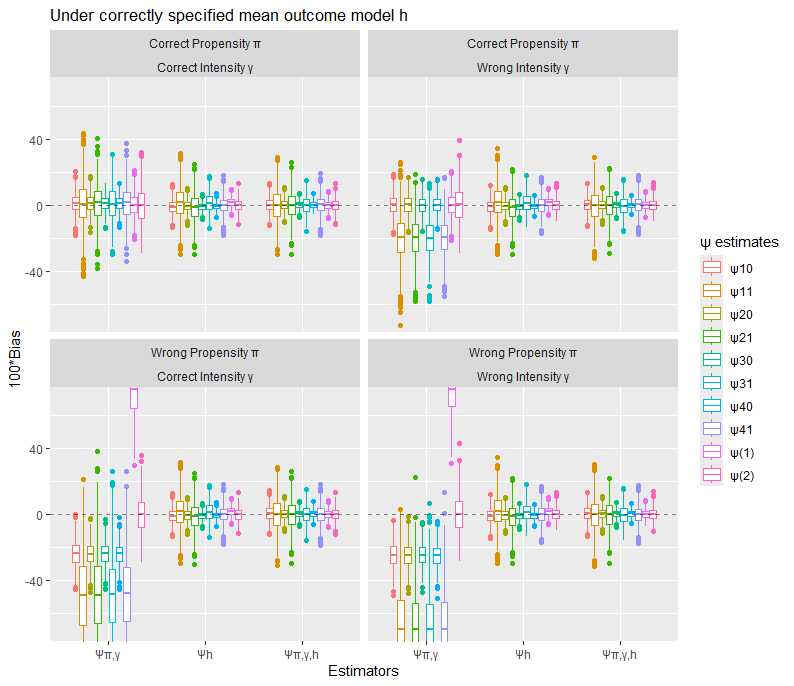} 
   \includegraphics[width=5in]{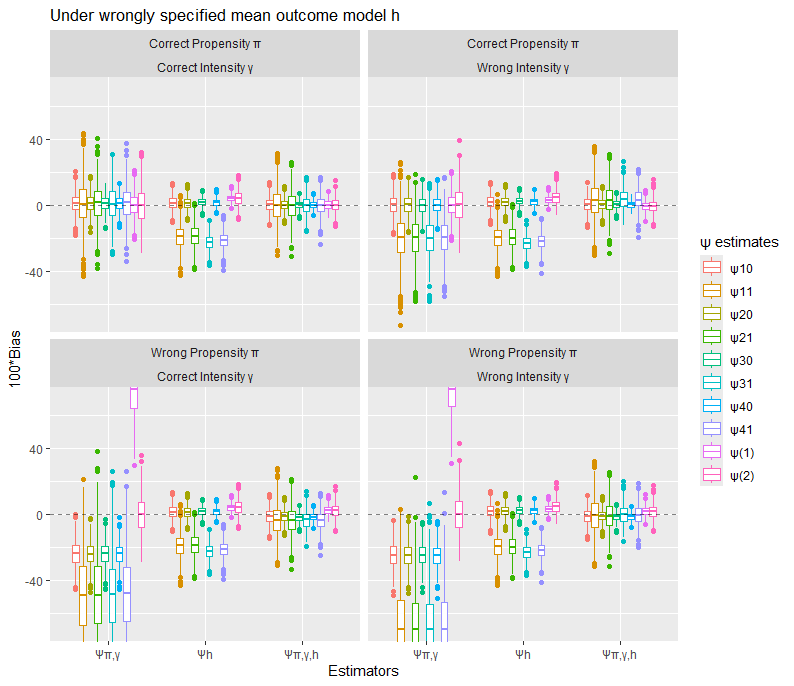} 
\caption{\label{simfig:revm2n600tau200} Boxplots of the estimators for (\ref{P2}), $n=600$ and $T=200$. 
Our proposed estimator is $\Psi_{\pi,\gamma,h}$. If both propensity model $\pi$ and intensity parameter $\gamma$ are correctly specified, $\Psi_{\pi,\gamma}$ and $\Psi_{\pi,\gamma,h}$ are consistent. If $h$ is correctly specified, $\Psi_{h}$ and $\Psi_{\pi,\gamma,h}$ are consistent. 
}
\end{figure}

You can find the detailed summary result corresponding to Figure \ref{simfig:revm1n600tau200} in Table \ref{sim:revm1n600tau200}. Due to space constraints, we cannot include the simulation results for $n=200$ and $T=100$ under (\ref{P1}) in the manuscript. Therefore, we have included the corresponding results (Table \ref{sim:revm1n200tau100} and Figure \ref{simfig:revm1n200tau100}) here. Note that under (\ref{P2}), where the signs of $\psi^{p}_{0}$ and $\psi^{y}_{0}$ are the same, or equivalently, the intervention affects the zero-inflation probability and the conditional mean outcome in the opposite direction. The corresponding simulation results are provided in Table \ref{sim:revm2n200tau100} - \ref{sim:revm2n600tau200} and Figure \ref{simfig:revm2n200tau100} - \ref{simfig:revm2n600tau200}. 
\begin{table}[H]
\caption{Simulation results for (\ref{P1}) under correctly/wrongly specified conditional mean outcome model $h(\Olit)$ for $n=200$, $T=100$ based on 1000 simulated datasets. Note: All figures are multiplied by $10^2$. Abbreviations: BIAS, Mean of the estimates minus true value; SSD, Sample standard deviation of the estimates; ESE, Mean of estimated standard error; CP, Empirical 95\% coverage probability.\label{sim:revm1n200tau100}}
\begin{center}
\scriptsize
\begin{tabular}{cc|ccc|ccc}
  \multicolumn{2}{c|}{\multirow{2}{*}{\shortstack{C=Correct\\ Specification}}} &  \multicolumn{1}{c}{$\Psi_{\pi,\gamma}$} & \multicolumn{1}{c}{$\Psi_{h}$} & \multicolumn{1}{c|}{$\Psi_{\pi,\gamma,h}$} &  \multicolumn{1}{c}{$\Psi_{\pi,\gamma}$} & \multicolumn{1}{c}{$\Psi_{h}$} & \multicolumn{1}{c}{$\Psi_{\pi,\gamma,h}$}   \\ 
  &  & BIAS(SSD) &  BIAS(SSD) & BIAS(SSD/ESE/CP) & BIAS(SSD) &  BIAS(SSD) & BIAS(SSD/ESE/CP) \\ 
  \cline{3-8}
 \multicolumn{2}{c|}{\multirow{2}{*}{\shortstack{W=Wrong\\ Specification}}}  & \multicolumn{3}{c|}{Correctly specified intensity parameter $\gamma_{C}$ } & \multicolumn{3}{c}{Wrongly specified intensity parameter $\gamma_{W}$ }\\
  \cline{3-8}
   &   & \multicolumn{6}{c}{Correctly specified mean conditional outcome model $h_{C}(\Olit)$}  \\
\hline
 \multirow{10}{*}{ $\pi_{C}$ } & $\psi_{10}$ & 1.4 (12.3) & -0.5 (9.6) & 0 (9.1/9.2/94.4) & 0 (13.4) & -0.6 (9.4) & -0.1 (9/9.2/94.9) \\ 
   & $\psi_{11}$ & 1.3 (31.6) & 1.7 (23.6) & -0.3 (21.5/20.8/94) & -19 (34.2) & 1.5 (23.4) & -0.7 (21.2/22.4/96.2) \\ 
   & $\psi_{20}$ & 1.8 (12.2) & -0.4 (9.6) & 0.5 (9.5/9.8/95.7) & 0.3 (13.4) & -0.8 (9.8) & 0.2 (9.8/10/95) \\ 
   & $\psi_{21}$ & -0.3 (37) & 2.6 (29.3) & -1.5 (28.6/27.3/93.9) & -20.4 (39.2) & 3.1 (28.7) & -1.6 (27.9/27.4/93.9) \\ 
   & $\psi_{30}$ & 1.4 (9.9) & 0.2 (5.7) & 0 (5.4/5.3/94) & 0 (11.4) & -0.1 (5.3) & -0.1 (5.1/5.1/94.5) \\ 
   & $\psi_{31}$ & 1.5 (25.8) & -0.4 (13.4) & 0.1 (13/12.3/93.2) & -18.8 (28.9) & -0.4 (13.5) & 0.1 (13.1/13.1/94.5) \\ 
   & $\psi_{40}$ & 1.3 (9.5) & -0.3 (5.3) & 0 (5.1/5/93.7) & -0.1 (10.9) & -0.6 (4.9) & -0.1 (4.8/4.8/93.8) \\ 
   & $\psi_{40}$ & 2.1 (26) & 3.3 (13) & 0.8 (13/12.9/94.5) & -18.2 (29.7) & 3.2 (13.3) & 0.5 (13.3/13.2/95) \\ 
   & $\psi_{(1)}$ & -0.7 (18.7) & -0.2 (7.6) & -0.3 (7.4/7.1/93.5) & -0.8 (18.4) & -0.1 (7.7) & -0.3 (7.4/7.2/94) \\ 
   & $\psi_{(2)}$ & -1.4 (25.3) & -1 (10.6) & -0.3 (10.1/9.7/94.3) & -0.8 (25.6) & -0.4 (10.3) & 0.1 (9.9/9.8/94.2) \\  \hline
   \multirow{10}{*}{ $\pi_{W}$ }  & $\psi_{10}$ & -23.7 (13.8) & -0.5 (9.6) & 0.3 (9.4/9.9/95.5) & -25 (14.9) & -0.6 (9.4) & 0.1 (9.2/10/96.9) \\ 
   & $\psi_{11}$ & -49.8 (44.6) & 1.7 (23.6) & 0 (21.9/20.3/92.8) & -70.2 (46.4) & 1.5 (23.4) & -0.3 (21.7/21.8/94.9) \\ 
   & $\psi_{20}$ & -23.3 (13.7) & -0.4 (9.6) & 0.9 (9.6/10.1/95.8) & -24.8 (15) & -0.8 (9.8) & 0.6 (9.7/10.3/95.3) \\ 
   & $\psi_{21}$ & -51.4 (49.6) & 2.6 (29.3) & -1.1 (28.9/27.1/92.9) & -71.6 (51.2) & 3.1 (28.7) & -1.3 (28.2/27/93.7) \\ 
   & $\psi_{30}$ & -23.7 (11.8) & 0.2 (5.7) & 0.1 (5.6/5.7/94.4) & -25.1 (13.2) & -0.1 (5.3) & -0.1 (5.2/5.5/96) \\ 
   & $\psi_{31}$ & -49.7 (40.2) & -0.4 (13.4) & 0.6 (13/11.5/91.6) & -70 (42) & -0.4 (13.5) & 0.5 (13.1/12.2/92.7) \\ 
   & $\psi_{40}$ & -23.8 (11.5) & -0.3 (5.3) & 0.3 (5.2/5.2/93.9) & -25.1 (12.9) & -0.6 (4.9) & 0.1 (4.8/5/94.8) \\ 
   & $\psi_{41}$ & -49.1 (40.4) & 3.3 (13) & 1.2 (12.9/12.2/94.4) & -69.4 (42.6) & 3.2 (13.3) & 0.8 (13.1/12.4/94.4) \\ 
   & $\psi_{(1)}$ & 75.7 (27.4) & -0.2 (7.6) & -0.6 (7.4/6.7/91.8) & 75.6 (27.3) & -0.1 (7.7) & -0.6 (7.5/6.7/91.8) \\ 
   & $\psi_{(2)}$ & -1.4 (25.7) & -1 (10.6) & -0.5 (10.3/9.8/94.4) & -0.7 (26.1) & -0.4 (10.3) & 0 (10.1/9.9/94) \\    \hline
   \multicolumn{2}{c|}{$\pi=$Propensity}  & \multicolumn{6}{c}{Wrongly specified mean conditional outcome model $h_{W}(\Olit)$}  \\
  \hline
   \multirow{10}{*}{ $\pi_{C}$ }  & $\psi_{10}$ & 1.4 (12.3) & 0.8 (8.1) & 0 (9.1/9.3/95.2) & 0 (13.4) & 1.2 (8.1) & 0.1 (9.1/9.4/95.4) \\ 
   & $\psi_{11}$  & 1.3 (31.6) & -20.3 (12.2) & -0.5 (21.4/21.4/95.7) & -19 (34.2) & -20.4 (12.5) & 3 (23.6/23.5/94.7) \\ 
   & $\psi_{20}$  & 1.8 (12.2) & 1 (9.3) & 0.4 (9.6/9.9/95.9) & 0.3 (13.4) & 1.4 (9.5) & 0.3 (9.9/10.1/95) \\ 
   & $\psi_{21}$  & -0.3 (37) & -19.7 (26.2) & -1.9 (28.5/27.7/94.2) & -20.4 (39.2) & -20.2 (25.4) & 1.6 (28.6/28.1/94.2) \\ 
   & $\psi_{30}$  & 1.4 (9.9) & 1.2 (5.5) & -0.1 (5.6/5.6/94.8) & 0 (11.4) & 1.7 (5.2) & 0 (5.4/5.4/94.8) \\ 
   & $\psi_{31}$  & 1.5 (25.8) & -21.8 (8.8) & -0.2 (13.3/13.4/94.3) & -18.8 (28.9) & -22.5 (8.9) & 3.5 (14.7/14.6/94.7) \\ 
   & $\psi_{40}$  & 1.3 (9.5) & 1.1 (5.1) & -0.1 (5.2/5.2/93.7) & -0.1 (10.9) & 1.6 (4.8) & 0 (5/5/93.8) \\ 
   & $\psi_{41}$  & 2.1 (26) & -20.2 (10.9) & 0.4 (13.1/13.5/95.3) & -18.2 (29.7) & -20.9 (11.1) & 3.9 (13.9/14.1/93.8) \\ 
   & $\psi_{(1)}$  & -0.7 (18.7) & 4.6 (6.1) & -0.2 (7.7/7.5/94.5) & -0.8 (18.4) & 3.7 (6) & -1.1 (7.9/7.6/93.2) \\ 
   & $\psi_{(2)}$ & -1.4 (25.3) & 4.1 (10.3) & -0.1 (10.4/10.2/94.7) & -0.8 (25.6) & 4.7 (10.1) & -0.5 (10.3/10.4/94.9) \\   \hline
   \multirow{10}{*}{ $\pi_{W}$ } & $\psi_{10}$ & -23.7 (13.8) & 0.8 (8.1) & -1.4 (9.4/10/95.7) & -25 (14.9) & 1.2 (8.1) & -1.2 (9.3/10.1/95.7) \\ 
   & $\psi_{11}$ & -49.8 (44.6) & -20.3 (12.2) & -3.8 (20.5/20.5/94.8) & -70.2 (46.4) & -20.4 (12.5) & -0.8 (22.5/22.4/95.2) \\ 
   & $\psi_{20}$ & -23.3 (13.7) & 1 (9.3) & -0.7 (9.6/10.2/96.2) & -24.8 (15) & 1.4 (9.5) & -0.8 (9.9/10.5/95.4) \\ 
   & $\psi_{21}$ & -51.4 (49.6) & -19.7 (26.2) & -5.4 (28.2/27.3/94) & -71.6 (51.2) & -20.2 (25.4) & -2.3 (28.1/27.4/94.3) \\ 
   & $\psi_{30}$ & -23.7 (11.8) & 1.2 (5.5) & -1.7 (5.8/5.9/95) & -25.1 (13.2) & 1.7 (5.2) & -1.6 (5.6/5.9/96.3) \\ 
   & $\psi_{31}$ & -49.7 (40.2) & -21.8 (8.8) & -3.3 (12.3/12.3/93.1) & -70 (42) & -22.5 (8.9) & -0.1 (13.5/13.4/93.8) \\ 
   & $\psi_{40}$ & -23.8 (11.5) & 1.1 (5.1) & -1.4 (5.3/5.4/94.7) & -25.1 (12.9) & 1.6 (4.8) & -1.3 (5.1/5.2/95.4) \\ 
   & $\psi_{41}$ & -49.1 (40.4) & -20.2 (10.9) & -3 (12.3/12.6/94.2) & -69.4 (42.6) & -20.9 (11.1) & 0.1 (13/13.1/95) \\ 
   & $\psi_{(1)}$ & 75.7 (27.4) & 4.6 (6.1) & 2 (7.1/7/93.7) & 75.6 (27.3) & 3.7 (6) & 1.2 (7.3/7/93.4) \\ 
   & $\psi_{(2)}$ & -1.4 (25.7) & 4.1 (10.3) & 2 (10.4/10.2/94.3) & -0.7 (26.1) & 4.7 (10.1) & 1.7 (10.2/10.4/94.3) \\  
   
\end{tabular}
\end{center}
\end{table}

\begin{table}[H]
\caption{Simulation results for (\ref{P1}) under correctly/wrongly specified conditional mean outcome model $h(\Olit)$ for $n=600$, $T=200$ based on 1000 simulated datasets. Note: All figures are multiplied by $10^2$. Abbreviations: BIAS, Mean of the estimates minus true value; SSD, Sample standard deviation of the estimates; ESE, Mean of estimated standard error; CP, Empirical 95\% coverage probability.\label{sim:revm1n600tau200}}
\begin{center}
\scriptsize
\begin{tabular}{cc|ccc|ccc}
  \multicolumn{2}{c|}{\multirow{2}{*}{\shortstack{C=Correct\\ Specification}}} &  \multicolumn{1}{c}{$\Psi_{\pi,\gamma}$} & \multicolumn{1}{c}{$\Psi_{h}$} & \multicolumn{1}{c|}{$\Psi_{\pi,\gamma,h}$} &  \multicolumn{1}{c}{$\Psi_{\pi,\gamma}$} & \multicolumn{1}{c}{$\Psi_{h}$} & \multicolumn{1}{c}{$\Psi_{\pi,\gamma,h}$}   \\ 
  &  & BIAS(SSD) &  BIAS(SSD) & BIAS(SSD/ESE/CP) & BIAS(SSD) &  BIAS(SSD) & BIAS(SSD/ESE/CP) \\ 
  \cline{3-8}
 \multicolumn{2}{c|}{\multirow{2}{*}{\shortstack{W=Wrong\\ Specification}}}  & \multicolumn{3}{c|}{Correctly specified intensity parameter $\gamma_{C}$ } & \multicolumn{3}{c}{Wrongly specified intensity parameter $\gamma_{W}$ }\\
  \cline{3-8}
   &   & \multicolumn{6}{c}{Correctly specified mean conditional outcome model $h_{C}(\Olit)$}  \\
\hline
 \multirow{10}{*}{ $\pi_{C}$ } & $\psi_{10}$ & 0.9 (5.6) & -0.4 (3.9) & 0.1 (3.8/3.8/94.1) & 0 (6.1) & -0.5 (3.9) & 0 (3.9/3.8/94.2) \\ 
   & $\psi_{11}$ & 0.8 (11.9) & 0.8 (9) & -0.3 (8.8/8.4/94.5) & -20.2 (13) & 0.9 (8.9) & -0.4 (8.7/9.1/96) \\ 
   & $\psi_{20}$ & 0.7 (5.5) & -0.9 (3.8) & 0 (3.8/4/96.8) & -0.1 (6.2) & -1 (3.9) & 0 (4/4.1/96.3) \\ 
   & $\psi_{21}$ & 1 (13.5) & 2.9 (10.8) & 0.1 (10.7/11.1/95.5) & -20 (14.6) & 2.9 (10.5) & -0.3 (10.4/11.2/95.9) \\ 
   & $\psi_{30}$ & 0.8 (4.5) & -0.1 (2.4) & 0 (2.3/2.3/94.6) & -0.1 (5.1) & -0.1 (2.3) & 0 (2.3/2.2/94.7) \\ 
   & $\psi_{31}$ & 0.9 (10) & -1.4 (5.2) & -0.2 (5.3/5/92.7) & -20.1 (11.3) & -1.5 (5.3) & -0.2 (5.3/5.4/94.4) \\ 
   & $\psi_{40}$ & 0.8 (4.4) & -0.6 (2.2) & 0 (2.2/2.2/95.1) & -0.1 (5.2) & -0.7 (2.2) & 0 (2.2/2.1/94.2) \\ 
   & $\psi_{41}$ & 1 (9.9) & 1.8 (5.4) & 0.1 (5.4/5.3/93.5) & -20 (11.5) & 1.8 (5.4) & -0.3 (5.4/5.4/95) \\ 
   & $\psi_{(1)}$ & 0.1 (6.8) & 0.7 (3) & 0.2 (3.1/2.9/93.3) & 0.1 (6.9) & 0.8 (3) & 0.2 (3/2.9/94.1) \\ 
   & $\psi_{(2)}$ & 0 (10.8) & -0.4 (4.1) & -0.3 (4.1/4.1/94.7) & 0.3 (11.2) & -0.2 (4.1) & -0.1 (4.1/4.1/95) \\   \hline
   \multirow{10}{*}{$\pi_{W}$}& $\psi_{10}$ & -24.1 (7) & -0.4 (3.9) & 0.5 (3.9/4/95.3) & -25 (7.4) & -0.5 (3.9) & 0.4 (3.9/4.1/95.6) \\ 
   & $\psi_{11}$ & -49.7 (24.4) & 0.8 (9) & -0.4 (8.9/8.2/93.7) & -70.7 (25) & 0.9 (8.9) & -0.5 (8.8/8.8/95.6) \\ 
   & $\psi_{20}$ & -24.2 (6.9) & -0.9 (3.8) & 0.3 (3.9/4.1/97.1) & -25 (7.5) & -1 (3.9) & 0.3 (4/4.2/97.3) \\ 
   & $\psi_{21}$ & -49.4 (25) & 2.9 (10.8) & 0.1 (10.7/10.9/95.1) & -70.5 (25.9) & 2.9 (10.5) & -0.3 (10.5/11/95.4) \\ 
   & $\psi_{30}$ & -24.2 (6.2) & -0.1 (2.4) & 0 (2.4/2.4/95.7) & -25 (6.6) & -0.1 (2.3) & -0.1 (2.3/2.4/95.8) \\ 
   & $\psi_{31}$ & -49.5 (23.5) & -1.4 (5.2) & -0.1 (5.3/4.7/90.7) & -70.5 (24.3) & -1.5 (5.3) & -0.1 (5.3/5/92.1) \\ 
   & $\psi_{40}$ & -24.2 (6.1) & -0.6 (2.2) & 0.2 (2.3/2.2/94.8) & -25 (6.7) & -0.7 (2.2) & 0.1 (2.2/2.2/95) \\ 
   & $\psi_{41}$ & -49.5 (23.6) & 1.8 (5.4) & 0.2 (5.4/5/92.4) & -70.5 (24.4) & 1.8 (5.4) & -0.2 (5.4/5.1/92.9) \\ 
   & $\psi_{(1)}$ & 75.6 (15.8) & 0.7 (3) & 0.1 (3/2.7/91.3) & 75.6 (15.7) & 0.8 (3) & 0.1 (3/2.7/91.4) \\ 
   & $\psi_{(2)}$ & 0.1 (11) & -0.4 (4.1) & -0.3 (4.2/4.1/94.9) & 0.3 (11.5) & -0.2 (4.1) & -0.2 (4.2/4.1/94.7) \\   \hline
   \multicolumn{2}{c|}{$\pi=$Propensity}  & \multicolumn{6}{c}{Wrongly specified mean conditional outcome model $h_{W}(\Olit)$}  \\
  \hline
   \multirow{10}{*}{$\pi_{C}$} & $\psi_{10}$  & 0.9 (5.6) & 1.8 (3.5) & 0.1 (3.9/3.9/94.9) & 0 (6.1) & 2.2 (3.5) & 0.3 (4/3.9/94.9) \\ 
   & $\psi_{11}$ & 0.8 (11.9) & -21 (4.8) & -0.5 (9/8.8/94.7) & -20.2 (13) & -21.3 (4.9) & 3.1 (9.9/9.8/94) \\ 
   & $\psi_{20}$ & 0.7 (5.5) & 1.4 (3.8) & 0 (3.9/4.1/96.5) & -0.1 (6.2) & 2 (3.9) & 0.2 (4.1/4.2/96.2) \\ 
   & $\psi_{21}$ & 1 (13.5) & -18.8 (9.9) & -0.2 (10.8/11.3/95.5) & -20 (14.6) & -19.4 (9.7) & 3.1 (10.8/11.6/94.8) \\ 
   & $\psi_{30}$ & 0.8 (4.5) & 2.2 (2.3) & 0 (2.4/2.4/95.4) & -0.1 (5.1) & 2.7 (2.2) & 0.2 (2.4/2.4/95.8) \\ 
   & $\psi_{31}$ & 0.9 (10) & -23.6 (3.6) & -0.3 (5.6/5.7/93.9) & -20.1 (11.3) & -24.1 (3.6) & 3.6 (6.1/6.3/92.9) \\ 
   & $\psi_{40}$ & 0.8 (4.4) & 2 (2.2) & 0 (2.3/2.3/94.2) & -0.1 (5.2) & 2.5 (2.2) & 0.2 (2.3/2.2/94) \\ 
   & $\psi_{41}$ & 1 (9.9) & -22 (4.6) & -0.3 (5.7/5.7/94.2) & -20 (11.5) & -22.6 (4.5) & 3.5 (6/6.1/90.3) \\ 
   & $\psi_{(1)}$ & 0.1 (6.8) & 4.1 (2.3) & 0.2 (3.2/3.1/93.8) & 0.1 (6.9) & 3.3 (2.3) & -0.7 (3.3/3.2/93.7) \\ 
   & $\psi_{(2)}$ & 0 (10.8) & 3.3 (4.3) & -0.1 (4.3/4.3/95.5) & 0.3 (11.2) & 3.7 (4.3) & -0.8 (4.3/4.4/95.3) \\  \hline
   \multirow{10}{*}{$\pi_{W}$} & $\psi_{10}$  & -24.1 (7) & 1.8 (3.5) & -1.7 (4/4.1/93.9) & -25 (7.4) & 2.2 (3.5) & -1.5 (4/4.2/94.4) \\ 
   & $\psi_{11}$ & -49.7 (24.4) & -21 (4.8) & -3.7 (8.6/8.5/91.6) & -70.7 (25) & -21.3 (4.9) & -0.5 (9.4/9.3/94.1) \\ 
   & $\psi_{20}$ & -24.2 (6.9) & 1.4 (3.8) & -1.3 (4/4.2/95.5) & -25 (7.5) & 2 (3.9) & -1.2 (4.1/4.3/95.4) \\ 
   & $\psi_{21}$ & -49.4 (25) & -18.8 (9.9) & -3.8 (10.6/11.1/94.1) & -70.5 (25.9) & -19.4 (9.7) & -0.8 (10.7/11.3/95.8) \\ 
   & $\psi_{30}$ & -24.2 (6.2) & 2.2 (2.3) & -2 (2.5/2.6/88.7) & -25 (6.6) & 2.7 (2.2) & -1.8 (2.5/2.6/88.9) \\ 
   & $\psi_{31}$ & -49.5 (23.5) & -23.6 (3.6) & -3.4 (5.3/5.2/89.1) & -70.5 (24.3) & -24.1 (3.6) & 0.1 (5.8/5.8/94.5) \\ 
   & $\psi_{40}$ & -24.2 (6.1) & 2 (2.2) & -1.7 (2.3/2.4/89.1) & -25 (6.7) & 2.5 (2.2) & -1.6 (2.3/2.3/90.1) \\ 
   & $\psi_{41}$ & -49.5 (23.6) & -22 (4.6) & -3.6 (5.4/5.3/88.8) & -70.5 (24.4) & -22.6 (4.5) & -0.2 (5.8/5.7/94.6) \\ 
   & $\psi_{(1)}$ & 75.6 (15.8) & 4.1 (2.3) & 2.5 (3/2.9/85.2) & 75.6 (15.7) & 3.3 (2.3) & 1.8 (3/2.9/89.7) \\ 
   & $\psi_{(2)}$ & 0.1 (11) & 3.3 (4.3) & 2 (4.3/4.3/92.8) & 0.3 (11.5) & 3.7 (4.3) & 1.6 (4.3/4.4/94.4) \\ 
   
\end{tabular}
\end{center}
\end{table}

\begin{table}[H]
\caption{Simulation results for (\ref{P2}) under correctly/wrongly specified conditional mean outcome model $h(\Olit)$ for $n=200$, $T=100$ based on 1000 simulated datasets. Note: All figures are multiplied by $10^2$. Abbreviations: BIAS, Mean of the estimates minus true value; SSD, Sample standard deviation of the estimates; ESE, Mean of estimated standard error; CP, Empirical 95\% coverage probability. \label{sim:revm2n200tau100}}
\begin{center}
\scriptsize
\begin{tabular}{cc|ccc|ccc}
  \multicolumn{2}{c|}{\multirow{2}{*}{\shortstack{C=Correct\\ Specification}}} &  \multicolumn{1}{c}{$\Psi_{\pi,\gamma}$} & \multicolumn{1}{c}{$\Psi_{h}$} & \multicolumn{1}{c|}{$\Psi_{\pi,\gamma,h}$} &  \multicolumn{1}{c}{$\Psi_{\pi,\gamma}$} & \multicolumn{1}{c}{$\Psi_{h}$} & \multicolumn{1}{c}{$\Psi_{\pi,\gamma,h}$}   \\ 
  &  & BIAS(SSD) &  BIAS(SSD) & BIAS(SSD/ESE/CP) & BIAS(SSD) &  BIAS(SSD) & BIAS(SSD/ESE/CP) \\ 
  \cline{3-8}
 \multicolumn{2}{c|}{\multirow{2}{*}{\shortstack{W=Wrong\\ Specification}}}  & \multicolumn{3}{c|}{Correctly specified intensity parameter $\gamma_{C}$ } & \multicolumn{3}{c}{Wrongly specified intensity parameter $\gamma_{W}$ }\\
  \cline{3-8}
   &   & \multicolumn{6}{c}{Correctly specified mean conditional outcome model $h_{C}(\Olit)$}  \\
\hline
 \multirow{10}{*}{ $\pi_{C}$ } & $\psi_{10}$ & 1.3 (13.1) & -1.2 (10.8) & 0.1 (10.6/10.3/94.4) & -0.2 (14.3) & -1.3 (10.9) & 0.1 (10.9/10.5/93.6) \\ 
   & $\psi_{11}$ & -1 (32.5) & 1.3 (26.9) & -2.3 (25.4/24.5/92.8) & -21.2 (35.8) & 1.3 (26.8) & -2.9 (25.2/25.8/95.1) \\ 
   & $\psi_{20}$ & 1 (11.6) & -0.7 (8.3) & -0.3 (8.1/7.9/93) & -0.8 (13.2) & -1 (8.3) & -0.6 (8.1/8/93.1) \\ 
   & $\psi_{21}$ & 0.3 (29.1) & -0.7 (21.7) & -0.6 (20.8/19.6/93.5) & -19.3 (31.7) & -0.4 (21.4) & -0.3 (20.4/20.2/94.2) \\ 
   & $\psi_{30}$ & 1.2 (10.4) & -0.8 (6.1) & 0 (5.9/5.6/93.3) & -0.5 (11.8) & -1 (5.8) & -0.1 (5.7/5.5/93.5) \\ 
   & $\psi_{31}$ & 0.6 (25.2) & 2 (15.3) & -0.4 (14.8/13.5/92.5) & -19.6 (28.9) & 1.9 (15.2) & -0.9 (14.6/14.2/93.7) \\ 
   & $\psi_{40}$ & 1.2 (9.8) & -0.5 (5.1) & 0 (4.9/4.9/94.3) & -0.5 (11.3) & -0.8 (4.9) & -0.2 (4.8/4.8/94.5) \\ 
   & $\psi_{41}$ & 0.5 (24.8) & 0.2 (14.2) & -0.8 (13.7/13.1/93.3) & -19.5 (28.5) & 0.3 (14.2) & -1 (13.6/13.3/93.5) \\ 
   & $\psi_{(1)}$ & 0.6 (17.6) & 1.3 (7.5) & 0.5 (7.2/6.8/92.6) & 0.5 (17.3) & 1.3 (7.5) & 0.4 (7.1/6.9/92.9) \\ 
   & $\psi_{(2)}$ & -1.4 (24.8) & -0.6 (9.6) & 0 (9.3/9/94.7) & -0.3 (25.1) & 0.1 (9.4) & 0.5 (9.1/9.1/94.6) \\   \hline
   \multirow{10}{*}{ $\pi_{W}$ } & $\psi_{10}$  & -23.4 (15) & -1.2 (10.8) & 0.4 (10.7/10.9/95.3) & -24.9 (16.1) & -1.3 (10.9) & 0.4 (10.8/11.2/95.6) \\ 
   & $\psi_{11}$ & -49.6 (46.2) & 1.3 (26.9) & -2 (25.6/24.1/92.6) & -69.8 (49.1) & 1.3 (26.8) & -2.7 (25.4/25.3/94.6) \\ 
   & $\psi_{20}$ & -23.7 (13.3) & -0.7 (8.3) & -0.1 (8.2/8.3/93.7) & -25.5 (14.8) & -1 (8.3) & -0.4 (8.1/8.4/94.4) \\ 
   & $\psi_{21}$ & -48.4 (43.1) & -0.7 (21.7) & -0.3 (21.1/19.1/92.7) & -68 (45.5) & -0.4 (21.4) & -0.1 (20.7/19.6/93) \\ 
   & $\psi_{30}$ & -23.5 (12.3) & -0.8 (6.1) & 0.3 (6/5.9/94.9) & -25.1 (13.6) & -1 (5.8) & 0.2 (5.7/5.9/95.5) \\ 
   & $\psi_{31}$ & -48.2 (41.1) & 2 (15.3) & -0.1 (14.9/12.8/91) & -68.3 (44) & 1.9 (15.2) & -0.6 (14.7/13.4/92.7) \\ 
   & $\psi_{40}$ & -23.5 (11.8) & -0.5 (5.1) & 0.4 (5.1/5.1/94.3) & -25.2 (13.2) & -0.8 (4.9) & 0.2 (4.8/5/94.9) \\ 
   & $\psi_{41}$ & -48.2 (40.6) & 0.2 (14.2) & -0.7 (13.8/12.5/92.1) & -68.2 (43.6) & 0.3 (14.2) & -0.9 (13.7/12.7/92) \\ 
   & $\psi_{(1)}$ & 74.9 (27.6) & 1.3 (7.5) & 0.2 (7.3/6.3/89.8) & 74.7 (27.4) & 1.3 (7.5) & 0.2 (7.3/6.4/90) \\ 
   & $\psi_{(2)}$ & -1.4 (25.1) & -0.6 (9.6) & -0.1 (9.4/9.1/94) & -0.2 (25.6) & 0.1 (9.4) & 0.4 (9.1/9.2/95.3) \\    \hline   \multicolumn{2}{c|}{$\pi=$Propensity}  & \multicolumn{6}{c}{Wrongly specified mean conditional outcome model $h_{W}(\Olit)$}  \\
  \hline
   \multirow{10}{*}{ $\pi_{C}$ } & $\psi_{10}$  & 1.3 (13.1) & 0.2 (9.9) & 0.1 (10.7/10.4/93.9) & -0.2 (14.3) & 0.8 (10.1) & 0.3 (10.9/10.6/94.2) \\ 
   & $\psi_{11}$ & -1 (32.5) & -20.3 (17.8) & -2.2 (25.4/24.9/93.8) & -21.2 (35.8) & -20.8 (17.9) & 0.9 (27/26.6/94) \\ 
   & $\psi_{20}$ & 1 (11.6) & 0.3 (7.7) & -0.2 (8.1/8.1/94.6) & -0.8 (13.2) & 0.8 (7.8) & -0.2 (8.2/8.3/94.7) \\ 
   & $\psi_{21}$ & 0.3 (29.1) & -19.6 (16.3) & -0.9 (20.9/20.4/94.2) & -19.3 (31.7) & -20 (16) & 2.8 (22/21.3/93.8) \\ 
   & $\psi_{30}$ & 1.2 (10.4) & 0.8 (5.7) & -0.1 (6.1/5.8/93.7) & -0.5 (11.8) & 1.4 (5.5) & 0 (5.9/5.8/94.5) \\ 
   & $\psi_{31}$ & 0.6 (25.2) & -21.9 (10.4) & -0.6 (14.9/14.1/92.8) & -19.6 (28.9) & -22.7 (10.4) & 2.8 (15.9/15.2/93.5) \\ 
   & $\psi_{40}$ & 1.2 (9.8) & 0.8 (5.1) & -0.1 (5.2/5.3/94.9) & -0.5 (11.3) & 1.4 (5) & 0 (5.1/5.2/95.3) \\ 
   & $\psi_{41}$ & 0.5 (24.8) & -21.1 (11.7) & -0.8 (14/14/93.9) & -19.5 (28.5) & -21.9 (11.6) & 2.7 (14.7/14.5/93.7) \\ 
   & $\psi_{(1)}$ & 0.6 (17.6) & 5.4 (5.6) & 0.5 (7.6/7.3/93.6) & 0.5 (17.3) & 4.4 (5.5) & -0.4 (7.7/7.4/94) \\ 
   & $\psi_{(2)}$ & -1.4 (24.8) & 4.7 (9.7) & 0 (9.7/9.6/93.7) & -0.3 (25.1) & 5.2 (9.5) & -0.4 (9.7/9.8/94.5) \\   \hline
   \multirow{10}{*}{ $\pi_{W}$ } & $\psi_{10}$  & -23.4 (15) & 0.2 (9.9) & -1.2 (10.8/11/95.1) & -24.9 (16.1) & 0.8 (10.1) & -0.9 (11/11.2/95.5) \\ 
   & $\psi_{11}$ & -49.6 (46.2) & -20.3 (17.8) & -5.6 (24.5/24.2/92.8) & -69.8 (49.1) & -20.8 (17.9) & -3 (25.9/25.7/94.2) \\ 
   & $\psi_{20}$ & -23.7 (13.3) & 0.3 (7.7) & -1.2 (8.2/8.5/95.2) & -25.5 (14.8) & 0.8 (7.8) & -1.3 (8.3/8.7/94.6) \\ 
   & $\psi_{21}$ & -48.4 (43.1) & -19.6 (16.3) & -4.5 (20.2/19.6/92.8) & -68 (45.5) & -20 (16) & -1.2 (21.2/20.4/93.9) \\ 
   & $\psi_{30}$ & -23.5 (12.3) & 0.8 (5.7) & -1.7 (6.2/6.1/93.1) & -25.1 (13.6) & 1.4 (5.5) & -1.5 (6/6.1/94.9) \\ 
   & $\psi_{31}$ & -48.2 (41.1) & -21.9 (10.4) & -3.8 (14/13.2/92) & -68.3 (44) & -22.7 (10.4) & -0.9 (14.8/14.1/93.5) \\ 
   & $\psi_{40}$ & -23.5 (11.8) & 0.8 (5.1) & -1.4 (5.3/5.4/94.4) & -25.2 (13.2) & 1.4 (5) & -1.3 (5.2/5.4/94.6) \\ 
   & $\psi_{41}$ & -48.2 (40.6) & -21.1 (11.7) & -4.2 (13.2/13.2/93.1) & -68.2 (43.6) & -21.9 (11.6) & -1.1 (13.8/13.6/94.4) \\ 
   & $\psi_{(1)}$ & 74.9 (27.6) & 5.4 (5.6) & 2.6 (7/6.7/92.1) & 74.7 (27.4) & 4.4 (5.5) & 1.8 (7/6.8/92.9) \\ 
   & $\psi_{(2)}$ & -1.4 (25.1) & 4.7 (9.7) & 2.1 (9.7/9.6/93.5) & -0.2 (25.6) & 5.2 (9.5) & 1.9 (9.6/9.8/94.5) \\ 
   
\end{tabular}
\end{center}
\end{table}

\begin{table}[H]
\caption{Simulation results for (\ref{P2}) under correctly/wrongly specified conditional mean outcome model $h(\Olit)$ for $n=600$, $T=200$ based on 1000 simulated datasets. Note: All figures are multiplied by $10^2$. Abbreviations: BIAS, Mean of the estimates minus true value; SSD, Sample standard deviation of the estimates; ESE, Mean of estimated standard error; CP, Empirical 95\% coverage probability.\label{sim:revm2n600tau200}}
\begin{center}
\scriptsize
\begin{tabular}{cc|ccc|ccc}
  \multicolumn{2}{c|}{\multirow{2}{*}{\shortstack{C=Correct\\ Specification}}} &  \multicolumn{1}{c}{$\Psi_{\pi,\gamma}$} & \multicolumn{1}{c}{$\Psi_{h}$} & \multicolumn{1}{c|}{$\Psi_{\pi,\gamma,h}$} &  \multicolumn{1}{c}{$\Psi_{\pi,\gamma}$} & \multicolumn{1}{c}{$\Psi_{h}$} & \multicolumn{1}{c}{$\Psi_{\pi,\gamma,h}$}   \\ 
  &  & BIAS(SSD) &  BIAS(SSD) & BIAS(SSD/ESE/CP) & BIAS(SSD) &  BIAS(SSD) & BIAS(SSD/ESE/CP) \\ 
  \cline{3-8}
 \multicolumn{2}{c|}{\multirow{2}{*}{\shortstack{W=Wrong\\ Specification}}}  & \multicolumn{3}{c|}{Correctly specified intensity parameter $\gamma_{C}$ } & \multicolumn{3}{c}{Wrongly specified intensity parameter $\gamma_{W}$ }\\
  \cline{3-8}
   &   & \multicolumn{6}{c}{Correctly specified mean conditional outcome model $h_{C}(\Olit)$}  \\
\hline
 \multirow{10}{*}{ $\pi_{C}$ } & $\psi_{10}$ & 1 (5.7) & -1 (4.3) & 0.1 (4.2/4.2/94.6) & 0.1 (6.3) & -1.1 (4.3) & 0 (4.3/4.3/95) \\ 
   & $\psi_{11}$ & 0.9 (13.1) & 1.5 (10.1) & -0.1 (9.9/9.8/95) & -19.9 (13.9) & 1.7 (10.1) & -0.4 (9.9/10.4/95.6) \\ 
   & $\psi_{20}$ & 1 (5.2) & -0.6 (3.5) & 0 (3.4/3.3/93.1) & 0.1 (5.9) & -0.6 (3.5) & 0 (3.5/3.3/93.6) \\ 
   & $\psi_{21}$ & 1.2 (11.8) & -1.1 (8.2) & 0.1 (8.2/7.9/93.3) & -19.8 (12.5) & -1.3 (8.1) & 0.1 (8.1/8.2/95.5) \\ 
   & $\psi_{30}$ & 1 (4.5) & -0.9 (2.3) & 0.1 (2.3/2.4/95.8) & 0.1 (5.1) & -0.9 (2.3) & 0.1 (2.3/2.3/95.3) \\ 
   & $\psi_{31}$ & 1 (10.3) & 1.1 (5.5) & 0 (5.4/5.5/95) & -19.9 (11.3) & 1.1 (5.5) & -0.4 (5.4/5.8/95.9) \\ 
   & $\psi_{40}$ & 0.9 (4.3) & -0.8 (2.1) & -0.1 (2.1/2.1/95.9) & 0 (5.1) & -0.8 (2) & -0.1 (2/2.1/95.7) \\ 
   & $\psi_{41}$ & 1.2 (10) & -0.3 (5.3) & 0.1 (5.3/5.3/93.7) & -19.7 (11.3) & -0.2 (5.4) & 0.1 (5.3/5.4/94.1) \\ 
   & $\psi_{(1)}$ & -0.1 (6.8) & 1.7 (2.8) & 0 (2.7/2.8/95.3) & 0 (6.7) & 1.7 (2.8) & 0 (2.7/2.8/95.3) \\ 
   & $\psi_{(2)}$ & -0.3 (10.6) & -0.1 (3.8) & 0 (3.8/3.8/94.5) & 0 (11) & 0 (3.8) & 0.1 (3.8/3.8/94.7) \\  \hline
   \multirow{10}{*}{ $\pi_{W}$ } & $\psi_{10}$  & -24 (7.2) & -1 (4.3) & 0.4 (4.3/4.5/95.2) & -25 (7.7) & -1.1 (4.3) & 0.3 (4.3/4.6/95.8) \\ 
   & $\psi_{11}$ & -49.4 (26.3) & 1.5 (10.1) & -0.3 (10/9.7/94.1) & -70.2 (26.6) & 1.7 (10.1) & -0.5 (10.1/10.1/94.7) \\ 
   & $\psi_{20}$ & -24.1 (6.9) & -0.6 (3.5) & 0.2 (3.5/3.4/94.2) & -25 (7.4) & -0.6 (3.5) & 0.1 (3.5/3.5/94.6) \\ 
   & $\psi_{21}$ & -49.1 (25.5) & -1.1 (8.2) & -0.1 (8.3/7.7/93) & -70.2 (25.9) & -1.3 (8.1) & -0.2 (8.2/8/94.3) \\ 
   & $\psi_{30}$ & -24.1 (6.3) & -0.9 (2.3) & 0.4 (2.4/2.5/96.4) & -25 (6.8) & -0.9 (2.3) & 0.3 (2.3/2.5/96.5) \\ 
   & $\psi_{31}$ & -49.3 (24.9) & 1.1 (5.5) & -0.2 (5.5/5.2/93.9) & -70.3 (25.3) & 1.1 (5.5) & -0.5 (5.5/5.4/95) \\ 
   & $\psi_{40}$ & -24.2 (6.3) & -0.8 (2.1) & 0.3 (2.1/2.2/95.1) & -25.1 (6.9) & -0.8 (2) & 0.2 (2/2.1/96.7) \\ 
   & $\psi_{41}$ & -49.1 (24.7) & -0.3 (5.3) & -0.2 (5.3/5/92.8) & -70.1 (25.2) & -0.2 (5.4) & -0.2 (5.3/5.1/93) \\ 
   & $\psi_{(1)}$ & 75.5 (16) & 1.7 (2.8) & 0.1 (2.8/2.6/93.3) & 75.6 (15.9) & 1.7 (2.8) & 0.1 (2.8/2.6/92.9) \\ 
   & $\psi_{(2)}$ & -0.2 (10.9) & -0.1 (3.8) & -0.1 (3.9/3.8/94.3) & 0.1 (11.4) & 0 (3.8) & 0.1 (3.8/3.8/94.8) \\  \hline   \multicolumn{2}{c|}{$\pi=$Propensity}  & \multicolumn{6}{c}{Wrongly specified mean conditional outcome model $h_{W}(\Olit)$}  \\
  \hline
   \multirow{10}{*}{ $\pi_{C}$ } & $\psi_{10}$  & 1 (5.7) & 1.1 (3.9) & 0.1 (4.3/4.3/95.1) & 0.1 (6.3) & 1.6 (3.9) & 0.3 (4.3/4.4/95.5) \\ 
   & $\psi_{11}$ & 0.9 (13.1) & -19.4 (7.1) & -0.4 (9.9/10.1/94.8) & -19.9 (13.9) & -19.7 (7) & 3.1 (10.8/10.8/95) \\ 
   & $\psi_{20}$ & 1 (5.2) & 1.2 (3.4) & 0.1 (3.6/3.4/93.6) & 0.1 (5.9) & 1.8 (3.4) & 0.3 (3.6/3.5/93.3) \\ 
   & $\psi_{21}$ & 1.2 (11.8) & -18.8 (6.7) & -0.1 (8.4/8.3/94.3) & -19.8 (12.5) & -19.5 (6.7) & 3.2 (8.8/8.8/93.2) \\ 
   & $\psi_{30}$ & 1 (4.5) & 1.8 (2.4) & 0.1 (2.5/2.5/95.6) & 0.1 (5.1) & 2.3 (2.4) & 0.3 (2.5/2.5/95.6) \\ 
   & $\psi_{31}$ & 1 (10.3) & -22.7 (4.2) & -0.3 (5.5/6/96) & -19.9 (11.3) & -23.3 (4.2) & 3.5 (6/6.5/93.7) \\ 
   & $\psi_{40}$ & 0.9 (4.3) & 1.5 (2.2) & 0 (2.2/2.3/95.8) & 0 (5.1) & 2.1 (2.2) & 0.2 (2.2/2.3/95.8) \\ 
   & $\psi_{41}$ & 1.2 (10) & -21.2 (4.8) & -0.1 (5.7/5.9/95.3) & -19.7 (11.3) & -22 (4.7) & 3.5 (6/6.2/91.9) \\ 
   & $\psi_{(1)}$ & -0.1 (6.8) & 4 (2.2) & 0 (2.9/3/95.7) & 0 (6.7) & 3.2 (2.2) & -0.8 (2.9/3.1/94.5) \\ 
   & $\psi_{(2)}$ & -0.3 (10.6) & 4.2 (4.1) & 0.1 (4.1/4/94.8) & 0 (11) & 4.4 (4.1) & -0.7 (4.2/4.1/94.1) \\  \hline
   \multirow{10}{*}{ $\pi_{W}$ } & $\psi_{10}$  & -24 (7.2) & 1.1 (3.9) & -1.5 (4.3/4.5/94.4) & -25 (7.7) & 1.6 (3.9) & -1.4 (4.4/4.6/95.1) \\ 
   & $\psi_{11}$ & -49.4 (26.3) & -19.4 (7.1) & -3.7 (9.7/9.8/92.8) & -70.2 (26.6) & -19.7 (7) & -0.6 (10.5/10.4/94.6) \\ 
   & $\psi_{20}$ & -24.1 (6.9) & 1.2 (3.4) & -1.2 (3.6/3.5/92.3) & -25 (7.4) & 1.8 (3.4) & -1.1 (3.7/3.6/93.2) \\ 
   & $\psi_{21}$ & -49.1 (25.5) & -18.8 (6.7) & -3.7 (8.1/8/92.8) & -70.2 (25.9) & -19.5 (6.7) & -0.8 (8.5/8.4/94.6) \\ 
   & $\psi_{30}$ & -24.1 (6.3) & 1.8 (2.4) & -1.9 (2.5/2.6/89) & -25 (6.8) & 2.3 (2.4) & -1.7 (2.5/2.6/91) \\ 
   & $\psi_{31}$ & -49.3 (24.9) & -22.7 (4.2) & -3.5 (5.3/5.6/90.8) & -70.3 (25.3) & -23.3 (4.2) & -0.2 (5.7/6/95) \\ 
   & $\psi_{40}$ & -24.2 (6.3) & 1.5 (2.2) & -1.7 (2.3/2.4/90.6) & -25.1 (6.9) & 2.1 (2.2) & -1.5 (2.2/2.4/91.7) \\ 
   & $\psi_{41}$ & -49.1 (24.7) & -21.2 (4.8) & -3.6 (5.5/5.5/90.3) & -70.1 (25.2) & -22 (4.7) & -0.4 (5.8/5.8/93.4) \\ 
   & $\psi_{(1)}$ & 75.5 (16) & 4 (2.2) & 2.4 (2.7/2.7/87.7) & 75.6 (15.9) & 3.2 (2.2) & 1.6 (2.7/2.8/91.9) \\ 
   & $\psi_{(2)}$ & -0.2 (10.9) & 4.2 (4.1) & 2.2 (4.2/4/90) & 0.1 (11.4) & 4.4 (4.1) & 1.7 (4.2/4.1/92.3) \\  
   
\end{tabular}
\end{center}
\end{table}

In conclusion, all the simulation results align with the manuscript, indicating that the proposed estimator, which is rate double robust, meaning it is consistent if either $h(\Olit)$ or $\{\pi(\overline{G}_{t}),\gamma\}$ is correctly specified, is the best choice across all simulation scenarios.

\section{Application}
\label{SecC}

\subsection{Google mobility data}
\begin{figure}[H]
\begin{center}
       \includegraphics[width=5in]{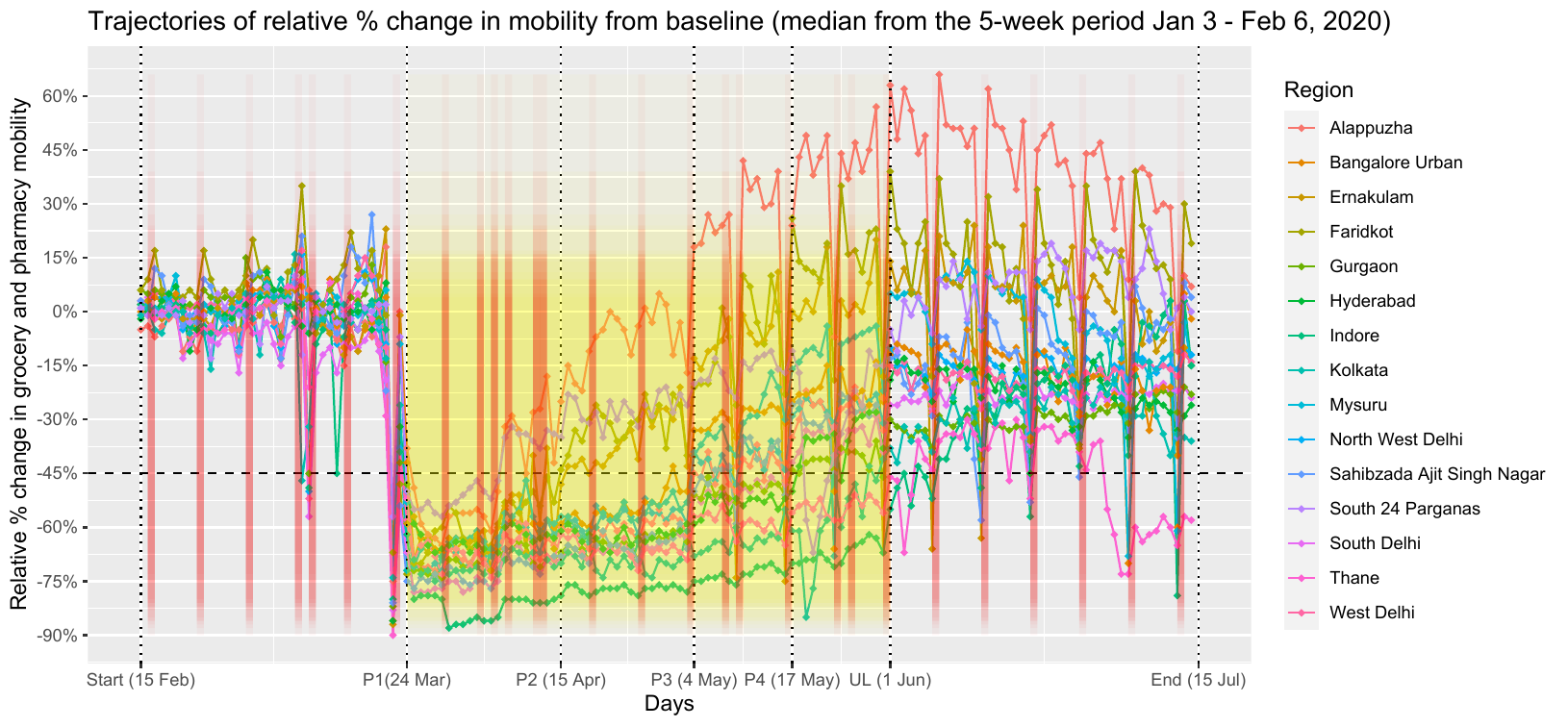}
\end{center}
\caption{\label{fig:gmob} The Google mobility data corresponding to Figure \ref{fig:mob}. The yellow-shaded region corresponds to the nationwide lockdown in India, declared on the evening of March 24th and subsequently extended through multiple phases until May 31st in 2020. In contrast, the blue-shaded region indicates periods characterized by a reduction in mobility exceeding 45$\%$, denoted as $A_{t}=1$. The red-shaded areas are gazetted holidays (10 Mar, Holi; 2 Apr, Ram Navami; 6 Apr, Mahavir Jayanti; 10 Apr, Good Friday; 7 May, Budha Purnima; 25 May, Id-ul-Fitr) and Sundays during the research period. Abbreviations: P1-P4, gradual phases in India's nationwide lockdown in 2020; UL, unlock phase.
}
\end{figure}

We utilize the Google Community Mobility Reports (CMRs) to consider mobility as our treatment. This data presents movement trends compared to the pre-COVID baseline (median from the 5-week period Jan 3 - Feb 6, 2020) across different regions in five categories: (i) percentage change in the total number of visitors in retail (grocery and pharmacy), parks, workplaces, transit, and (ii) percentage change in residential duration; refer to Figure \ref{fig:gmob}. The regular spikes on weekends are worth noting. As our focus is on analyzing market transactions, we find it sufficient to use the relative mobility change in grocery and pharmacy. A comprehensive overview of the data is available at the following link: (https://support.google.com/covid19-mobility/answer/9824897?hl=en).

\subsection{Extra details on the Indian transaction data}
\label{app:data}

From the transaction data (Feb 15th - Jul 15th), we have identified 22,666 eligible consumers with complete and reliable data on their gender information, who made significant expenditures before the research commencement, i.e., January 1st to February 14th, 2020. There are no overlapping occurrences across different regions. We have not considered age and marital status due to substantial missing data and unreliability. It's important to note that we excluded a single consumer with exceptionally high expenditures (exceeding $y>50,000$) in a single category, as such outliers can significantly affect the suitability of the nonparametric model. Our analysis is limited to customers with reliable gender information who visited the markets at least once from January 1st to February 14th before the start of our research.

The data includes transaction attributes such as market region, customer ID, transaction timing, and billing details for specific item categories, including general merchandise (consumer durables, IT devices, fashion items), non-food groceries, food groceries, fresh products, super fresh items, staples, and others (coupons, gift cards). The categories are defined internally within the specified market chain; we highlight notable subclasses within each category.
\begin{enumerate}
    \item \textbf{General merchandise}: LCD monitor, Bluethooth speaker, Shirts, Plastic container, Curtain, etc.
    \item \textbf{Non-food groceries}: Skin care products, Deodorant, Hair accessory, Floor cleaner, Sunscreen, etc.  
    \item \textbf{Food groceries}: Snack, Sandwich, Cookie, Tea, Jam, etc.
    \item \textbf{Fresh products}: Corn, Carrots,  Cucumbers, Oats, Brown rice, Barley, etc.
    \item \textbf{Super fresh items}: Avocados, Blueberries, Kale, Spinach, Beet greens, Fish, Collard greens, etc.
    \item \textbf{Staples}: Spices, Masala powder, Dry fruits, Flours, Cereals, Processed nuts, etc.
    \item \textbf{Others}: Saving coupons, Gift items,  Packing material, etc.
\end{enumerate}
Note that ``super fresh" foods typically refer to produce harvested at peak ripeness, ensuring optimal quality and nutrient density. This designation often emphasizes a swift transition from harvest to sale. In contrast, ``fresh" foods generally denote produce that is unprocessed or unpreserved but may not necessarily be at the peak of freshness.

Visits made on the same day are aggregated for analysis. The study incorporates temporal variables such as the nationwide lockdown in India from March 24th to May 31st \citep{Jeffrey2020}, as well as gazetted holidays with Sundays during the research period. 
\begin{figure}[H]
\begin{center}
\includegraphics[width=5.3in]{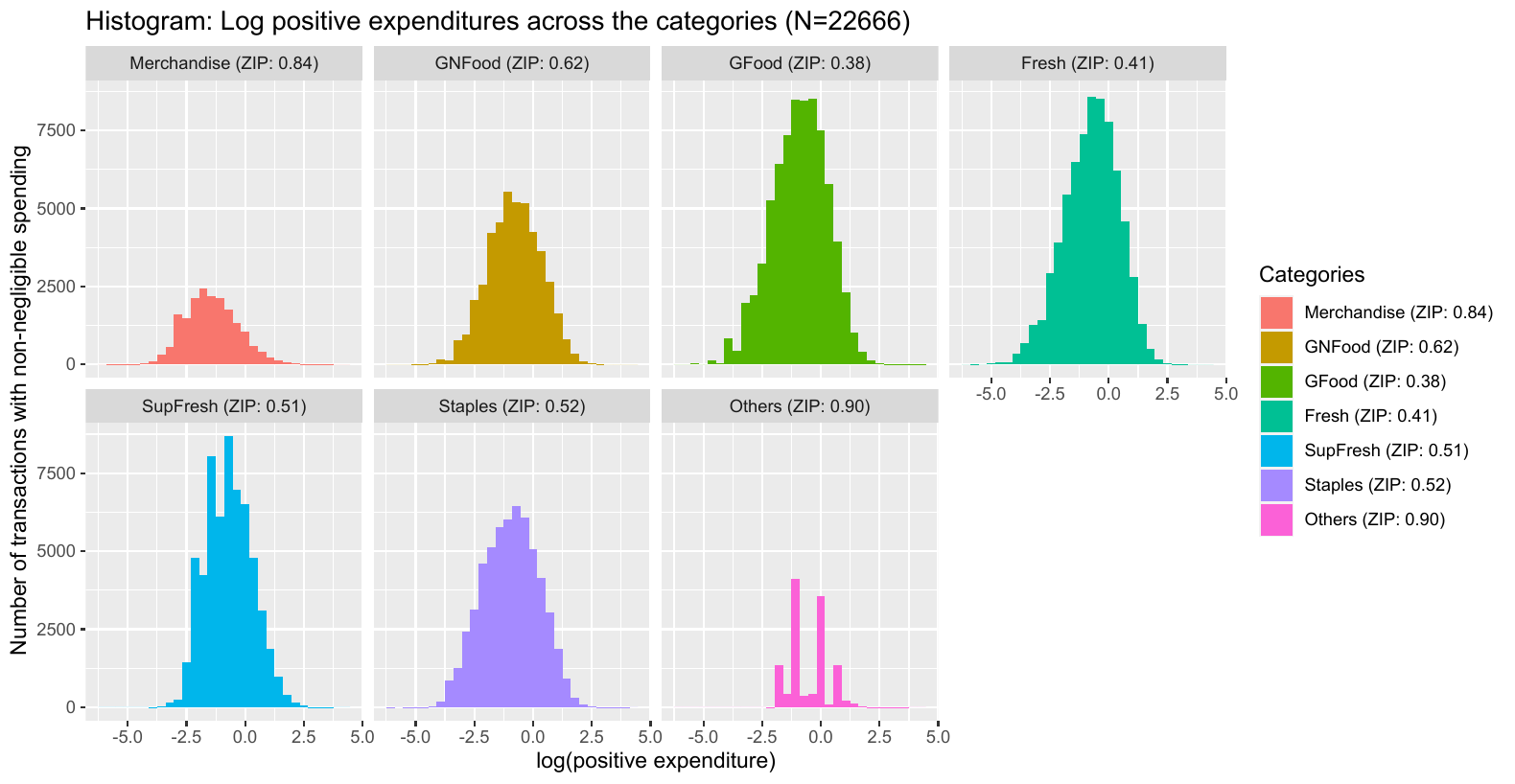}
\end{center}
\caption{\label{data:lognorm} Log normal modeling on the expenditures across the categories. Abbreviations: Merchandise, General merchandise; GNFood, non-food grocery; GFood, food grocery; Fresh, fresh food; SupFresh, super fresh food; ZIP, sample zero-inflation probability of the given category.}
\end{figure}

\begin{figure}[H]
\begin{center}
\includegraphics[width=5.3in]{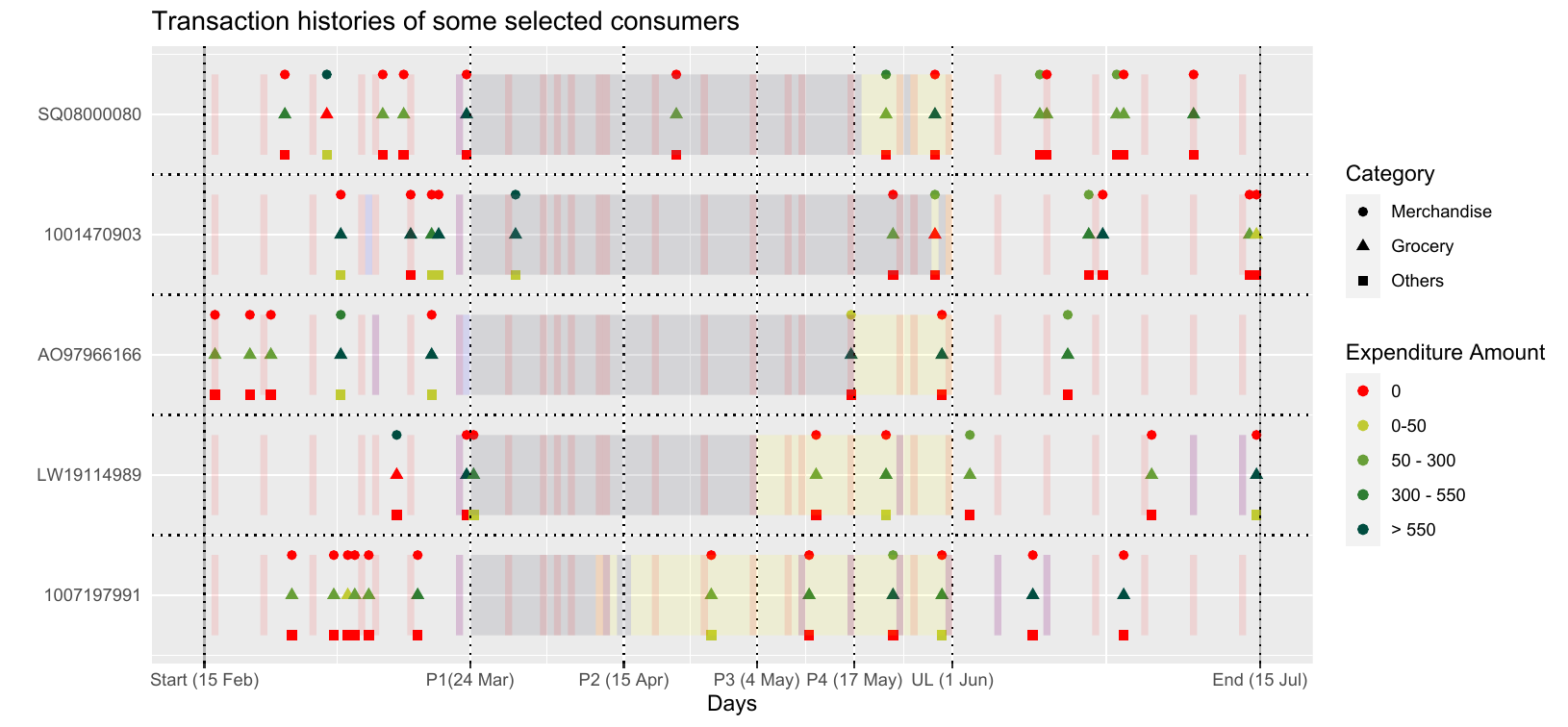}
\end{center}
\caption{ \label{data:trans} Visualization of the transaction histories of five consumers during the year 2020. Yellow and red-shaded areas are defined in Figure \ref{fig:mob}. The blue-shaded region indicates periods characterized by a reduction in mobility exceeding 45$\%$, denoted as $A_{t}=1$. The gray-shaded area represents the overlap between the yellow and blue-shaded intervals. Abbreviations: P1-P4, gradual phases in India's nationwide lockdown in 2020; UL, unlock phase.}
\end{figure}
During the initial exploratory data analysis, two key characteristics of the expenditure data were identified. Firstly, there is a notable presence of zero-inflated probabilities within the expenditures, with varying degrees of inflation across categories. Secondly, the log-normal modeling approach for positive expenditures is generally applicable, as depicted in Figure \ref{data:lognorm}. It is worth noting that positive expenditures within the  `Others' category may not adhere to a log-normal distribution; however, we intend to include `Others' category expenditures in our analysis due to the potential for undisclosed relationships between spending behaviors across different item categories.

Figure \ref{data:trans} illustrates the observed shopping behaviors of five selected consumers across diverse regions. Notably, nearly half of the data points are marked in red, indicating instances where consumers refrained from making purchases in specific categories. The presence of a nationwide lockdown, depicted in yellow shading, also seems to have negatively impacted consumer store visits. However, it is important to note that the blue shading, representing significant mobility reduction, provides a more accurate reflection of consumer visiting patterns. This is evident from the substantial decline in consumer visits at the onset of the lockdown, followed by a gradual recovery over time. These shifts in visitation patterns are closely linked to time-varying variables, including changes in mobility trends.

\subsection{Intensity estimation result}
\label{app:int}

\begin{table}[H]
\setlength\tabcolsep{2pt} 
\caption{Intensity parameter $\gamma$, their standard errors, and the corresponding p-value with stars. Note: *** if the p-value is less than 0.001, ** if it is less than 0.01, * if it is less than 0.05, and . if it is less than 0.1.\label{tab:Int}}
\begin{center}
\scriptsize
\begin{tabular}{ccccccccc}
  \multicolumn{9}{c}{Intensity Parameter $\gamma$ Estimation}\\
  \hline
    \multicolumn{9}{c}{Covariates} \\
      & $\hat{\gamma}$ (SE) & p-value & & $\hat{\gamma}$ (SE) & p-value & & $\hat{\gamma}$ (SE) & p-value  \\
  \hline
  $I_{\text{BLR Urban}}$ & -0.04 (0.04) & 0.368 () &
  $I_{\text{Ernakulam}}$ & 0.06 (0.05) & 0.233 () &
  $I_{\text{Faridkot}}$ & -0.36 (0.04) & $<$0.001 (***) \\ 
  $I_{\text{Gurgaon}}$ & 0.24 (0.04) & $<$0.001 (***)  & 
  $I_{\text{Hyderabad}}$ & -0.59 (0.03) & $<$0.001 (***) & 
  $I_{\text{Indore}}$ & -1.11 (0.05) & $<$0.001 (***) \\ 
  $I_{\text{Kolkata}}$ & 0.19 (0.12) & 0.118 ()& 
  $I_{\text{Mysuru}}$ & -0.36 (0.04) & $<$0.001 (***) &
  $I_{\text{NW Delhi}}$ & -0.62 (0.05) & $<$0.001 (***) \\ 
  $I_{\text{SAS Nagar}}$ & -0.29 (0.05) & $<$0.001 (***) &
  $I_{\text{24 PGS (S)}}$ & 0.25 (0.11) & 0.027 (*) & 
  $I_{\text{S Delhi}}$ & 0.18 (0.04) & $<$0.001 (***) \\ 
  $I_{\text{Thane}}$ & -0.19 (0.07) & 0.008 (**) &
  $I_{\text{W Delhi}}$ & -0.57 (0.08) & $<$0.001 (***) &
  $R(\overline{A}_{t-})$ & -0.03 (0.03) & 0.364 ()  \\ 
    $I_{\text{Male},i}$ & 0.12 (0.02) & $<$0.001 (***) &
  $E_{\text{Long},i,t-}$ & 0.01 (0.01) & 0.102 () & 
  $E_{\text{Short},i,t-}$ & 0.26 (0.03) & $<$0.001 (***) \\  
  \hline
    \multicolumn{9}{c}{Intervention and Modifiers} \\
      & $\hat{\gamma}$ (SE) & p-value & & $\hat{\gamma}$ (SE) & p-value & & $\hat{\gamma}$ (SE) & p-value  \\
  \hline
  Intercept & -0.22 (0.07) & 0.002 (**) &
  $\times I_{\text{Hday},t}$ & 0.37 (0.04) & $<$0.001 (***) &
  $\times D_{\text{LkDn},t}$ & -1.47 (0.09) & $<$0.001 (***) \\ 
  $\times I_{\text{Hm}}$ & 0.5 (0.03) & $<$0.001 (***) &
  $\times R(\overline{A}_{t-})$ & 0.19 (0.06) & 0.001 (**) &  
  $\times I_{\text{LkDn},t}$ & 0.68 (0.06) & $<$0.001 (***) \\ 
    $\times I_{\text{Male},i}$ & 0.07 (0.04) & 0.051 (.) & 
  $\times E_{\text{Long},i,t-}$ & 0.04 (0.01) & 0.009 (**) &  
  $\times E_{\text{Short},i,t-}$ & -0.56 (0.05) & $<$0.001 (***)  \\ 
\end{tabular}
\end{center}
\end{table}

Table \ref{tab:Int} presents estimates for intensity and causal models. Note that these estimates do not have causal interpretation as we have not set any causal assumptions on them; however, they may provide us some insights regarding the level of association between the factors and the visiting moments. Negative coefficients in the intensity estimation indicate a decrease in visit frequency or a delay in the next visit. Each region has a unique average visiting pattern compared to the default region, `Alappuzha.' For example, consumers in Gurgaon, South 24 Parganas, and South Delhi tend to visit stores more frequently, while some other regions visit less frequently. Individualized variables also impact consumers' visiting patterns, with males and those who failed to purchase a variety of categorical items in the previous shopping tending to visit the shop more frequently. However, the average spending in all past visits is not a significant factor associated with shopping frequency, suggesting that only short-term consumer behavior influences visiting patterns. The intervention is associated with a reduction in shopping frequency, although the level diminishes for holidays. This trend intensifies as time progresses following the initiation of the nationwide lockdown, indicating a further decrease in average shopping frequency. However, continuous mobility reduction in the area, aligned with the nationwide lockdown, tends to be associated with attracting more frequent visits to hypermarkets.

\end{document}